\documentclass{pasj00}

\usepackage{lscape}

\begin{document}
\SetRunningHead{Akiyama et al.}{X-ray Selected AGNs in SXDS}

\title{THE SUBARU - XMM-NEWTON DEEP SURVEY (SXDS) VIII.: MULTI-WAVELENGTH
IDENTIFICATION, OPTICAL/NIR SPECTROSCOPIC PROPERTIES, AND PHOTOMETRIC REDSHIFTS OF X-RAY SOURCES\thanks{Based in part on data collected at Subaru Telescope, which is operated by the National Astronomical Observatory of Japan.}}

\author{Masayuki \textsc{Akiyama},\altaffilmark{1}
  Yoshihiro \textsc{Ueda},\altaffilmark{2}
  Mike G. \textsc{Watson},\altaffilmark{3}
  Hisanori \textsc{Furusawa},\altaffilmark{4}
  Tadafumi \textsc{Takata},\altaffilmark{4} 
  Chris \textsc{Simpson},\altaffilmark{5}
  Tomoki \textsc{Morokuma},\altaffilmark{6}
  Toru \textsc{Yamada},\altaffilmark{1} 
  Kouji \textsc{Ohta},\altaffilmark{2}
  Fumihide \textsc{Iwamuro},\altaffilmark{2}
  Kiyoto \textsc{Yabe},\altaffilmark{4}
  Naoyuki \textsc{Tamura},\altaffilmark{7}
  Yuuki \textsc{Moritani},\altaffilmark{8}
  Naruhisa \textsc{Takato},\altaffilmark{4}
  Masahiko \textsc{Kimura},\altaffilmark{9}
  Toshinori \textsc{Maihara},\altaffilmark{2}
  Gavin \textsc{Dalton},\altaffilmark{10,11}
  Ian \textsc{Lewis},\altaffilmark{10}
  Hanshin \textsc{Lee},\altaffilmark{10,12}
  Emma \textsc{Curtis Lake},\altaffilmark{10,13}
  Edward \textsc{Macaulay},\altaffilmark{10,14}
  Frazer \textsc{Clarke},\altaffilmark{10}
  John D. \textsc{Silverman},\altaffilmark{7}
  Scott \textsc{Croom},\altaffilmark{15}
  Masami \textsc{Ouchi},\altaffilmark{16}
  Hitoshi \textsc{Hanami},\altaffilmark{17}
  J.Diaz \textsc{Tello},\altaffilmark{18}
  Tomohiro \textsc{Yoshikawa},\altaffilmark{19}
  Naofumi \textsc{Fujishiro},\altaffilmark{20}
  Kazuhiro \textsc{Sekiguchi},\altaffilmark{4}
}
\altaffiltext{1}{Astronomical Institute, Tohoku University, Aramaki, Aoba-ku, Sendai, 980-8578, Japan}
\email{akiyama@astr.tohoku.ac.jp}
\altaffiltext{2}{Department of Astronomy, Kyoto University, Kitashirakawa-Oiwake-cho, Sakyo-ku, Kyoto, 606-8502, Japan}
\altaffiltext{3}{XROA Group, Department of Physics and Astronomy, University of Leicester, Leicester LE1 7RH, UK}
\altaffiltext{4}{National Astronomical Observatory of Japan, Mitaka, Tokyo 181-8588, Japan}
\altaffiltext{5}{Astrophysics Research Institute, Liverpool John Moores University, ic2
Building, 146 Brownlow Hill, Liverpool L3 5RF, UK}
\altaffiltext{6}{Institute of Astronomy, Graduate School of Science, University of Tokyo, 2-21-1, Osawa, Mitaka, Tokyo, 181-0015, Japan}
\altaffiltext{7}{Kavli Institute for the Physics and Mathematics of the Universe, The University of Tokyo, Kashiwa, 277-8583, Japan}
\altaffiltext{8}{Hiroshima Astrophysical Science Center, Hiroshima University, Higashi-Hiroshima, Hiroshima, 739-8526, Japan}
\altaffiltext{9}{Institute of Astronomy and Astrophysics, Academia Sinica, No.1, Sec.4, Roosevelt Rd, Taipei, 10617, Taiwan, R.O.C.}
\altaffiltext{10}{Department of Astrophysics, University of Oxford, Keble Road, Oxford, OX1 3RH, UK}
\altaffiltext{11}{STFC Rutherford Appleton Laboratory, HSIC, Didcot, Oxfordshire OX11 0QX, UK}
\altaffiltext{12}{McDonald Observatory, University of Texas at Austin, 1 University Station C1402, Austin, TX 78712, USA}
\altaffiltext{13}{Institute for Astronomy, University of Edinburgh, Royal Observatory, Edinburgh EH9 3HJ, UK}
\altaffiltext{14}{School of Mathematics and Physics, University of Queensland, Brisbane, QLD 4072, Australia}
\altaffiltext{15}{Sydney Institute for Astronomy, School of Physics, University of Sydney, NSW 2006, Australia}
\altaffiltext{16}{Institute for Cosmic Ray Research, University of Tokyo, Kashiwa, Chiba, 277-8582, Japan}
\altaffiltext{17}{Physics Section, Faculty of Humanities and Social Sciences, Iwate University, 020-8550, Morioka, Japan}
\altaffiltext{18}{IATE, Observatorio Astronomico de Cordoba, Universidad Nacional de Cordoba, Argentina}
\altaffiltext{19}{Kyoto-Nijikoubou Co., Ltd., 17-203, Iwakura Minamiosagicho, Sakyo-ku, Kyoto-shi, Kyoto, 606-0003, Japan}
\altaffiltext{20}{Koyama Astronomical Observatory, Kyoto Sangyo University, Motoyama,Kamigamo Kita-ku,Kyoto, 603-8047, Japan}


\maketitle

\begin{abstract}
We report the multi-wavelength identification of the X-ray sources found in the
Subaru $-$ {\it XMM-Newton} Deep Survey (SXDS) using deep imaging data covering the
wavelength range between the far-UV to the mid-IR. We select a primary counterpart of
each X-ray source by applying the likelihood ratio method to $R$-band, 3.6{\micron}, near-UV, 
and 24{\micron} source catalogs as well as matching catalogs of AGN candidates selected
in 1.4GHz radio and $i'$-band variability surveys. Once candidates of Galactic stars,
ultra-luminous X-ray sources in a nearby galaxy, and clusters of galaxies are removed
there are 896 AGN candidates in the sample. We conduct spectroscopic
observations of the primary counterparts with multi-object spectrographs in the
optical and NIR; 65\% of the X-ray AGN candidates are spectroscopically-identified. 
For the remaining X-ray AGN candidates, we evaluate their photometric redshift 
with photometric data in 15 bands. Utilising the multi-wavelength photometric 
data of the large sample of X-ray selected AGNs, we evaluate the stellar masses, $M_{*}$,
of the host galaxies of the narrow-line AGNs. The distribution of the stellar
mass is remarkably constant from $z=0.1$ to $4.0$. The relation between $M_{*}$
and 2--10~keV luminosity can be explained with strong cosmological evolution of
the relationship between the black hole mass and $M_{*}$.
We also evaluate the scatter of the UV-MIR spectral energy distribution (SED)
of the X-ray AGNs as a function of X-ray luminosity and absorption to the nucleus.
The scatter is compared with galaxies which have  redshift and
stellar mass distribution matched with the X-ray AGN. The UV-NIR SEDs of obscured
X-ray AGNs are similar to those of the galaxies in the matched sample. In the
NIR-MIR range, the median SEDs of X-ray AGNs are redder, but the scatter
of the SEDs of the X-ray AGN broadly overlaps that of the galaxies
in the matched sample.
\end{abstract}

\section{Introduction}

After the discovery of super-massive black holes (SMBHs) at the 
center of every massive galaxy in the local Universe (e.g., \cite{kormendy95}), the 
issue of how these SMBHs formed and evolved over cosmic history 
has become one of the major unanswered questions in observational cosmology. 
Luminosities of AGNs reflect the mass accretion rates of their SMBHs, 
therefore the luminosity function of AGNs and its cosmological 
evolution reflects the growth history of SMBHs through accretion. 
Hard X-ray selection is an effective selection method for AGNs,
it can detect obscured AGNs without serious bias, 
except for Compton-thick AGNs, and can efficiently sample AGNs
without contamination by a large number of star-forming galaxies.
The hard X-ray luminosity
function of AGNs and its evolution as a function of redshift
have been examined with combining X-ray selected AGN samples with
various survey depths and areas \citep{ueda03, barger05, lafranca05, 
hasinger05, silverman08, aird08, yencho09, ebrero09, aird10, ueda14}.
Based on the cosmological evolution of the hard X-ray luminosity
function, the average accretion growth curves of SMBHs are
inferred \citep{marconi04, ueda14}.

In order to construct a large sample of X-ray-selected AGNs,
we have conducted deep multi-wavelength imaging
observations covering the FUV to MIR and deep optical and NIR multi-object
spectroscopic observations of the X-ray sources found in the 
Subaru {\it XMM-Newton} Deep Survey (SXDS) \citep{sekiguchi05}.
The X-ray images obtained with {\it XMM-Newton} cover a 1.3 deg$^{2}$ area 
down to $1\times10^{-15}$ erg s$^{-1}$
cm$^{-2}$ in the 0.5--2~keV band and $3\times10^{-15}$ erg
s$^{-1}$ cm$^{-2}$ in the 2--10~keV band \citep{ueda08}.
The $\log N - \log S$ relation of the X-ray sources has a knee
at a hard X-ray flux of $1\times10^{-14}$ erg s$^{-1}$ cm$^{-2}$
and the contribution to the cosmic X-ray background is maximum
in this flux range \citep{cowie02}.
Several hard X-ray surveys covering $\sim1$deg$^{2}$ area
in this flux range have been conducted
\citep{fiore03, kim04, yang04, chiappetti05, murray05,
laird09, elvis09, cappelluti09, civano12}.
Among the X-ray surveys with similar depth and area, 
SXDS is notable in having deep multi-wavelength data 
in the wavelength range from far-UV to mid-IR.

In this paper, we describe the X-ray sample in Section 2, 
and the multi-wavelength imaging data in Section 3. 
The 3 $\sigma$ detection limits of the imaging data 
in the $i'$-, $K$-, and 3.6{\micron} bands
reach 27.4 mag, 25.3 mag, and 24.5 mag in the AB magnitude system, 
respectively. They are expected to be deep enough to detect
counterparts of almost all of the X-ray sources at the flux limit:
90\% of the X-ray sources found in a wide area {\it Chandra} survey
at a flux limit of $1\times10^{-15}$
erg s$^{-1}$ cm$^{-2}$ in the 0.5--2.0~keV band have a counterpart brighter than
26.4 mag ($i'$-band), 23.6 mag ($K$-band), and 22.8 mag (3.6{\micron}-band)
\citep{civano12}.
The positional uncertainties of the {\it XMM-Newton} X-ray sources
are larger than those of {\it Chandra} sources, and they are not 
small enough to securely identify the faintest optical counterparts
(\cite{loaring05}; \cite{chiappetti05}). In order to 
quantitatively select the counterparts, we apply the
likelihood ratio method \citep{deruiter77, wolstencroft86, sutherland92}.
We apply the method not only to the deep optical source catalog
but also to the deep 3.6{\micron}, NUV, and 24{\micron} source
catalogs to improve the reliability of the counterpart 
selection. The identification procedure with the multi-wavelength 
source catalogs is described in Section 4.

In order to reveal the nature of the X-ray sources, 
we have been conducting a large campaign of spectroscopic 
observations in the NIR as well as in the optical
wavelength range. Because obscured AGNs at $z=1-2$ can be significant
contributor to the X-ray source population, NIR spectroscopy
is effective in revealing their nature by detecting strong 
rest-frame optical emission lines. The spectroscopic follow-up observations
are described in Section 5.
Even with the deep optical and NIR spectroscopic observations, 
a significant fraction of the X-ray sources are fainter than the 
spectroscopic limit. In order to overcome the limitation
of the spectroscopic identification, photometric redshift
estimation with a deep imaging data set covering wide 
wavelength range is important \citep{zheng04}. The multiwavelength photometry and photometric redshift evaluation
are described in Section 6. In Section 7, we discuss the
X-ray properties and host galaxy stellar mass of the
X-ray selected AGNs. Utilising the multi-wavelength
photometry, we examine the scatter of the SEDs of the
X-ray AGNs, and compare the scatter with that of
non-X-ray galaxies with similar redshifts and stellar masses.
Throughout the paper, cosmological parameters 
H$_0$ = 70 km s$^{-1}$ Mpc$^{-1}$, $\Omega_{\rm M}=0.3$, and
$\Omega_{\Lambda}=0.7$ are used. The magnitudes are based on
AB magnitude system \citep{oke74}.

Using the large sample of X-ray selected AGNs, \citet{hiroi12}
derive the comoving space density and obscured fraction 
of AGNs at $3<z<5$. Furthermore, \citet{ueda14} examine the
cosmological evolution of the X-ray AGN luminosity function 
including our sample. 
The black hole mass and Eddington ratio distribution
functions of the X-ray selected AGNs at $z\sim1.4$ are
discussed in \citet{nobuta12}, using the optical and
NIR spectra of the X-ray selected AGNs.

\section{X-ray Sample Definition}

\begin{figure*}
 \begin{center}
  \includegraphics{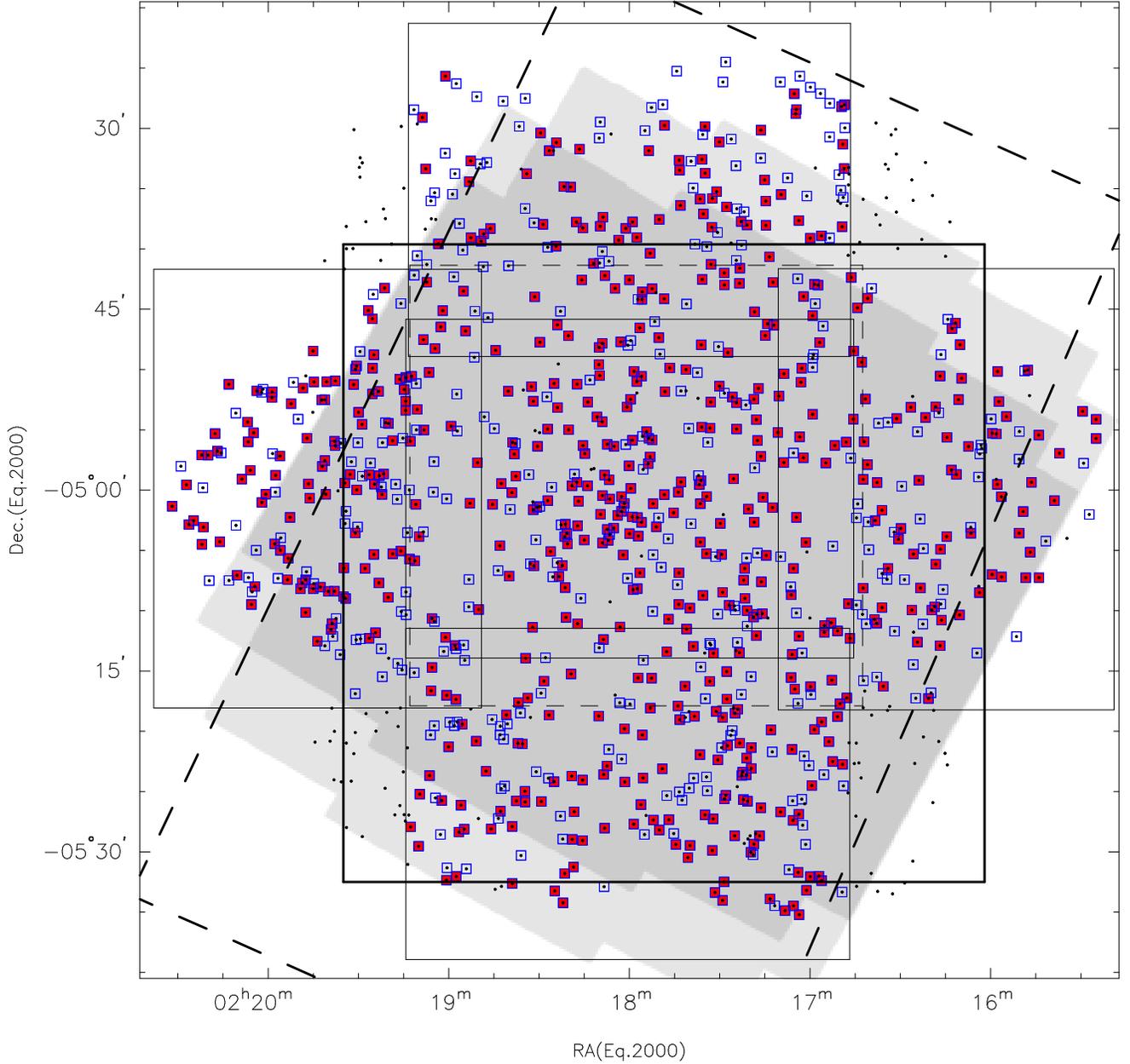} 
 \end{center}
\caption{
The areas covered by $u$-band, optical, NIR, and MIR imaging observations are shown overlaid on
a map of the detected X-ray sources from the
7 {\it XMM-Newton} images \citep{ueda08}. The 5 thin solid rectangles
indicate the area covered by the deep Suprime-cam
imaging observations. The thin dashed
line shows the area covered by the deep $u$-band observations
with the Mosaic-II camera.
The thick solid square indicates the region covered
by the deep $J$-, $H$-, $K$-band imaging observations of
UKIDSS/UDS.
The gray region indicates
the region covered by SpUDS IRAC observations.
The lower (upper) light gray region is only covered
by 3.6{\micron} and 5.8{\micron} (4.5{\micron} and 8.0{\micron})
observations. 
The thick dashed line represents
the region covered by the SpUDS MIPS observations.
The black dots show positions of the 
X-ray sources with likelihood larger than 7 either
in the soft or hard band from \citet{ueda08}.
The blue squares indicate X-ray sources included
in the current sample which are covered by
the Suprime-cam observations and not affected by
halos of very bright stars.
Red filled symbols indicate spectroscopically
identified sources.}\label{fig:SXDS_fov}
\end{figure*}

\begin{figure*}
 \begin{center}
  \includegraphics[width=170mm]{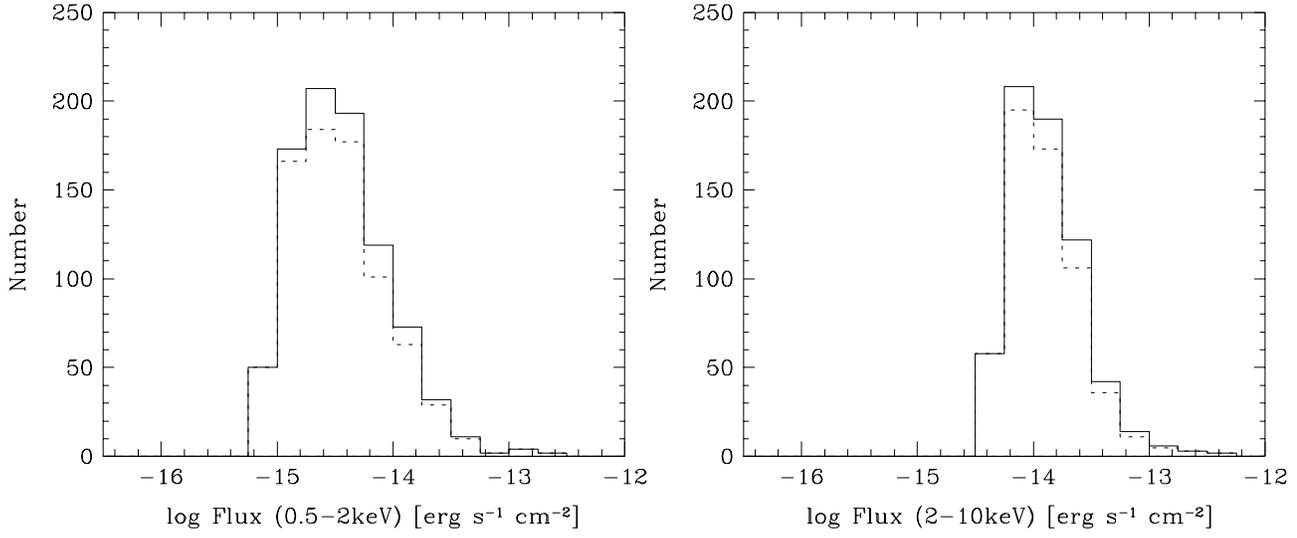} 
 \end{center}
\caption{
Left) The 0.5--2.0~keV band flux distribution of the X-ray sources
in the SXDS. The solid histogram represents the number density of
the entire soft-band sample of \citet{ueda08}, and the dotted histogram 
shows the number density of the sample considered in this paper.
Right) The 2.0--10.0~keV band flux distribution with details as for left panel.}\label{fig:flux_dist}
\end{figure*}

\begin{figure*}
 \begin{center}
  \includegraphics[width=170mm]{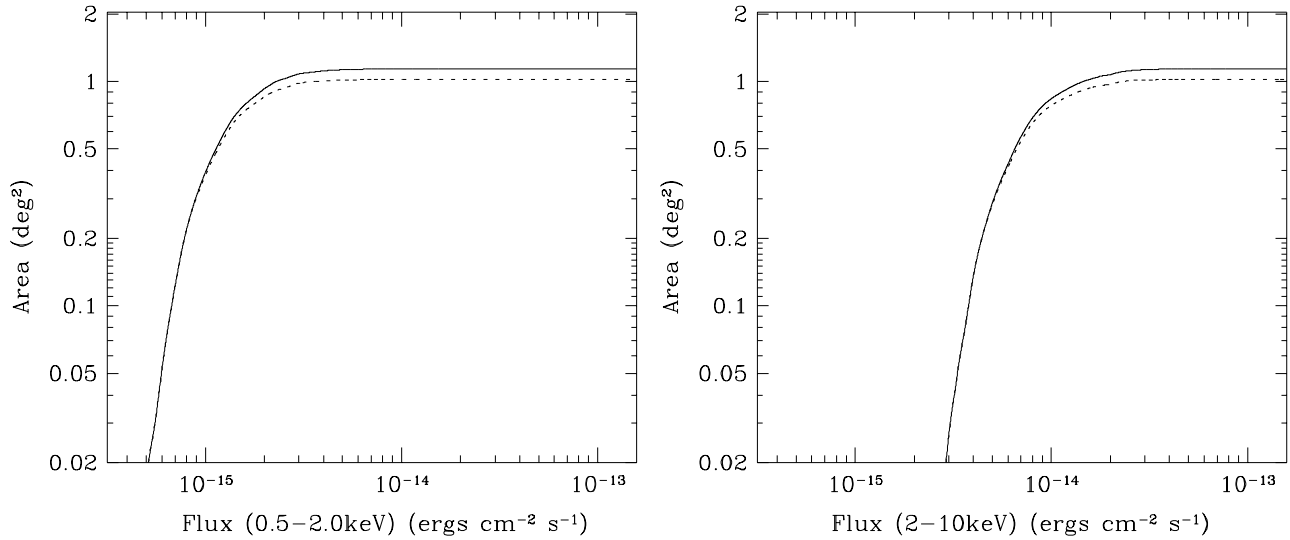} 
 \end{center}
\caption{
Left) Survey area as a function of 0.5-2.0~keV X-ray flux
for the soft-band sample.
The solid line represents the survey area of the entire soft-band sample of
\citet{ueda08}, and the dotted line shows the survey area of the
sample considered in this paper. The flux is determined by
multiplying the soft-band countrate by
$0.167\times10^{-11}$ (erg s$^{-1}$ cm$^{-2}$)
/ (cts ks$^{-1}$) assuming power-law spectrum with $\Gamma$=1.8.
Right) Survey area for the hard-band sample. The flux is determined by
multiplying the hard-band countrate by
$0.838\times10^{-11}$ (erg s$^{-1}$ cm$^{-2}$) / (cts ks$^{-1}$) 
assuming the same power-law.}\label{fig:SXDS_areacurve}
\end{figure*}

The SXDS field is centered on \timeform{02h18m} and \timeform{-05D00'}. 
The field is selected to be observable with 8-10m class 
optical telescopes in both of the northern and the southern 
hemispheres. The field was observed with {\it XMM-Newton} 
with one central $30^{\prime}$ diameter field with a 100ks 
exposure and six flanking fields with 50ks exposure time 
each \citep{ueda08}. These seven pointings  cover 
a 1.3 deg$^{2}$ field. Details of the observations and data 
reduction procedures are described in \citet{ueda08}. 

In total, 866 and 645 sources are detected above likelihood 
of 7 in the 0.5--2~keV (soft) and 2--10~keV (hard) bands,
respectively, through source detection utilising a maximum
likelihood fit method. The spatial distribution of the sources is 
shown in Figure~\ref{fig:SXDS_fov} (small dots). A likelihood 
($L$) of 7 corresponds to the probability of being a random 
background noise, $P_{\rm random}$ of $9.1\times10^{-4}$ 
($L=-{\rm ln} \ P_{\rm random}$). The 0.5--2~keV and 2--10~keV 
band X-ray flux distributions of the X-ray sources are shown 
in Figure~\ref{fig:flux_dist}. The 0.5--2~keV (2--10~keV) flux 
is calculated by multiplying the 0.5--2~keV (2--10~keV) count 
rate with 0.167 (0.838)$\times10^{-14}$ [erg s$^{-1}$ cm$^{-2}$]/[cts ksec$^{-1}$] \citep{ueda08}. 
These are the conversion factors for a power-law spectrum with $\Gamma=1.8$,
which is the typical apparent X-ray spectral slope of a non-obscured AGN.

In this paper, only X-ray sources in the region covered with deep
optical imaging data described in the next section are considered.
The region is shown as 5 rectangles with thin solid lines.
Some of the X-ray sources whose optical images are affected by halo
or electron leakage from very bright objects are removed as well.
A total of 781 (584) soft (hard)-band X-ray sources with
likelihood above 7 are covered. Hereafter these sources are referred as the
soft-(hard-)band sample. Considering sources common to both samples, there
are 945 unique X-ray sources, we refer the combined sample as the total sample. 
The flux distributions of the samples are
shown with thick histograms in Figure~\ref{fig:flux_dist}.
The corresponding survey area for the soft-(hard-)band sample
is shown in the left (right) panel of Figure~\ref{fig:SXDS_areacurve}. 
The solid line represents
the survey area of the entire sample of \citet{ueda08}, and the
dotted line shows the survey area of the sample considered in this
paper.

\section{Multi-wavelength Imaging Dataset}

\subsection{Deep Optical Imaging Data}

Deep optical imaging observation in the field was conducted with the
optical wide-field $34^{\prime}\times27^{\prime}$ mosaic CCD camera, Suprime-Cam, attached to the
prime-focus of the 8.2m Subaru Telescope. The observation was conducted
as a Subaru observatory project from September 2002 to January 2004.
Imaging data in five overlapping FoVs in the $B$, $V$, $R$, $i'$, and $z'$ bands was
obtained. The thin solid rectangles in Figure~\ref{fig:SXDS_fov} indicate the area covered. The total exposure times for each field range 5.5-5.8h ($B$), 
4.9-5.3h ($V$), 3.9-4.1h ($R$),
5.2-10.8h ($i'$), 3.1-5.2h ($z'$). For details of the deep imaging data, 
see \citet{furusawa08}.
In these images the FWHM of stellar objects is $0.\!^{\prime\prime}7-0.\!^{\prime\prime}9$.
We apply a Gaussian convolution to make the resulting PSFs in the 5 bands
constant for each region.
We use the photometric zero-point provided in \citet{furusawa08}.
The typical 2 $\sigma$ detection limits in $2^{\prime\prime}$ diameter
apertures are determined by measuring the standard deviation of fluxes
within a fixed aperture at positions where no object is detected.
The limits are 
28.7($B$), 28.1($V$), 28.1($V$), 28.0($i'$), and 26.9($z'$) mag.
Correcting for the aperture loss for point sources, the typical magnitude limits are
28.1($B$), 27.6($V$), 27.6($R$), 27.6($i'$), and
26.6($z'$) mag for the total magnitudes of point sources.
The limiting magnitudes in the aperture and for the total magnitudes
are summarised in Table~\ref{tab:imagelist}. For other bands, we derive 
the limiting magnitudes for aperture and total magnitudes in the same way.

We use the Suprime-cam $R$-band 
astrometry as the astrometric reference frame in this paper.
Astrometric calibration is carried out in the $R$-band images using the 2MASS
point source catalog. The rms residual of the fitting is typically
$0.\!^{\prime\prime}03$. The
multi-wavelength data described below are aligned to the $R$-band
image using commonly detected stellar objects. 

\subsection{U-band Imaging Data}

The entire survey region is covered with
the Canada France Hawaii Telescope Legacy Survey
(CFHTLS\footnote{See http://www.cfht.hawaii.edu/Science/CFHTLS}). 
We use the $U$-band data from the CFHTLS archive.
The typical 3.0$\sigma$ total-magnitude limit for a point source is
26.11 mag.
Part of the survey region, the southern FoV out of the 5 Suprime-cam FoVs,
is covered by a deeper $U-$band imaging observation with Suprime-cam
with 3.7 hours integration.
Details of these data are described in \citet{yoshida08}.
The 3.0$\sigma$ limiting total magnitude of a point
sources is 26.46 mag.
The central $37^{\prime}\times37^{\prime}$ area of the survey region
is also covered by a deep imaging data obtained with the Mosaic-II
camera attached to the CTIO Blanco 4m telescope (\cite{diaztello13}; PI. N.Padilla).
In this observation, the Sloan Digital Sky Survey (SDSS) $u$-band filter is used.
The region covered  is shown with a thin dashed line
in Figure~\ref{fig:SXDS_fov}. The total integration time is 10h and
the 3.0$\sigma$ limiting total magnitude of
a point source is 26.29 mag.
Because the filter systems
used in the three $U$-band images are different, we
adjust the Suprime-cam and Mosaic-II magnitudes
to the photometric system used in the MegaCam.
Using stars in the overlapping regions, we fit
the magnitude difference between the Suprime-cam (Mosaic-II) 
and MegaCam photometry with a linear function of the $U-B$ color,
and apply the function to the Suprime-cam (Mosaic-II) photometry.
We check the photometric zero-point of the MegaCam data
using the $B$ and $R$ photometry from the Suprime-cam data
and the $U-B$ and $B-R$ color-color diagram
of stellar objects. The distribution 
of the stellar objects on the diagram is consistent with
the colors derived with the SEDs of stars from \citet{gunn83}.

\subsection{Shallow Optical and U-band Imaging Data}

For bright objects which are saturated in the deep optical images,
their optical magnitudes are determined with shallow optical images
taken with Wide Field Camera attached to Issac Newton Telescope.
$U$-, $G$-, $R$-, $i$-, $z$-bands data are available.
The $G$, $R$, $i$, $z$ optical magnitudes are adjusted to the
Suprime-Cam $B$, $R$, $i'$, $z'$ system with magnitude
offset and its color dependence derived with stars
in the overlapping region.

\subsection{NIR Imaging Data}

A large part of the SXDS region is covered by the Ultra Deep Survey
of UKIRT Infrared Deep Sky Survey (UKIDSS UDS; \cite{lawrence07}).
The coverage of this survey is shown with the thick solid rectangle
in Figure~\ref{fig:SXDS_fov}. Archival image data from data 
release 8 (UKIDSS DR8) is used 
in this paper\footnote{See http://surveys.roe.ac.uk/wsa}. The stellar image
size is $0.7^{\prime\prime}$. We use the photometric zero-point
provided in the data release.
The 3 $\sigma$ detection limit of the image for a point source
is measured to be 25.40, 24.82, 25.30 mag for $J$-, $H$-, and $K$
bands, respectively, with a $2.0^{\prime\prime}$
diameter aperture photometry and aperture correction for point
sources. For X-ray sources outside of the UKIDSS UDS region, 
we do not have NIR photometric data.

\subsection{MIR Imaging Data}

\subsubsection{{\it Spitzer} IRAC Data}

Almost the entire FoV of the SXDS is covered
by the {\it Spitzer} observation with the Infra-Red Array Camera (IRAC)
instrument obtained in the course of the {\it Spitzer} Wide-area
Infra-Red Extragalactic (SWIRE) legacy survey project
(\cite{lonsdale03}; \cite{lonsdale04}). Seven sources in the
total sample are outside of the filed covered. We use publicly
available mosaiced images of IRAC (DR2 2005-06-03; \cite{surace05}).

The FWHM of the PSFs in the IRAC images are determined to be
1.9, 2.0, 1.9, and 2.2$^{\prime\prime}$ by combining images
of $>10$ stars in the FoV, and the corresponding
aperture corrections for a $3.8^{\prime\prime}$ diameter aperture
are 0.33, 0.36, 0.54, and 0.66 mag for 3.6{\micron}, 4.5{\micron}, 
5.8{\micron}, and 8.0{\micron} bands, respectively \citep{surace05}. 
We determine the photometric zero-points of the mosaiced IRAC 
images using
$K-3.6\micron$, $K-4.5\micron$, $K-5.8\micron$, and $K-8.0\micron$ colors
of bright blue stars following \citet{lacy05}; values of 0.04, 0.02, 0.00, and
0.00 mag are used for stars with $J-K$ colors bluer than 0.3 mag 
based on a theoretical stellar SED model
(Table 5 of \cite{lacy05}). Near-infrared photometry of 
stellar objects in the SXDS field is taken from the
Two Micron All-Sky Survey Point Source Catalog (2MASS PSC; \cite{skrutskie06}).
$K$-band magnitudes in the 2MASS PSC are determined with a
profile fitting method.
We select blue stars with $J-K<0.3$ mag and examine their
average $K-3.6\micron$, $K-4.5\micron$, $K-5.8\micron$, and $K-8.0\micron$ 
colors. Photometry in the IRAC bands are determined 
with a $3.8^{\prime\prime}$ diameter aperture with an aperture correction 
derived from the PSF. We determine the photometric zero-points to match 
the expected colors from the theoretical stellar SED model.

The 3 $\sigma$ detection limit of the image for point sources
is measured to be 22.77, 22.07, 19.97, and 19.91 mag 
for the 3.6{\micron}, 4.5{\micron}, 5.8{\micron}, and 8.0{\micron} bands,
respectively, with a $3.8^{\prime\prime}$
 aperture photometry diameter and assumption for the aperture correction for point
sources. 

Comparing the coordinates of objects detected in both of the $R$-band
and 3.6{\micron}-band, we found a $0.\!^{\prime\prime}36$ offset in the
right ascension direction between the deep Suprime-Cam $R$-band and the
IRAC 3.6{\micron}-band images. We shift the IRAC image to match the
coordinates of the Suprime-Cam image. After the correction, the difference
between the $R$-band and 3.6{\micron}-band coordinates is
$0.\!^{\prime\prime}4$ r.m.s.,  which is thought to be dominated by the
positional uncertainty of the 3.6{\micron} source. 

A significant part of the SXDS region is covered by one of
the {\it Spitzer} legacy programs, {\it Spitzer} UKIDSS Ultra Deep Survey (SpUDS).
IRAC observations cover the region shown in gray in Figure~\ref{fig:SXDS_fov}.
The photometric zero-point of the SpUDS data is adjusted to
SWIRE photometry with overlapping objects. 
The typical 2.0$\sigma$ detection limit for $3.\!^{\prime\prime}8$ aperture
is 25.26, 24.83, 22.99, and 23.19 mag
and the
detection limits for the point source are estimated to be
24.49 ($3.0\sigma$), 24.03 ($3.0\sigma$), 21.69 ($3.0\sigma$), 21.77 ($4.0\sigma$) mag
for 3.6{\micron}, 4.5{\micron}, 5.8{\micron}, and 8.0{\micron} bands, 
respectively.

\subsubsection{{\it Spitzer} MIPS Data}

The {\it Spitzer} observation with the Multiband
Imaging Photometer (MIPS) instrument 
in the SWIRE legacy survey (\cite{lonsdale03}; \cite{lonsdale04})
also covers almost the entire area of the SXDS. We use publicly
available mosaiced images of MIPS data in 24{\micron} band
(DR5 2006-12-23).

The FWHM of PSF of the MIPS image is $5.\!^{\prime\prime}9$,
and aperture correction for a $6.\!^{\prime\prime}0$ diameter aperture 
is 1.23 mag \footnote{Taken from Multiband Imaging Photometer for {\it Spitzer} Data Handbook Version 3.2}. 
We use the photometric zero-point provided in the
data release.
The 2$\sigma$ detection limit of the data in the
aperture is 20.21 mag, and the 4$\sigma$ detection limit for
the total magnitude of stellar objects is 18.22mag.

A large part of the SXDS region is also covered by 
the MIPS observation of SpUDS. The coverage is shown
with the thick dashed line in Figure~\ref{fig:SXDS_fov}.
The 2$\sigma$ detection limit in a $6.\!^{\prime\prime}0$ diameter
aperture is 22.06 mag, and the 4$\sigma$ detection limit
in total magnitude of stellar objects is 20.07mag.

\subsection{UV Imaging Data}

Deep Imaging Survey (DIS) of the Galaxy Evolution Explorer ({\it GALEX};
\cite{morrissey07}) 
covers almost the entire area of the SXDS in the FUV (effective wavelength
of 1516\AA) and NUV (effective wavelength of 2267\AA) bands.
We use the reduced data obtained from the {\it GALEX}
archive\footnote{See http://galex.stsci.edu/GR6/}.
The tilenames of XMMLSS\_03 (29189s, 30289s), 
XMMLSS\_04 (27306s, 28561s), XMMLSS\_07 (27440s, 27520s), 
and XMMLSS\_09 (30606s, 32354s) overlap with the SXDS.
The numbers in the parenthesis show the total exposure times
in FUV and NUV bands, respectively, for each tile. 

The FWHM of the reduced images are $5.\!^{\prime\prime}0$ and
$5.\!^{\prime\prime}1$ in the FUV and NUV images, respectively.
We use the photometric zero-point provided in \citet{morrissey07};
18.82mag and 20.08mag for FUV and NUV bands. 
The aperture corrections for point sources are
0.36 and 0.59 mag for FUV and NUV bands for a $7.\!^{\prime\prime}.6$
diameter aperture \citep{morrissey07}. The typical point source limiting magnitudes
estimated with the $7.\!^{\prime\prime}6$ aperture
photometry reach down to 25.52 (4$\sigma$) and 25.15 (3$\sigma$)
mag in the FUV and NUV bands, respectively.

\section{Multi-wavelength Identification of the X-ray Sources}

\subsection{Identification with $R$-band Objects and Its Reliability}

\begin{figure*}
 \begin{center}
  \includegraphics[width=170mm]{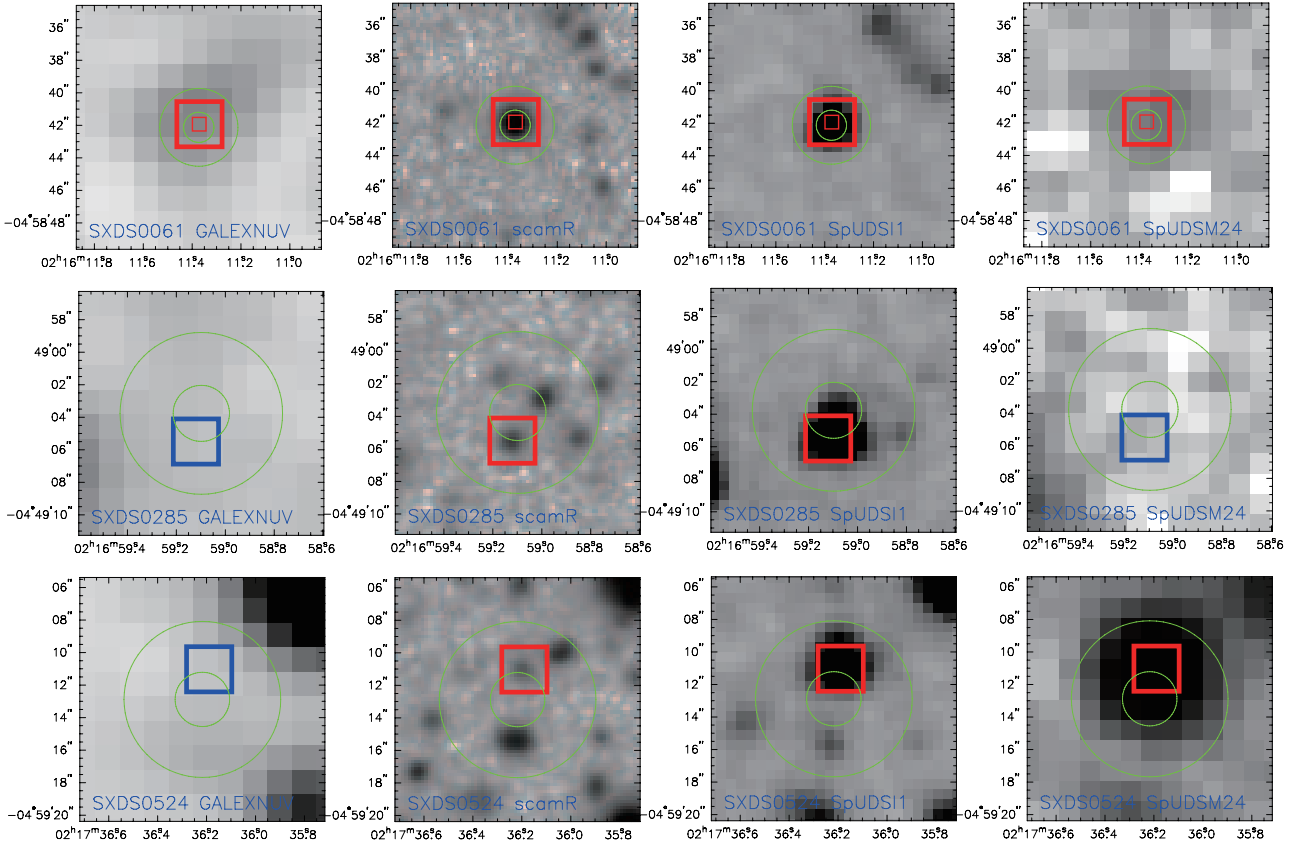} 
 \end{center}
 \caption{Sample identification images.
From left to right, {\it GALEX} NUV, Suprime-cam $R$-band,
{\it Spitzer} 3.6{\micron}-band, and {\it Spitzer} 24{\micron}-band
images with $15^{\prime\prime}\times15^{\prime\prime}$ FoV. 
Top row for SXDS0061, whose primary candidates in the four
band match, middle and bottom rows for SXDS0285 and SXDS0524, respectively,
whose primary counterpart
is based on the {\it Spitzer} 3.6{\micron} identification.
Green circles indicate the X-ray source position; 
the inner and outer circles have radii of 1.0$\times \sigma_{{\rm X}i}$
and 3.0$\times \sigma_{{\rm X}i}$. The position of the
primary counterpart is indicated with large open square. 
The small open square indicates the object selected in a
AGN variability search.
}\label{fig:Identification_sample}
\end{figure*}

\begin{figure*}
 \begin{center}
  \includegraphics[width=170mm]{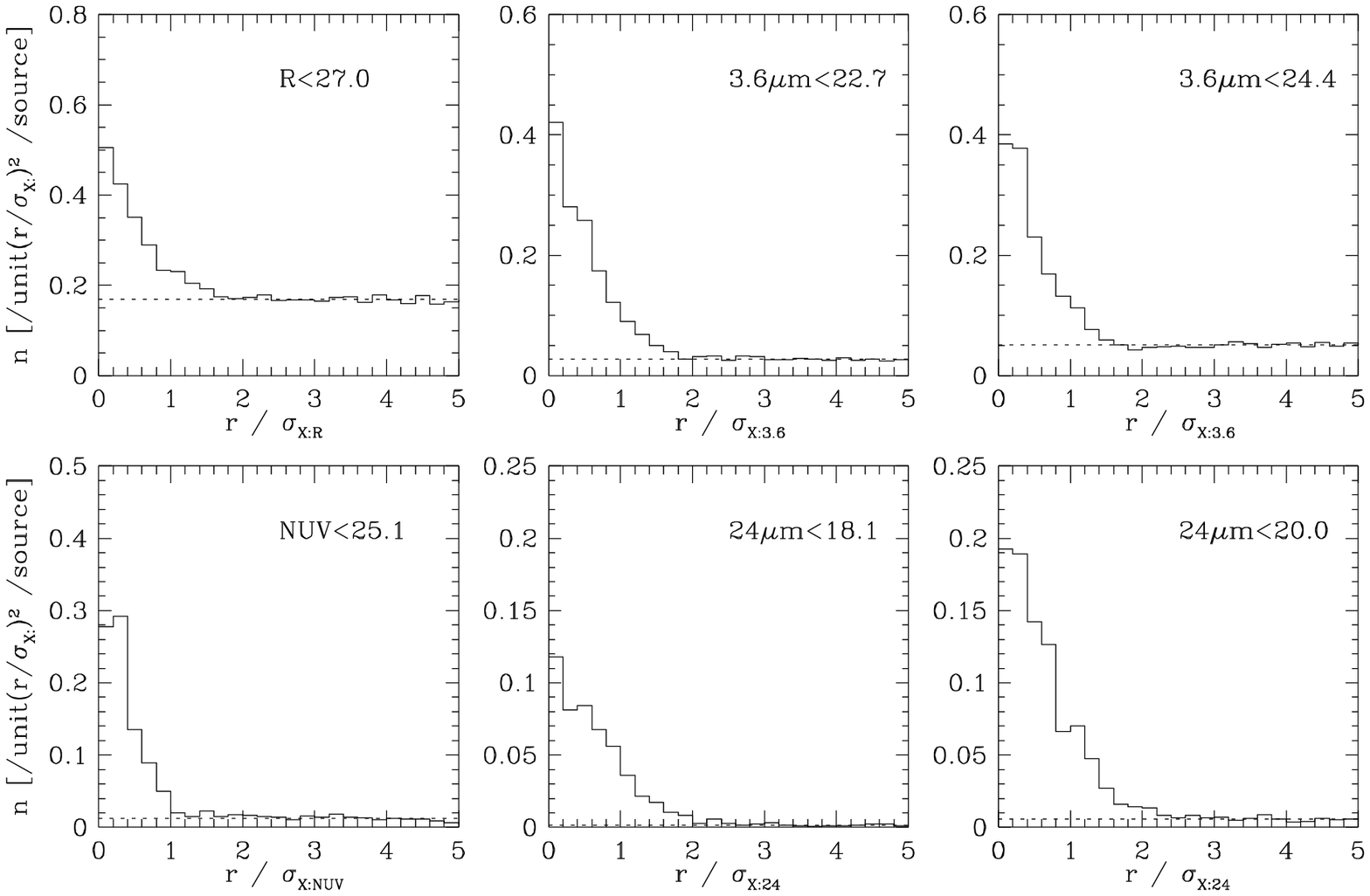} 
 \end{center}
\caption{
Radial distribution of sources from the centers of the
X-ray sources. Radial distance is normalized by 
the uncertainty radius, which is a X-ray positional
uncertainty convolved with a positional uncertainty in 
each band. Number per unit area is divided by 
the number of X-ray sources covered in each dataset.
Dotted lines indicate the average source density outside
of X-ray uncertainty area determined at 
$3.5-4.0 \sigma_{\rm X:band}$.
Upper left) $R$-band sources down to $R=27.0$mag,
upper middle) $3.6\mu$m sources down to $22.7$mag
determined with SWIRE coverage, upper right)
$3.6\mu$m sources down to $24.4$mag determined
with SpUDS coverage, lower left) NUV source down
to $25.1$mag, lower middle) $24\mu$m sources
down to $18.1$mag determined with SWIRE coverage, 
and lower right) $24\mu$m sources
down to $20.0$mag determined with SpUDS coverage.}\label{fig:coord_diff}
\end{figure*}

For each X-ray source the statistical error in the position
estimated through the maximum likelihood fit, $\sigma_{{\rm X}i}$, is given in \citet{ueda08}.
When we match the X-ray and $R$-band sources, we use an
uncertainty radius, $\sigma_{{\rm X:R}i}$, which is determined
by the square of
the $\sigma_{{\rm X}i}$ and $1.\!^{\prime\prime}1$ added in quadrature.
The latter term represents any positional uncertainties other than
$\sigma_{{\rm X}i}$, i.e. any
systematic uncertainty in the positions of the $R$-band sources and X-ray sources. The term is estimated
through the scatter of positional differences of X-ray sources
whose $\sigma_{{\rm X}i}$ is less than $0.\!^{\prime\prime}5$
and their candidate $R$-band counterparts.

Sample images of three X-ray sources in the $R$-band are
shown in the second column of Figure~\ref{fig:Identification_sample}.
Almost all of the X-ray sources have 
at least a $R$-band source in the $3\times\sigma_{{\rm X:R}i}$
radius above the detection limit of the $R$-band images.
Some X-ray sources have multiple $R$-band sources within their error radii.
Upper left panel of Figure~\ref{fig:coord_diff} shows radial
distribution of $R$-band sources down to 27.0 mag from the X-ray positions.
The dotted line shows the average number density of sources
outside of X-ray uncertainty area.
The excess sources above the line
statistically represent the counterparts of the X-ray sources. 
The average number density
corresponds to the expected number of the contaminating sources.
If we consider the region within $2\times\sigma_{{\rm X:R}i}$
down to $R=27.0$ mag, the ratio between the number of excess
and contaminating sources is 0.30.
It is thus important to select the most probable
counterpart for each X-ray source.

At first, we identify the X-ray sources with sources
detected in the $R$-band. We construct a catalog of $R$-band sources by
applying Sextractor \citep{bertin96} to the deep Suprime-cam $R$-band images.
In order to quantitatively select the most probable
identification, we utilise the likelihood ratio
defined as the probability of
an X-ray source $i$ with X-ray flux $f_{Xi}$ having a
counterpart $j$ with magnitude $m_{j}$ and 
distance $r_{ij}$ divided by the probability of it being 
a chance match
(\cite{deruiter77}; \cite{wolstencroft86}; \cite{sutherland92}).
The likelihood ratio is calculated as:

\[L_{ij} \equiv \frac{Q(<m_{j}, f_{Xi}) \exp(-r_{ij}^{2}/2\sigma_{{\rm X:R}i}^{2})}
{2\pi \sigma_{{\rm X:R}i}^{2} N(<m_{j}) }. \]

assuming the spatial distribution of the counterparts in the X-ray
error circles follows the Gaussian distribution.
$N(<m_{j})$ is a cumulative number density of optical objects
brighter than $m_{j}$. 
$Q(<m_{j}, f_{Xi})$ is the probability of the $i$-th X-ray source
has its counterpart brighter than $m_{j}$. 
If the probability of being a counterpart for the X-ray source
exceeds the probability of being a chance match, the likelihood
ratio is greater than 1. 
The number density of optical objects
is calculated from the Suprime-Cam $R$-band image directly without
correction for detection incompleteness.
We determine $Q(<m_{j}, f_{Xi})$ by examining the 
distribution of $\log f_{X}/f_{R}$ for the
X-ray sources, and assume that the distribution
is constant in the SXDS X-ray flux range.
We consider $R$-band sources within 5.0$\times \sigma_{{\rm X:R}i}$ of
each X-ray source.

Firstly, we calculate $L_{ij}$ with $Q(<m_{j}, f_{Xi})=1.0$
and select the $R$-band object with the highest likelihood ratio for
each X-ray source as the initial guess for the $R$ primary candidate. 
Based on the $R$-band magnitudes of the
identified objects, we determine $Q(<m_{j}, f_{Xi})$. 
The $\log f_{Xi}/f_{Rj}$ value
is calculated with $\log f_{Xi}/f_{Rj} = \log f_{Xi} + 5.53 + 0.4 Rj$
in the AB magnitude system. The half-maximum wavelength
range of the $R$-band filter response (5950{\AA}--7085{\AA})
of the Suprime-Cam is used in the calculation.
We use the 0.5--10~keV flux, which is derived by summing the flux
in the 0.5--2~keV and 2--10~keV bands, as the X-ray flux $f_{Xi}$.
We call the $R$-band object with the highest likelihood ratio
for a given X-ray source the $R$ primary candidate for the source.

The reliability of the identification is evaluated as
the probability that a $R$-band object $j$ is 
the counterpart of an X-ray source $i$.
The reliability is given by

\[R_{ij} = \frac{L_{ij}}{\sum_{j} L_{ij} + (1-Q(<m_{\rm limit}, f_{Xi}))} \]

\begin{figure}
 \begin{center}
  \includegraphics[width=75mm]{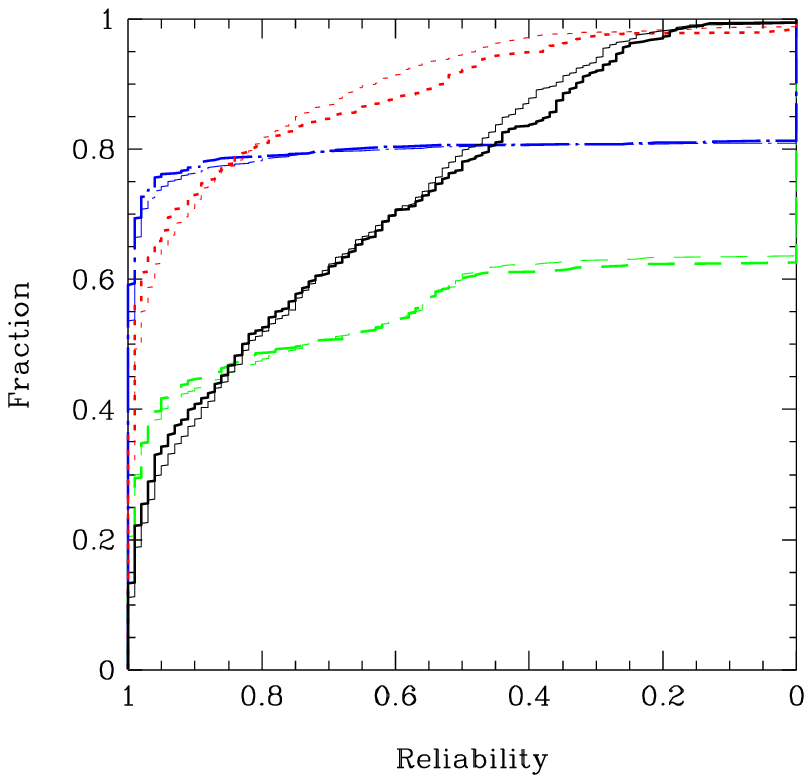} 
 \end{center}
\caption{
Distribution of reliability $R_{ij}$ for the primary candidate
identified in each band. Black solid curves are for the $R$-band
identification. Red dotted lines are for the 3.6{\micron} identification.
Blue dot-dashed lines are for 24{\micron} identification. Green 
dashed lines are for NUV identification.
The thick (thin) lines are for the hard-band (soft-band) sample.}\label{fig:Reliable_hist_all}
\end{figure}

\citep{sutherland92}.
$Q(<m_{\rm limit}, f_{Xi})$ represents the probability of finding
an $R$-band counterpart above the detection limit $m_{\rm limit}$.
We use a value of $Q(<m_{\rm limit}, f_{Xi})$ of 0.8 for all X-ray
sources for simplicity. The cumulative
distribution of the reliability of the $R$ primary
candidate is shown in Figure~\ref{fig:Reliable_hist_all}
with thick and thin black solid lines for the soft- and hard-band samples,
respectively.
A significant fraction of the X-ray sources have
low reliability; 30\% of the X-ray sources have primary optical
counterpart with reliability smaller than 0.6.
X-ray sources that have the $R$ primary candidate with
low-reliability have a few $R$-band objects with similar
$R$-band magnitudes and distances from the X-ray position.
These are the problematic cases and the fraction of such 
cases in the total sample
is similar to those in {\it XMM-Newton} survey fields with similar depth
\citep{brusa10}.

In order to quantitatively examine the reliability of the optical
identification process, we conduct Monte-Carlo
simulations of the optical identification procedure. We use the large catalog of
the optical objects in the fields made with the deep optical imaging.
We regard each optical object as
an X-ray source and add pseudo X-ray flux information and simulate the
X-ray coordinates by adding positional uncertainty to the
optical coordinates. We use $1\times10^{-14}$ erg s$^{-1}$ cm$^{-2}$
in the total band for the X-ray flux. We assume that the positional
error follows a Gaussian distribution. We simulated
a positional error $\sigma_{{\rm X:R}i}$ of  $0.\!^{\prime\prime}0$
to $3.\!^{\prime\prime}0$ with a $0.\!^{\prime\prime}2$ step.
We apply the same optical identification process described above to
the pseudo-X-ray source catalog. The identification process recovers
more than 90\% of the counterparts for the pseudo-X-ray sources
with $\sigma_{{\rm X:R}i}<0.\!^{\prime\prime}8$, and 
the fraction decreases down to 40\% for the pseudo-X-ray
sources with $\sigma_{{\rm X:R}i}=2.\!^{\prime\prime}0$ and
counterparts of $R=26.0$mag. 

It is important to evaluate
the fraction of correct identifications among the $R$-band 
primary candidates. For pseudo-X-ray sources with 
$\sigma_{{\rm X:R}i}<0.\!^{\prime\prime}8$ and primary candidate
of $R<25.5$mag, 95\% of the primary candidates are the correct
identification. On the other hand, only 50\% of the primary
candidate with $R\sim25.0$ mag are correct for pseudo-X-ray
sources with $\sigma_{{\rm X:R}i}\sim2.\!^{\prime\prime}0-3.\!^{\prime\prime}0$.
The results also show that even among the optical identifications with
a bright optical counterpart, $R\sim20$ mag, there can be non-negligible
fraction of incorrect identifications. 

Based on the simulations, we can estimate the probability of
mis-identification for each X-ray source with $\sigma_{{\rm X:R}i}$
and the $R$-band magnitude of the primary candidate.
Summing the probability of the mis-identification for all
of the X-ray sources, we calculate the expected number of
mis-identification to be 102 among the 945 sources in the total sample.
The identification only with the $R$-band information
is clearly affected by mis-identification.

\subsection{Identification with 3.6{\micron} Objects and Its Reliability}

Most of the expected mis-identifications are thought to be
associated with X-ray sources with large $\sigma_{{\rm X}i}$ and
faint optical counterparts. Because optically-faint X-ray sources
have on-average a red optical-MIR color (\cite{yan03}; \cite{koekemoer04}; \cite{mignoli04}); 
identification 
in the MIR-band can be more reliable (\cite{brusa10}; \cite{civano12}).
Upper middle and upper right panels of Figure~\ref{fig:coord_diff}
show the radial distribution of 3.6{\micron} sources from the centers
of the X-ray sources. 
For the uncertainty radius, $\sigma_{{\rm X:3.6{\micron}}i}$, 
we use root sum squared of the $\sigma_{{\rm X}i}$ and 
$1.\!^{\prime\prime}1$. The latter term is derived same way 
for the $R$-band sources.
If we take the ratio between the numbers of excess
and contaminating sources within $2\times\sigma_{{\rm X:3.6}i}$, 
the ratios are 2.2 and 0.9 at the
depth of the SWIRE and SpUDS coverages, respectively. The ratios are higher
than the ratio in the $R$-band, and it is expected that more reliable
identification can be made in the 3.6{\micron}-band.
\citet{feruglio08} really show that 3.6{\micron} identification provides
a better counterpart assignment than optical only identification 
for {\it XMM-Newton} X-ray sources.
They checked the $R$ and 3.6{\micron} identifications of {\it XMM-Newton} X-ray sources
with the source positions measured with {\it Chandra}, and found  20\%
and 4\% different identifications from {\it Chandra} for the $R$ and
3.6{\micron} primary candidates, respectively.

In order to improve the reliability of the identification, 
we apply the 3.6{\micron} identification utilising the
SWIRE and SpUDS 3.6{\micron} images. First, we construct a
3.6{\micron} object catalog applying Sextractor \citep{bertin96} to the
3.6{\micron} images with detection criteria of 
3 connected pixels above 2$\sigma$ of the background.
We use MAG\_AUTO as the 3.6{\micron} magnitude.
Then, we identify the X-ray sources with the 3.6{\micron} objects
in the same way for the $R$-band objects using $L_{ij}$ and
$R_{ij}$ determined with the 3.6{\micron} object list.
We consider 3.6{\micron} sources within
5$\times \sigma_{{\rm X:3.6{\micron}}i}$ of each X-ray source.
The number count of the 3.6{\micron} objects are derived
from the source number density in the 3.6{\micron} source catalog. We select
3.6{\micron} primary candidates with $Q(<m_{{\rm 3.6 \micron}j}, f_{Xi})=1.0$, 
then we derive $Q(<m_{{\rm 3.6 \micron}j}, f_{Xi})$ with the
3.6{\micron} magnitudes of the primary candidates. 
Finally, we conduct the 3.6{\micron} identification with the
derived $Q(<m_{{\rm 3.6 \micron}j}, f_{Xi})$.

The distribution of the reliability of the 3.6{\micron} primary
candidate is shown in Figure~\ref{fig:Reliable_hist_all} with the red
dotted lines. The fraction of the high reliability identifications
among the 3.6{\micron} identifications is higher than that among 
the $R$-band identifications: 80\% of the sources have a primary candidate
with $R_{ij}>0.8$. We evaluate the expected number of 
mis-identifications among the 3.6{\micron} primary candidates
using a Monte Carlo simulation as for the $R$-band
identification. The mis-identification rate 
for all of the X-ray sources is 44 among the 945 X-ray sources.
The fraction 
of expected mis-identifications is consistent with that reported in 
\citet{feruglio08}. The expected
number of mis-identifications is significantly reduced from 
that of the $R$-band identification.

The PSF size of the 3.6{\micron}-band IRAC image is larger than 
that of the $R$-band image, but as can be seen in Figure~\ref{fig:Identification_sample},
sufficiently small to pinpoint which 
is more reliable counterpart among the multiple $R$-band sources that 
are in the X-ray uncertainty area. We use 3.6{\micron} identification
to pinpoint the optical counterpart, and we refer the position of the
selected optical counterpart as the coordinate of the counterpart.

Similar to the 3.6{\micron} identification, identifications in the
$K$-band can be effective in selecting optically-faint and red counterparts.
Additionally the PSF size of the UKIDSS UDS $K$-band image is significantly
smaller than that of the IRAC 3.6{\micron} image, therefore it is possible
that true counterparts that are blended with contaminating sources in the 
3.6{\micron} image can be resolved in the $K$-band image.
Using the UKIDSS UDS $K$-band image, we check the difference of the $K$-band
identifications from those in the $R$ and 3.6{\micron} bands. The $K$-band
identifications basically follow the $R$ and 3.6{\micron} identifications;
among 762 X-ray sources covered in the $K$-band image, 636 and 650 sources
have $K$-band primary candidate consistent with primary candidate in 
the $R$ and 3.6{\micron} identifications, respectively. If we remove $K$-band
primary candidate which are located far outside of the X-ray error circle,
there are 11 $K$-band primary candidates that are different from both of
the $R$-band and 3.6{\micron} identifications. All of the 11 $K$-band 
primary candidates have reliability lower than 0.5, i.e. there are other
$K$-band source with similar likelihood, and the identifications are
marginal. Considering the similarity to the $R$ and 3.6{\micron} identifications
and the limited coverage of the $K$-band image, we do not refer the
$K$-band identifications in the final selection of the primary counterparts.

\subsection{Primary Counterpart Selected with Multi-wavelength Identification}

\begin{figure}
 \begin{center}
  \includegraphics[width=75mm]{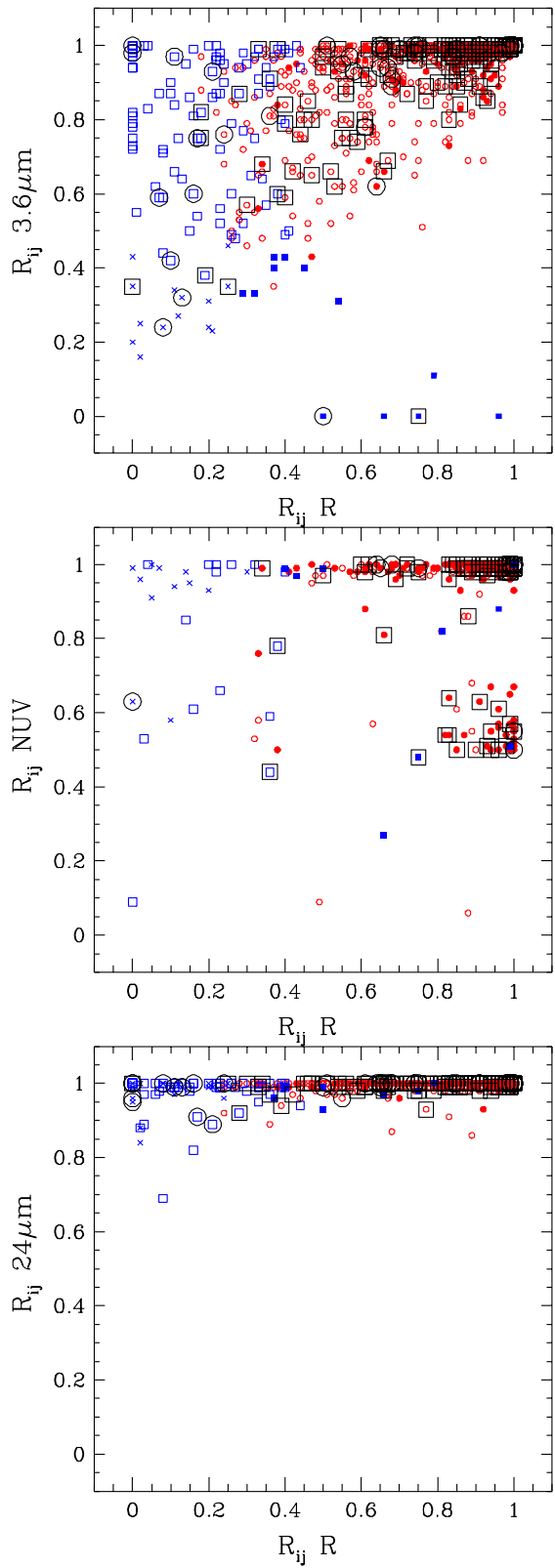} 
 \end{center}
\caption{
The reliability $R_{ij}$ determined in the 3.6{\micron} (top), 
NUV (middle), and 24{\micron} (bottom) bands identification vs. 
$R$-band identification. Red filled circles represent objects where all
4 identifications are the same. Red open circles have 
matched identifications in the $R$ and 3.6{\micron} bands.
Blue filled squares and blue open squares are 
identified either in $R$ or 3.6{\micron} bands. Blue crosses
are identified in the NUV or 24{\micron} bands.
Black open squares and black open circles indicate 
objects matched with radio and variability AGN candidates, 
respectively.
}\label{fig:Multi_reliability}
\end{figure}

Utilising the multi-wavelength dataset of the SXDS, we can
further improve the reliability of the identification. Luminous
blue broad-line AGNs at low to intermediate redshifts are
reliably identified with strong NUV emission. On the other
hand, obscured narrow-line AGNs show strong 
24{\micron} emission. 
In the bottom panels of Figure~\ref{fig:coord_diff},
radial distributions of NUV and 24{\micron} sources are shown.
The 1$\sigma$ uncertainty radius, $\sigma_{{\rm X:NUV}i}$
and $\sigma_{{\rm X:24{\micron}}i}$, are
calculated with root sum of squares
of $\sigma_{{\rm X}i}$ and the additional term associated with
the systematic X-ray source and NUV or 24{\micron} position uncertainty.
We use $2.\!^{\prime\prime}3$ and $1.\!^{\prime\prime}4$ as the
latter term for NUV and 24{\micron} identifications, respectively.
In the same way for $R$ and 3.6{\micron}-bands, if we consider
within $2\times\sigma_{X:NUV}i$ or $2\times\sigma_{X:24}i$, 
the ratio between the number of excess and contaminating sources
are 2.5, 22.0, and 8.3 for NUV down to 25.1 mag, 24{\micron}
down to 18.1 mag (SWIRE depth) and 24{\micron} down to
20.0 mag (SpUDS depth), respectively. Thus most of the 
associated 24{\micron} sources are expected to be counterparts
of the X-ray sources. Some contaminations are expected for
NUV sources, see discussions on the distributions of reliability
below.
Therefore, we utilise
NUV and 24{\micron} identifications as well as the $R$ and
3.6{\micron} identifications to determine the primary counterpart
of each X-ray source. We refer a primary object selected
in each identification as the primary candidate, and the
best candidate in the combined identifications as the 
primary counterpart.

We perform NUV and 24{\micron} identifications in the same way
as for the $R$ and 3.6{\micron} identifications. 
First, we construct object catalogs from the NUV and 24{\micron}
images with Sextractor. We use the NUV and 24{\micron} magnitude determined
in $7.\!^{\prime\prime}6$ and $6.\!^{\prime\prime}0$ diameter
apertures with aperture corrections of 0.59 and 1.23 mag
determined with average PSF, respectively. 
The number densities of NUV and 24{\micron} sources are
directly derived from the source list made by applying
Sextractor to the SXDS dataset. For the NUV and 24{\micron}
identifications, we do not consider
$Q(<m_{j}, f_{X_{i}})$ and set $Q(<m_{j}, f_{X_{i}})=1.0$ for
simplicity. We calculate $R_{ij}$ in the same way for $R$ and
3.6{\micron} identification.
As shown in Figure~\ref{fig:Identification_sample}, 
the PSF sizes of the NUV and 24{\micron}-band images are 
significantly larger than that of the $R$-band images.
The PSF sizes are still sufficiently small to pinpoint
one $R$-band counterpart in most of the cases. We apply eye-ball
check for the $R$-band images of the NUV and 24{\micron} sources.
Same for 3.6{\micron}
identification, even if we use NUV or 24{\micron} identification, 
we refer the position of the selected optical counterpart as 
the coordinate of the counterpart.

The distribution of the $R_{ij}$ of NUV and 24{\micron} 
identifications are shown with green dashed and blue dot-dashed
lines, respectively, in Figure~\ref{fig:Reliable_hist_all}.
The NUV identification is good for blue luminous broad-line AGNs, 
and higher reliability identification can be achieved than $R$
identification for more than 40\% of the X-ray sources. However, 
another 40\% of the sources have multiple NUV sources with similar
magnitude in the uncertainty area giving a reliability less than 0.5.
In the 24{\micron} band we achieve highly reliable identification,
$R_{ij}>0.8$, for 80\% of the sources. For the rest of the X-ray
sources, the 24{\micron} data is not deep enough to detect the counterparts
of the X-ray sources and the fraction is saturated at 80\%.
The higher reliability of the NUV and 24{\micron} identifications can partly
be caused by the blending effect due to the large PSF size in the two
bands. The counterparts of the X-ray sources are bright in the NUV and 24{\micron} bands, and
fainter NUV or 24{\micron} band sources that are contaminating and 
are located at the X-ray source position by chance can be hidden 
due to blending with the brighter true counterpart.
Such effect of blending is discussed for 3.6{\micron} sources in \citet{cardamone08}.

Additionally, candidates of AGNs are selected in 
radio and variability surveys in SXDS, and we utilise this
information in the final identification. 
A large part
(0.8 deg$^{2}$) of the SXDS field is covered by the
Very Large Array (VLA) radio survey observation at
1.4GHz down to 100$\mu$Jy \citep{simpson06, simpson12}. 
In total, 505 radio sources are detected, and 480 sources
are within the X-ray coverage region. Among the 945 X-ray sources
in the total sample, 789 of them are covered in the 100$\mu$Jy
sample.
If we consider 1$\sigma$ positional matching uncertainty, 
$\sigma_{{\rm X:radio}i}$
with the root sum of squares of $\sigma_{{\rm X}i}$ and $1.\!^{\prime\prime}0$,
there are 62 radio sources associated with the X-ray sources
within 4.0$\times \sigma_{{\rm X:radio}i}$. Most of the broad-line
AGNs detected in the radio survey are also detected in X-ray;
among 25 broad-line AGNs detected in the VLA survey \citep{simpson12}, 
21 of them are in the X-ray survey region, and 17 of them are
associated with X-ray sources. 
Three radio broad-line AGNs without X-ray detection
are fainter than $i'$=23.5 mag, and are likely to be below the
X-ray detection limit. 
The remaining radio broad-line AGN without X-ray detection
are located in a gap of the X-ray detectors, and the AGN can 
be missed due to bad sensitivity in the region.

Variable AGN candidates are selected with $i'$-band Suprime-cam
imaging data \citep{morokuma08a, morokuma08b}. There are 936
variable objects detected down to magnitude of the varying component 
of $i^{\prime}_{\rm vari}=25.5$
in the overlapping region with the X-ray images \citep{morokuma08b},
and 193 of them are matched with X-ray sources.

Finally the primary counterpart is determined utilising all of the
$R$, 3.6{\micron}, 24{\micron}, and NUV identifications.
Radio and variability AGN candidates are also considered 
during the selection of the primary counterpart.

Among the 945 X-ray sources, 213 sources have matched primary
candidate considering all 4 identifications, and we select this
object as the primary counterpart. 
404 X-ray sources have only one object which have
likelihood ratio larger than 1 either in the 4 identifications
within the X-ray uncertainty area, and we select the 
object as the primary counterpart.
For the remaining X-ray sources, there are more than 2 
objects with $L_{ij}>1$ within the X-ray uncertainty area.
We multiply the likelihood ratios of the 4 identifications
for each source, and if there is an object with a
likelihood ratio product larger than 1000, we select this object as the
primary counterpart. There are 112 X-ray sources 
that meet the criteria. 
Of the remaining 183 X-ray sources, 
90 X-ray sources have a primary candidate selected
both in the $R$ and 3.6{\micron} identifications. 
Among them, we select 77 of them as the primary
counterpart. For 13 X-ray sources, there is
another primary candidate selected at 24{\micron}
or in the NUV, and the candidate is selected as the
primary counterpart.
The other 93 X-ray sources have different primary
candidates in the $R$ and 3.6{\micron} identifications.
For such source, we determine the primary counterpart
comparing the likelihood ratios in the 4 identifications.
We select 52, 24, 20, and 6 primary candidates with the
3.6{\micron}, $R$, 24{\micron}, and NUV identifications,
respectively. 

The distribution of the reliability of the primary
counterparts in the $R$, 3.6{\micron}, 24{\micron}, and
NUV identifications are shown in Figure~\ref{fig:Multi_reliability}.
Almost all of the primary counterparts have high reliability 
 in one of the 4 identifications. 
The list of the primary counterparts is shown in Table~\ref{tab:XrayID}.
In the list, the X-ray identification number (column 1), 
X-ray RA and Dec. position (column 2 and 3), X-ray 
1$\sigma$ uncertainty radius, $\sigma_{{\rm X}i}$ (column 4), 
and likelihood in the 0.5-2~keV
and 2-10~keV bands (column 5 and 6) are shown. 
Columns 7 to 10 provide the information from the 
multi-wavelength identification process in the
$R$ (7), NUV (8), 3.6{\micron} (9), and 24{\micron} (10)
bands. Flag 1 means the primary counterpart is
the primary candidate in the band, flag $-1$ means
the counterpart is the primary candidate but
the likelihood is less than 1.0, and flag $0$ means
the counterpart is not the primary candidate in the band.
Column 11 and 12 show the matching with the 1.4GHz radio
source and variability AGN candidates.
The RA and Dec. of the primary counterpart is given in 
columns 13 and 14.
The multi-wavelenth images of the X-ray sources
are available on-line\footnote{See http://www.astr.tohoku.ac.jp/\~{}akiyama/SXDS/post\_stamp.html}.

\subsection{Candidates of Clusters of Galaxies}

The  identifications discussed above assume the X-ray source
is a single object. This is not the case for
clusters of galaxies, which are non-negligible population
among extra-galactic X-ray sources. Some of the identifications
with low-reliability can be caused by the multiple galaxies
associated with an X-ray source that is associated with a cluster of galaxies. 

\citet{finoguenov10} constructed catalogs of groups and
clusters of galaxies in the SXDS field through a wavelet
detection of extended X-ray emission for the {\it XMM-Newton}
imaging dataset. They cross-match their candidates with
the extended sources in \citet{ueda08}, and 14 sources
in \citet{ueda08} are matched. Among them, we regard 10
extended X-ray sources as candidate of clusters of galaxies as listed
in Table~\ref{tab:CLUSTER}.
Among the remaining 4 sources, we regard 3 sources
as candidates of AGNs.
SXDS0051 is cross-identified with SXDF47XGG at $z=0.348$, but
SXDS0051 is significantly detected in the 2.0--10~keV band and
the observed 0.5--2~keV and 2--4.5~keV hardness ratio, HR2,
is relatively large ($-0.50$).
Thus we regard the X-ray source as an AGN candidate.
SXDS0396 is cross-identified with SXDF64XGG at $z=1.030$, but
SXDS0396 is significantly detected in the 2.0--10~keV band 
and HR2 ($-0.36$) implies a power-law with photon index of 1.6.
Therefore, we regard the X-ray source as an AGN candidate. The
source is also close to SXDS0386.
SXDS0934 is cross-identified with SXDF06XGG at $z=0.451$, but the
primary optical counterpart is identified with
a broad-line AGN at $z=1.669$, therefore we
regard the AGN as the X-ray source.
SXDS0784 matched with
SXDF07XGG has a detection likelihood less than 7
in both of the soft and hard X-ray bands, and it is
not included in the total sample.

We also found 4 additional candidates of clusters of
galaxies with an excess of objects with similar $B-z'$ and
$z'-K$ colors within $r=15^{\prime\prime}$ from the
X-ray source position. These  are also listed in Table~\ref{tab:CLUSTER}.
The median photometric redshift of objects
within $r=30^{\prime\prime}$ from the X-ray position is
used as the photometric redshift for the candidates.

In the table, the X-ray source ID (Column 1), HR2 (2),
spectroscopic redshift (3), and photometric redshift (4) are summarised.

\subsection{Candidates of ULXs}\label{sec:ULX}

Four X-ray sources, SXDS0416, SXDS0390, SXDS0404, and SXDS0402,
are associated with a bright nearby
galaxy KUG 0214$-$057 at $z=0.018$ \citep{watson05}.
They are strong ultra-luminous
X-ray sources (ULXs) candidates.

\section{Spectroscopic Identification}

\subsection{Optical Spectroscopic Observations}

\subsubsection{FOCAS/Subaru Observation in Selected Fields}

We conducted a spectroscopic campaign observation with a multi-slit
optical spectrograph, FOCAS on the Subaru telescope from
January 2001 to November 2005. 
We targeted the primary counterparts with the highest
priority, and in addition we observed secondary counterparts
if possible. We put the slits at the positions of the 
primary counterparts determined in the optical images,
even for primary counterparts selected by the identifications
in the 3.6$\mu$m, 
24$\mu$m, or NUV band. The situation is the same for following
spectroscopic observations.
The FoV of
the instrument is a 6$'$ diameter circle. In total 50 fields
were observed. About three quarters of the fields were selected
to cover as many as X-ray sources as possible. The remaining
fields were chosen to cover as many as $z\sim1$ red sequence
galaxies as possible \citep{yamada05} and to cover a
candidate $z=5.7$ cluster of galaxies
found by narrow-band surveys \citep{ouchi05, ouchi08}.
In such fields, X-ray sources are observed as a filler
targets for the masks.

In the observations, we use a 150R grating with SY47 order-cut
filter to cover 4700{\AA} to 9400{\AA}. A slit width of
$0.\!^{\prime\prime}8$ is used and the resulting spectral resolution
is 20{\AA}. Hereafter, we use the spectral resolution 
measured with night sky emission lines obtained simultaneously 
with the target spectrum. The spectral resolution for the
target can be higher because the slit
illumination by a stellar object can be more concentrated 
than the smooth illumination by night sky emission.
In later observations, we use 300B grating with order-cut
filter SY47 or without the filter. With the order-cut
filter the spectral coverage is 4700{\AA} to 9200{\AA}.
Without the filter the spectral coverage in the blue 
extends down to 3800{\AA}, but 2nd order light can contaminate
wavelength range above 7600{\AA}. A slit width of $0.\!^{\prime\prime}8$
is used and the resulting spectral resolution is 13{\AA}.
The central coordinate, observing mode, and the exposure time
of each field are summarised in Table~\ref{tab:FOCAS_FOVs}.
During the observation, a few additional objects were
observed in long-slit mode with 300B grating with SY47 order-cut filter.

The data are reduced with a script based on IRAF.
The procedure follows usual analysis of long-slit observations.
The bias counts are subtracted using the overscan area of the CCD,
and a flat-fielding correction is applied with a dome flat frame
taken after the observation. Wavelength calibration is performed using a
ThAr lamp spectrum. We did not dither the object position
along the slit during the observation, and the sky subtraction
is performed by using the sky spectra obtained on either side
of an object.
The flux calibration of the reduced spectra is carried out
utilising the spectra of spectroscopic standard stars
taken with a $2.\!^{\prime\prime}0$-width slit. 
Feige 110 and G191B2B are used as the standard stars.
The standard stars are obtained with and without SY47
order-cut filter, in order to evaluate the contamination
of the 2nd order light at the red end of the wavelength 
coverage.
The sensitivity curve for each night is determined
with the standard stars. Furthermore, we adjust the
normalization of the spectra of the targets using their 
photometric data.
After the sensitivity correction,
we also applied correction for atmospheric
absorption features, A-band and B-band, using 
the absorption profile of the bands derived with a
bright object in the same mask.

\subsubsection{2dF/AAT Observation for Bright Optical Counterparts}

Spectroscopic follow-up observations for bright optical
counterparts with $R<21.0$ mag were obtained with the 2dF
400 multi-fibre optical spectrograph on Anglo-Australian 3.9m
Telescope on 3-6 Nov. 2002 in director's discretionary time.
The integration time ranges from 6 min for the
brightest objects to 3 hours for the faintest objects.
The 2dF has 2 spectrographs, each spectrograph covers 200 fibres.
We used a 270R grating with wavelength range of 4400-9100{\AA}
for the first spectrograph, and 316R grating with wavelength range
of 4700-8500{\AA} for the second spectrograph.
The spectral resolutions are 14 and 12{\AA} and the 
spectral samplings are 4.8 and 4.1{\AA} pixel$^{-1}$ for the
former and latter spectrographs. We observed a
relatively bright star in each fibre configuration in order
to correct wavelength dependence of the sensitivity.
The data are reduced in the standard way for the 2dF data
using the {\it 2dfdr} software.

\subsubsection{Additional Spectroscopic Observations for Faint Optical Counterparts}

Some of the radio and X-ray sources in the SXDS field were observed
with the Visible Multi-Object Spectrograph (VIMOS) on the
Very Large Telescope (VLT) (P074.A-0333; \cite{simpson12}).
The MR-Orange grating and the GG475 order-cut filter were
used and a
spectral resolution of 10{\AA} in the
wavelength range between 4800 {\AA} and 10,000 {\AA} is
achieved.
27 slit masks were prepared to cover $\sim80$\% of
the radio and X-ray sources in the SXDS. $\sim$75\% 
of the 27 masks were observed with one 2700-s and
two 1350-s exposures.

Additionally, two FoVs of VIMOS were observed
in the course of spectroscopic observations of
extended Ly$\alpha$ emitters by \citet{saito08} (P074.A-0524). 
The HR-Orange grism and the GG435 order-cut filter
were used. The wavelength range between 5000 and 6800{\AA} 
was covered with 2.8{\AA} resolution.

A deep spectroscopic survey for $K$-band selected galaxies in the
UKIDSS UDS region was conducted with the VLT VIMOS and VLT 
FOcal Reducer and low dispersion Spectrograph 2 (FORS2)
instruments in the UDSz ESO Large Programme (180.A-0776, PI:Almaini;
\cite{bradshaw13}; \cite{mclure13}).
In the VIMOS observations, the LR-Blue and LR-Red grisms
were used for each mask and wavelength range between 3700-9500{\AA}
was covered. A spectral resolution of 35{\AA} was achieved
at 7000{\AA}. The FORS2 observations were conducted with a
300I grism covering 6000-10000{\AA} and with spectral
resolution of 12{\AA}.

In the course of the optical spectroscopic survey of
Lyman Alpha Emitters and Lyman Break Galaxies in the SXDS field, the
primary counterparts of the X-ray sources were observed
as filler objects in the masks.
A multi-object spectroscopic observation with the
DEep Imaging Multi-Object Spectrograph (DEIMOS) instrument 
attached to the Keck II telescope was also carried out \citep{ouchi09}.
The observations used the 830G grating and the GG495 filter. The 5700-9500{\AA}
wavelength range was covered with a spectral resolution of
2.6{\AA}. 

Further spectroscopic survey observations were made with the Inamori-Magellan Areal Camera and Spectrograph
(IMACS) on the Magellan 6.5m Baade telescope  \citep{ouchi09}.
In the IMACS observations, the F/2 camera
with $27.\!^{\prime}4$ diameter FoV and 300 lines mm$^{-1}$ grism
with central wavelength of either 6700{\AA} or 8000{\AA} 
were used. Wavelength ranges of from 4700{\AA} to 8000{\AA} 
(from 6200{\AA} to 9600{\AA}) were covered with 
7{\AA} (6{\AA}) resolution and 1.34{\AA} pixel$^{-1}$
(1.25{\AA} pixel$^{-1}$) by the former (latter) setup.

X-ray sources are also observed as filler targets during
spectroscopic observations of the Balmer-break galaxies in the 
field with IMACS on Magellan Baade telescope
(PI. N.Padilla; \cite{diaztello13}). In the observation, the F/2
camera with $27.\!^{\prime}4$ diameter FoV and 200 lines mm$^{-1}$
grism with central wavelength of 6600{\AA} was used. 
The wavelength range from 4500{\AA} to 9000{\AA} is covered
with 10{\AA} spectral resolution and 2.0{\AA} pixel$^{-1}$
sampling.

Some QSO candidates identified with the KX selection technique
were spectroscopically observed with the AAOmega spectrograph on
the AAT 3.9m telescope and X-ray AGNs are also observed as 
filler targets \citep{smail08}.
The 580V and 385R gratings were used in the dual-beam 
spectrograph of AAOmega and the wavelength range from 
3700{\AA} to 8600{\AA} was covered simultaneously. 
Spectral resolution is 5.0{\AA} with spectral
sampling of 3.4{\AA} pixel$^{-1}$.

\subsection{NIR Spectroscopic Observations with FMOS}

In addition to the optical spectroscopic observations, we conducted intensive NIR
spectroscopic observations with the Fiber Multi Object Spectrograph
(FMOS) on the Subaru telescope \citep{kimura10}. This instrument can observe up to 200 objects
simultaneously over a
30$^{\prime}$ diameter FoV in the cross-beam switching mode with two spectrographs, IRS1 and IRS2.
The fibers have an aperture of $1.\!^{\prime\prime}2$ diameter on the sky.
The spectrographs cover the wavelength range between 9000{\AA} and 18000{\AA}
with spectral resolution of $R\sim800$ at $\lambda\sim1.55 \mu$m in the low-resolution mode.
A total of 851 sources were observed in observations of 22 FoVs 
with this setup during guaranteed, engineering
and open-use (S11B-098 Silverman et al. and S11B-048 Yabe et al.) time observations.
The central coordinate, observation period, integration time, and observing mode
of each FoV are summarised in Table~\ref{tab:FMOS_FOVs}. 

The NIR spectra are reduced with the pipeline data reduction 
software, FIBRE-pac \citep{iwamuro12}. The resulting one-dimensional
spectra are corrected for atmospheric absorption and sensitivity
dependence on wavelength using relatively faint F-G type stars
observed simultaneously with the targets. 

\subsection{Spectroscopic Identification Summary}

The primary counterparts of 31 X-ray sources 
with a bright stellar object show optical spectra consistent
with Galactic stars. We identify them as Galactic stars. 
The ID number and HR2 of the X-ray sources are given 
in Table~\ref{tab:STAR}. Almost all of them have a small
HR2 value, which is consistent with the thermal X-ray 
emission from Galactic stars. 
Hereafter, we exclude X-ray sources that are 
selected as candidates of clusters of galaxies (Section 4.4), 
ULXs (Section 4.5), and Galactic stars, and
refer to the remaining X-ray sources as AGN candidates.
There are 896 AGN candidates in the total sample
as summarised in Table~\ref{tab:number_summary}.

Among the primary counterparts of the 896 AGN candidates,
597 counterparts are observed in the optical
spectroscopic observations discussed above, as is summarised in Table~\ref{tab:number_summary}.
As a first step, we determine the redshift
of each object by examining its optical spectrum and
fitting Gaussian profiles to the
observed emission and absorption lines.

A large fraction of objects show multiple features.
If a narrow emission line is detected, we determine
the redshift with that line. For example, 
if [O \emissiontype{II}]$\lambda$3727 emission 
line is detected, we determine the redshift of the
object with that emission line. For low-redshift 
objects, the optical spectrum covers the H$\alpha$ and H$\beta$
wavelength range and shows strong [O \emissiontype{III}]$\lambda$5007
emission line. We determine their redshifts with
the [O \emissiontype{III}]$\lambda$5007 line.
For high-redshift objects, optical spectrum covers the
rest-frame wavelength range below 
Mg \emissiontype{II}$\lambda$2800 emission line. 
If that emission line is detected, it is used to
determine the redshifts, otherwise, C \emissiontype{III}]$\lambda1909$ or
C \emissiontype{IV}$\lambda1550$
emission lines are used. For objects only showing 
absorption lines, we determine
their redshift with the Ca \emissiontype{II} HK doublet absorption lines 
at 3969{\AA} and 3934{\AA}, if these
absorption lines are detected. 

We regard a permitted emission line with velocity width 
larger than 1000 km s$^{-1}$ (after deconvolving the instrumental
profile) as a broad-emission line.
This threshold is narrower than the typical criteria
for broad-line AGNs (1500 or 2000 km s$^{-1}$). We set
the threshold considering the distribution of the FWHM
of the broad-line AGNs in the local Universe
\citep{hao05, stern12}. Although the threshold is 
narrower, most of the broad emission lines are broader than 
1500 km s$^{-1}$ and the distribution of the FWHM of the
broad-emission lines, especially the lack of the 
broad-emission lines below 1500 km s$^{-1}$, is similar 
to that of optically-selected broad-line AGNs in the SDSS
as discussed in \citet{nobuta12}.

Some objects only show one either broad or narrow
emission line in the observed optical wavelength range. 
For such objects, we identify most of the single narrow-emission 
lines with [O \emissiontype{II}]$\lambda$3727, considering
the absence of any other emission line. If the identification 
 is not conclusive
due to the narrow wavelength coverage of a spectrum, 
we identify the single narrow-emission line, utilising
the photometric redshift described in the next section.
For objects with one prominent broad emission line, 
 we mostly identify it with
Mg \emissiontype{II}$\lambda$2800,  consistent with the
absence of another broad emission line in the observed
wavelength range. The identifications with a single  broad-emission line in the optical are checked 
with the NIR spectra covering rest-frame optical
wavelength range. 

Intensive spectroscopic observations in the NIR
wavelength range cover 851 of the total X-ray sources
as summarised in Table~\ref{tab:number_summary}.
Spectroscopic identification in the NIR spectra is
conducted in the same manner as for the optical spectra. 
In the NIR spectra, 
if one narrow or broad emission line is detected, 
almost all of them are identified with the H$\alpha$ emission line.
We identify some single narrow-emission lines with 
[O \emissiontype{III}]$\lambda$5007, considering
the photometric redshift. In such cases, 
[O \emissiontype{III}]$\lambda$4959 is presumed to
be weaker than the detection limit.
Among the objects that are observed but not identified in the optical
spectroscopy, 23 objects are identified with NIR spectroscopic
observations with FMOS. They are AGNs at $z\sim0.9-2.5$, and
15 of them are narrow-line AGNs, which are mostly identified
through their strong H$\alpha$ or [O \emissiontype{III}]$\lambda$5007 lines.

Objects that show broad-emission in at least
one of the observed permitted emission lines are
referred with "broad-line" AGN, hereafter.
It needs to be noted that the criteria is not
homogeneous and depends on the rest-frame wavelength
coverage of the optical and NIR spectroscopic observations.
For example, some objects are identified as "broad-line" AGN
with one prominent broad H$\alpha$ emission line, but 
the same objects can be identified as "narrow-line" AGN,
if spectra covering H$\alpha$ line are not available.
Furthermore, an object with broad H$\alpha$ emission lines
can be type 1, 1.5, 1.8 or 1.9 Seyferts.
On the other hand, 
objects without broad-emission lines
are referred with "narrow-line" AGNs. Considering the
X-ray luminosity distribution of the sample, their 
X-ray emission is thought to be dominated by the AGN
component, and we ignore contributions to X-ray sources 
from star-forming galaxies in the AGN candidates.

\begin{figure}
 \begin{center}
  \includegraphics[width=75mm]{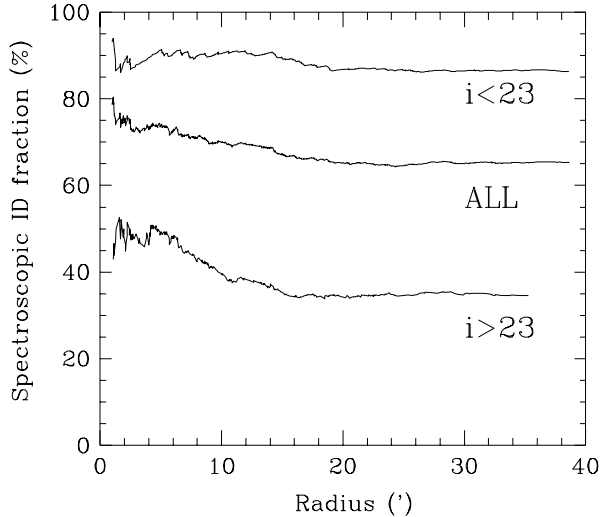}
 \end{center}
\caption{
Spectroscopic identification rate of the AGN candidates 
as a function of radius from the center of the survey field.
From top to bottom, the lines are for X-ray sources
with primary counterparts brighter than $i'=23$mag,
for all X-ray sources, and for X-ray sources with
$i'>23$ mag.}\label{fig:SXDS_idrate}
\end{figure}

Utilising the optical and NIR spectra, 586 out of the
896 AGN candidates (65\%) are spectroscopically identified.
The fraction of spectroscopically identified objects
in the entire AGN candidate sample is shown in Figure~\ref{fig:SXDS_idrate}
as a function of radius from the center of the survey
field. If we limit the sample to primary counterparts
brighter than $i'=23$ mag, the identification rate reaches
86\%. The spectroscopic identification of each AGN candidate
is summarised in Table~\ref{tab:Multi3}. Column 22 
shows the spectroscopic redshift, and column 23 describes the 
classification. BLA and NLA represent broad-line and narrow-line
AGNs, respectively. NID indicates the object is observed in the optical
but can not be identified with the optical spectrum.
NSP indicates the object is not observed in the optical.
NID and NSP do not refer the availability of NIR spectrum.


\section{Multi-wavelength Photometry and Photometric Redshift Determination}\label{sec:photometry}

\subsection{Multi-wavelength Photometry}

For X-ray sources without spectroscopic identification, 
we determine their photometric redshifts. In order to
determine photometric redshifts reliably, we 
conduct multi-wavelength photometry of each primary 
counterpart paying careful attention to the difference in the
PSF size and shape. 

In the photometry, the $i'$-band image taken with
Suprime-cam is used as the reference frame for the object detection.
Sextractor is used with a detection
threshold of 3 connected pixels above 2.0 $\sigma$. 
The aperture magnitude determined with a variable elliptical 
aperture (MAG\_AUTO) is used as the $i'$-band total magnitude of 
an object. We use PHOT\_AUTOPARAMS of (2.5, 3.5) to determine the
variable elliptical aperture. With the parameters, 94\% of 
total flux will be covered within the variable elliptical aperture
\citep{bertin96}.
Using the $i'$-band detected catalog, optical colors
of the object are measured in $2.\!^{\prime\prime}0$ diameter
aperture using the multi-image mode of Sextractor.
The total magnitude in
each optical band is determined with a fixed aperture color and
the aperture correction determined in the $i'$-band, i.e.
the difference between 
$i'$-band MAG\_AUTO and $i'$-band MAG\_APP.

For photometry in wavelengths other than the optical bands,
we measure a fixed aperture color between a selected band and
$i'$-band using the PSF-matched $i'$-band image whose PSF is matched
to the image in the selected band. The fixed aperture color
is converted into the total magnitude in the selected band
using the difference between the $i'$-band MAG\_AUTO in the
original $i'$-band image and the aperture magnitude measured
in the PSF-matched image. We try to conserve the fixed
aperture color, and assume the light profile in the 
selected band follows that in the $i'$-band. 
The details are described as follows.

At first we subtracted the background component of each image
using the object detection software, Sextractor. The background is
determined locally with median filtering in a 48$^{\prime\prime}$ box.
For very bright and extended objects with sizes larger than 30$^{\prime\prime}$, 
i.e. nearby galaxies,
the background removal subtracts the extended halo component, thus the
photometry is not reliable for such objects.
Background-subtracted images are magnified and aligned
to Suprime-cam $i'$-band imaging data. Coordinate conversions are
determined with positions of common bright stars using the {\bf geomap}
command in IRAF with a 2nd order polynomial and the images are
registered with the {\bf gregister} command. Fluxes of objects in the
original images are conserved.

The PSF size of the Suprime-cam $i'$-band image is the smallest
amongst the multi-wavelength data. We make the PSF-matched $i'$-band
images by convolving with a kernel which represents the PSF
difference between the $i'$-band and a given band. The determination
of the kernel and the convolution is performed with the
{\bf psfmatch} command in IRAF.

After matching the PSF size and shape, we measure colors of
objects with the multi-image mode of Sextractor. Colors in the
$i'$-band and a given band are measured from the images
in the band and the PSF-matched $i'$-band at the position in
the original $i'$-band image.
The total magnitude in the band is derived from the
aperture photometry and the aperture correction derived
from the difference between the MAG\_AUTO in the 
original $i'$-band image and the MAG\_APP in the PSF-matched
$i'$-band image.
Different aperture diameters are used for different 
datasets; $7.\!^{\prime\prime}6$ diameter for {\it GALEX} FUV and NUV,
$2.\!^{\prime\prime}0$ diameter for $U$-band images, and 
UKIDSS $J$, $H$, and $K$-bands, $3.\!^{\prime\prime}8$
diameter for the IRAC images.

The uncertainties in the photometric measurement are determined as
the 1$\sigma$ uncertainty of the flux measurement in the
aperture magnitude assuming a background limited regime.
For bright objects, the systematic uncertainty dominates the
measurement uncertainty, and constant values, 0.01-0.05 mag depending
on wavelength, are 
added to represent such systematic uncertainties.

In order to obtain photometric data for primary counterparts
that are not detected in the $i'$-band selected catalog,
we also construct
$B$-, $R$-, $V$-, $z'$- and 3.6{\micron} band selected catalogs
using the images in each band as the reference frame. We apply
the same procedures as for the $i'$-band selected catalog.

The results of the multi-wavelength photometry are summarised 
in Table~\ref{tab:Multi1}, \ref{tab:Multi2}, and \ref{tab:Multi3}.
The total magnitude and associated error for
each band are given. An error of 99.99 indicates the associated
magnitude is an upper limit. A total magnitude of 99.99 indicates 
the object is not covered in the imaging data of that band.
The values in the tables are corrected for Galactic extinction.
The applied corrections are summarised in Table~\ref{tab:imagelist}.

In the photometric redshift determination, we consider the
stellarity of the primary counterpart. The stellarity
is determined in the deep $i'$-band image.
For objects brighter than $i'=20.0$mag, the CLS\_STAR parameter 
from Sextractor is used. An object with CLS\_STAR
larger than 0.90 is classified as stellar. For fainter
objects, we use the FWHM in the $i'$-band image to define stellar
objects. If the $i'$-band FWHM is smaller than $0.\!^{\prime\prime}9$,
an object is classified as stellar.
For bright objects with saturation in their peak,
the stellarity measurements are not reliable. In such cases
we check the optical images by eye, and assign the corrected
stellarity in the catalog. The resulting stellarity is
also listed in Column 21 of Table~\ref{tab:Multi3}.

\subsection{Photometric Redshift Determination with Galaxy and QSO SED templates}\label{sec:photz}

\begin{figure*}
 \begin{center}
  \includegraphics[width=110mm]{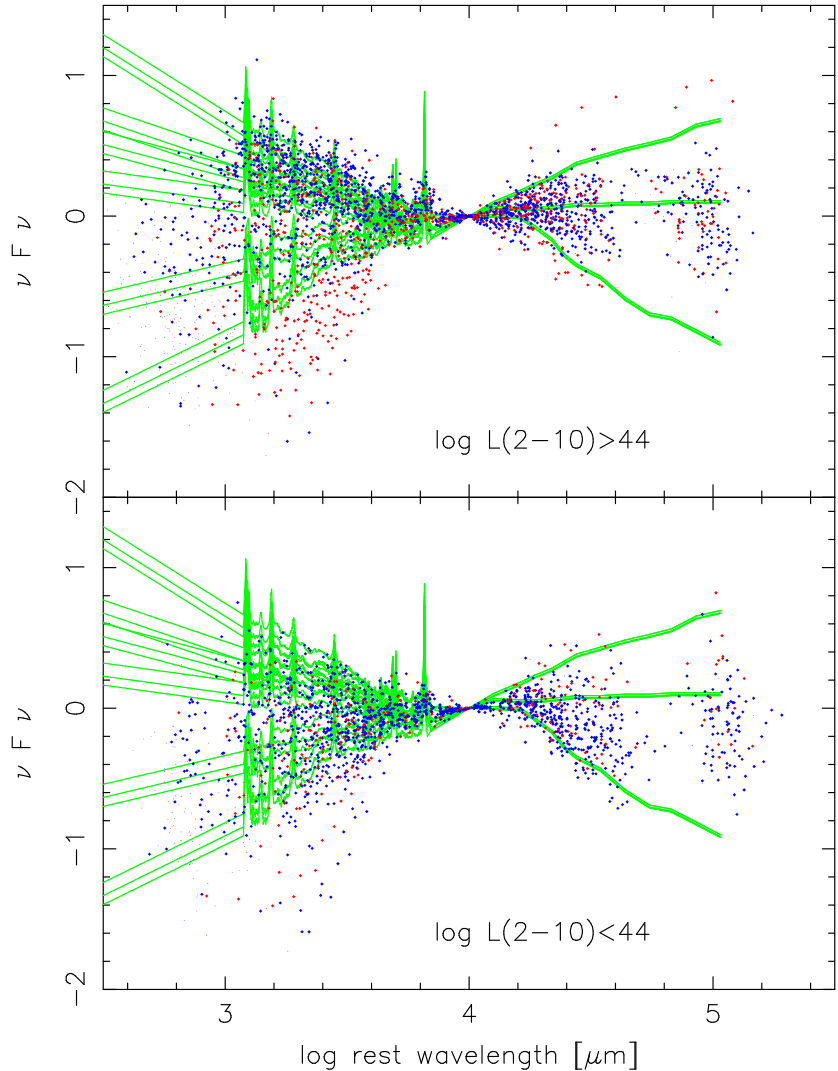} 
 \end{center}
\caption{
Template SEDs for QSOs (green solid lines) 
compared with
UV to MIR SEDs of spectroscopically-identified
broad-line AGNs. 
Blue and red symbols represent broad-line AGNs with X-ray
absorption smaller and larger than 
$\log N_{\rm H} = 22.0$, respectively.
Small dots indicate upper limits.}\label{fig:SED_plot_type1}
\end{figure*}

The photometric redshift of each primary counterpart is
determined with the hyperz
code \citep{bolzonella00}, which determines photometric
redshifts via template fitting with $\chi^2$ minimization
of the multi-wavelength photometry ranging from 1500{\AA}
to 8.0{\micron}. We use two sets of templates, one
represents galaxy SEDs and the other one represents
QSO SEDs.

For each object, we determine photometric redshifts 
from the galaxy and QSO templates separately.
For the galaxy templates, we use the spectral evolutionary
models of \citet{bruzual93} with Miller \& Scalo Initial
Mass Function (IMF) \citep{miller79} and evolving
metallicity. It needs to be noted that we assume a Salpeter
IMF \citep{salpeter55} in the determination of stellar mass of host galaxies
of narrow-line AGNs in Section 7.2 in order to compare with
results in literature. The difference in the estimated
photometric redshift caused by the difference in the 
applied IMF is expected to be negligible. In the hyperz code, Burst,
constant, and $\tau$ models with $\tau=1$Gyr to 30Gyr
are available. The dust extinction curve of star-forming
galaxies \citep{calzetti00} is utilised with $A_V=0.0$mag to
$A_{V}=2.5$mag in step of $\delta A_{V}=0.5$. 

The galaxy templates only represent the  stellar continuum
component, and neither the excess emission of an AGN hot dust component
nor PAH emission are included, although such emission can affect
the rest-frame wavelength range above 2$\mu$m.
Instead of including such a component in addition to the
galaxy SED model, we increase the IRAC photometric errors to compensate.

For QSO templates, we construct our own QSO template by
combining average UV-optical average SEDs from SDSS and
NIR-MIR SEDs of QSOs from Infrared Space Observatory ({\it ISO}) observations.
We use average QSO SEDs in \citet{richards03} in the
wavelength range between 1200{\AA} and 7000{\AA}. 
A total of 6 average QSO SEDs are constructed depending on the
UV-optical color. We extend the SED
to the shorter wavelength range by extrapolating the continuum component
in the above range with a power-law.
In the longer wavelength range above 1{\micron}, we used
three QSO IR SEDs from \citet{leipski05}.
The SEDs are constructed from the {\it ISO} observations of broad-line QSOs.
By adjusting the normalization, we connect the six optical and
the three IR SEDs smoothly and make 18 QSO templates covering
UV to MIR wavelength range. Because the observed slopes in 
the UV-optical and IR wavelength ranges are independent, 
we consider SEDs with blue UV-optical SED and red IR SED, 
and vice versa. The templates are compared with the observed SEDs of
spectroscopically-identified broad-line AGNs in Figure~\ref{fig:SED_plot_type1}.
All of the SEDs are normalized at 1{\micron} in the rest-frame.
The red and blue dots represents SEDs of X-ray AGNs with 
absorption characterised by a hydrogen column density, $N_{\rm H}$,
larger than and smaller than $10^{22}$ cm$^{-2}$, respectively.
The derivation of $N_{\rm H}$ is described in 
Section 7.1.
The templates enclose the entire range of SEDs of broad-line AGNs, 
although some AGNs with $\log N_{\rm H} > 22.0$ 
show redder SED in the UV to optical wavelength range.
Their SEDs are similar to the galaxy SEDs (see Section 7.3).
The templates also cover the variety of the SEDs observed
in the X-ray-selected broad-line AGNs in the COSMOS survey \citep{elvis12}.

\begin{figure*}
 \begin{center}
  \includegraphics[width=170mm]{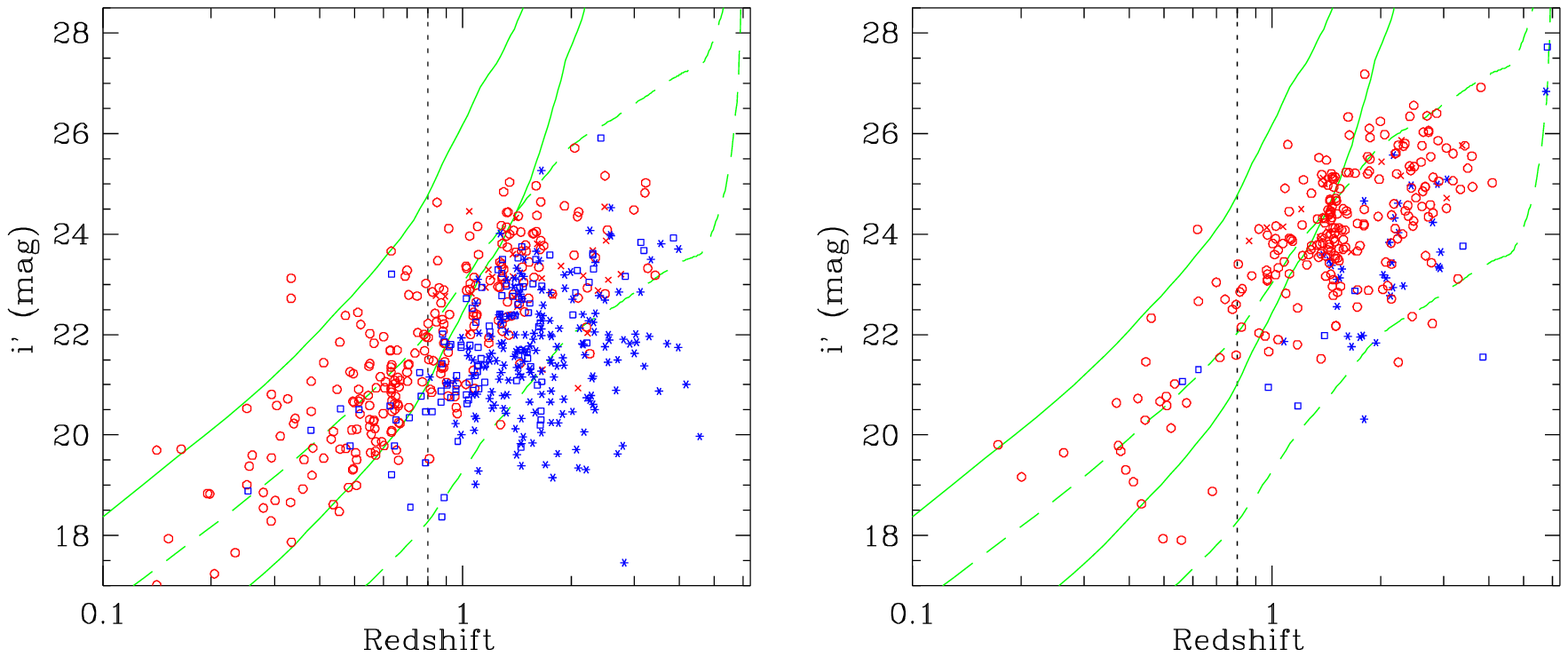} 
 \end{center}
\caption{
Left) Stellarity distribution of spectroscopically
identified AGNs plotted on the redshift vs. $i'$-band magnitude plane.
Blue asterisks and red crosses represent AGNs with stellar morphology
in the $i'$-band image.
Blue (square and asterisk) and red 
(circle and cross) symbols show broad-line and
narrow-line AGNs, respectively.
The vertical dotted line marks
$z=0.8$. The green lines 
indicate the redshift evolution of the $i'$-band
magnitude for a 10Gyr-old (solid) and 2Gyr-old (dashed) single burst stellar
population with stellar mass of $\log M_{\rm *}=10.5$ (upper)
and $\log M_{\rm *}=12.0$ (lower).
Right) Same for AGNs without spectroscopic identification.
In this panel, AGNs are classified photometrically as described
in Section 6.2.}\label{fig:SXDS_Cstar_i}
\end{figure*}

\begin{figure*}
 \begin{center}
  \includegraphics[width=170mm]{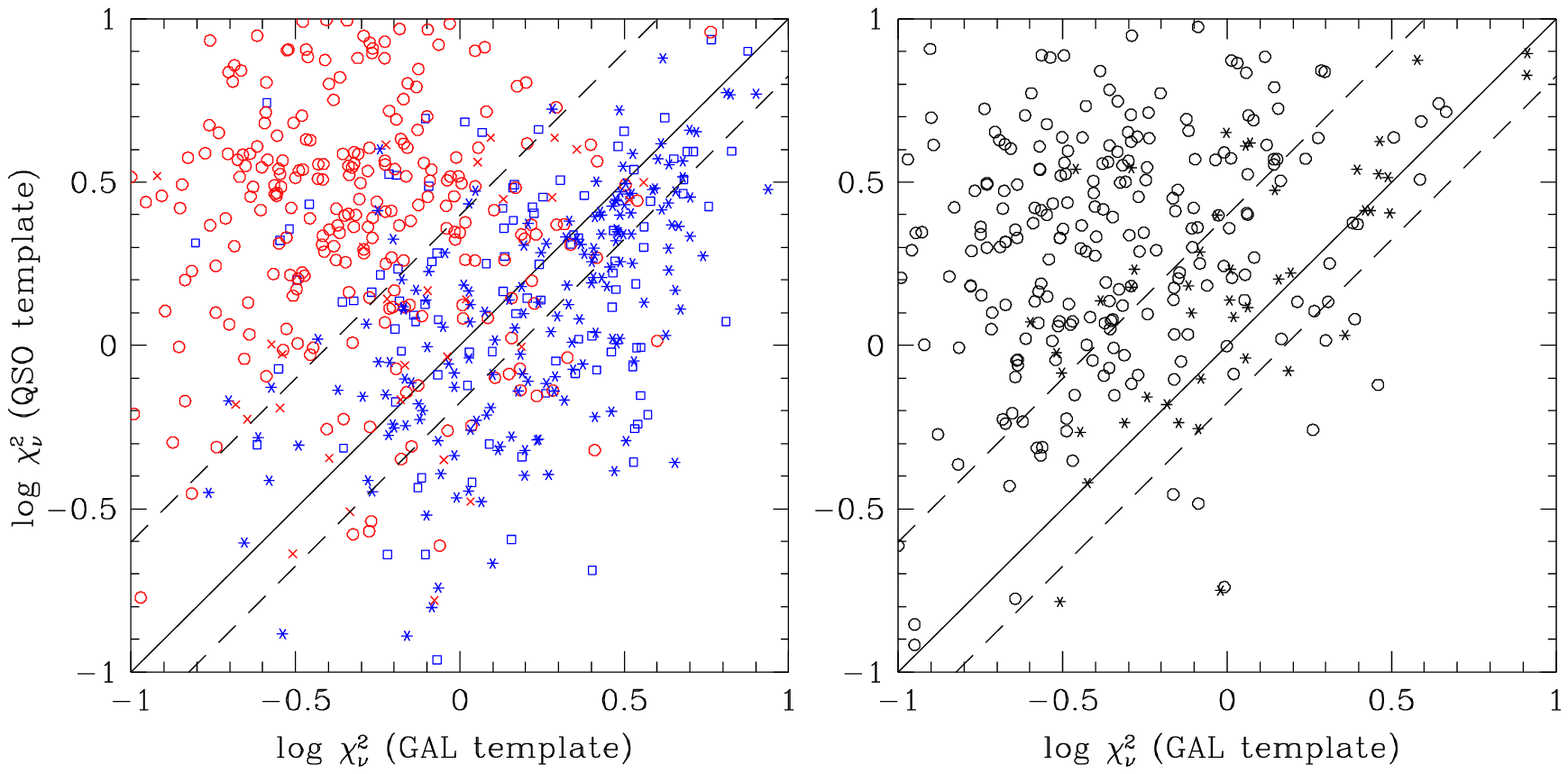} 
 \end{center}
\caption{
Left) Reduced $\chi^2$ values of fitting results with QSO-
and galaxy-templates for the spectroscopically-identified
sample. Blue (square and asterisk) and red (circle and cross)
symbols represent broad-line and narrow-line AGNs.
Asterisks and crosses are for stellar objects.
In order to evaluate the uncertainties in the type determination
due to the photometric redshift uncertainties, the redshift is
NOT fixed to the spectroscopic redshift during the fitting process.
Right) same figure for objects without spectroscopic identification.
Circles and asterisks represent non-stellar and stellar
objects.
In both panels, the solid line shows the equal $\chi^2_\nu$ line
and the upper and lower dashed lines represent the type determination
criteria for stellar and non-stellar objects, respectively.}\label{fig:Hyperz_chi2}
\end{figure*}

In the template fitting, two additional constraints
are applied in addition to the $\chi^2$ minimization.
The first one is that an object with
stellar morphology in the $i'$-band image is assumed to be
at $z>0.8$.
In the left panel of 
Figure~\ref{fig:SXDS_Cstar_i}, the $i'$-band magnitudes
of the spectroscopically-identified objects are 
shown as a function of redshift. Stellar objects
are shown with asterisks and crosses in this figure. 
Out of 261 stellar objects, 209 are spectroscopically
identified. A total of 182 and 27 of them are identified with 
broad-line and narrow-line AGNs at $z>0.8$, respectively.
Stellar broad-line AGNs show bright nuclei
and their observed optical light is dominated by
the nuclear component with stellar morphology. 
It should be noted that
this constraint is only applicable to the X-ray 
samples with similar depth and area to SXDS. As shown in 
Section 7.1, in the limited area of the SXDS, 
there is a correlation between redshifts 
and luminosities, therefore luminous ($\log L_{\rm 2-10keV}>44$)
broad-line AGNs, i.e. QSOs, whose optical morphologies are
dominated by the nuclear component, statistically appear
only above $z>0.8$. Such correlation is a typical of
flux-limited surveys. Less-luminous broad-line AGNs are 
found at lower-redshifts and they show extended morphologies
due to their host galaxies.
Stellar narrow-line AGNs are not resolved possibly due
to the fact that they are at high redshifts. 

An additional consideration is the absolute magnitude range of
the galaxy and QSO templates. 
Taking into account the absolute magnitude ranges of
spectroscopically identified AGNs, we
limit the $z$-band absolute magnitude range of the
galaxy (QSO) template to lie between $M_{z}=-20.0$ and $-25.0$ mag
($M_{z}=-22.0$ and $-26.5$ mag).
In Section 7.3, it will be shown that the
rest-frame optical to NIR continuum of narrow-line AGNs
are dominated by stellar continuum of their host galaxies.
Moreover, the range of the estimated stellar mass of the 
host galaxies of spectroscopically-identified narrow-line AGNs
is limited (see Section 7.2).
Such limited mass range of X-ray AGN host
galaxies has already been reported (e.g. \cite{yamada09}).
The absolute magnitude range of galaxy template corresponds
to the stellar mass range of $10^{10}\sim10^{12}$ M$_{\odot}$
with the galaxy SED templates used in Section 7.2.
We introduce the faint limit of the QSO absolute magnitude
range, because the host galaxy component will dominate
the total magnitudes if nuclear component is fainter than the limit.
These consideration is necessary to reduce the number of 
outliers in photometric redshift estimation.

Once we determine photometric redshifts with 
both of the QSO and galaxy templates for each AGNs, we compare the
minimum $\chi^2_{\nu}$ with QSO templates, $\chi^2_{\nu \ {\rm QSO}}$,
and that with galaxy templates, $\chi^2_{\nu \ {\rm GAL}}$, to 
chose the best photometric redshift for each AGNs.
In the left panel of Figure~\ref{fig:Hyperz_chi2}, 
$\chi^2_{\nu \ {\rm QSO}}$ and $\chi^2_{\nu \ {\rm GAL}}$ are
compared for spectroscopically-identified AGNs. 
The redshift is not fixed to their spectroscopic redshift 
during the fitting process, in order to include the
effect of redshift uncertainty.
Narrow-line AGNs (red circles and crosses) on average better fitted with galaxy templates, 
and broad-line AGNs (blue squares and asterisks) on average better fitted with QSO templates. 
Crosses and asterisks represent AGNs with stellar morphology.
Because the number of the galaxy templates is larger than
that of the QSO templates and the average $\chi^2_{\nu}$ value
for extended narrow-line AGNs with the galaxy templates
is smaller than that for stellar broad-line AGNs with
QSO-templates, we do not chose the best photometric
redshift just by comparing $\chi^2_{\nu \ {\rm QSO}}$ and $\chi^2_{\nu \ {\rm GAL}}$, 
but consider the stellarity as well as the $\chi^2_{\nu}$ values. 
Considering the distribution in the panel,
we regard the QSO templates as better if $\chi^2_{\nu \ {\rm QSO}}$
is smaller than $2.5 \times \chi^2_{\nu \ {\rm GAL}}$ for stellar
objects. On the contrary, for extended objects, we regard
the GAL-template is better if
$\chi^2_{\nu \ {\rm GAL}}$ is smaller than $1.5 \times \chi^2_{\nu \ {\rm QSO}}$.
These selection criteria are shown with the dashed line in
Figure~\ref{fig:Hyperz_chi2}.

Based on the comparison, we constrain not only their 
redshifts but also their AGN types. We classify
an object whose photometric redshift is determined
with the galaxy (QSO) templates as narrow-line 
(broad-line) AGN, hereafter. We can examine the reliability 
of the photometric classification with the template
fitting results for the AGNs with spectroscopic classification.
As described above, we do not fix their redshifts with 
spectroscopic redshifts during the fitting process.
Among the 302 narrow-line
(284 broad-line) AGNs with spectroscopic identification, 
267 (206) of them are correctly classified as a 
narrow-line (broad-line)
AGN in the photometric classification. 
Because less luminous broad-line AGNs have extended
morphology, a fraction of them are fitted better with the galaxy templates
and photometrically classified as narrow-line AGNs.
Their optical continuum can be dominated by the host
galaxy component, see Section 7.3.
Those mis-classifications lower the rate of
correct classification for the broad-line AGNs. 
If we concentrate on the 180 broad-line AGNs with stellar
morphology, 170 are correctly classified as a broad-line
AGN in the photometric classification.

The derived photometric redshifts and classification 
are summarised in Columns 24 and 25 of
Table~\ref{tab:Multi3}, respectively. Column 26 describes flags for the 
photometric redshift determination:
flag=1 means the reduced $\chi^2$ is
larger than 10, flag=2 means photometric redshift is determined
with unreliable photometric data for a bright and extended galaxy,
and flag=3 means photometric redshift is determined with less than
four bands. The number of X-ray AGNs with each flag are
summarised in Table~\ref{tab:number_summary}.

\subsection{Photometric Redshift Results}

\begin{figure*}
 \begin{center}
  \includegraphics[width=150mm]{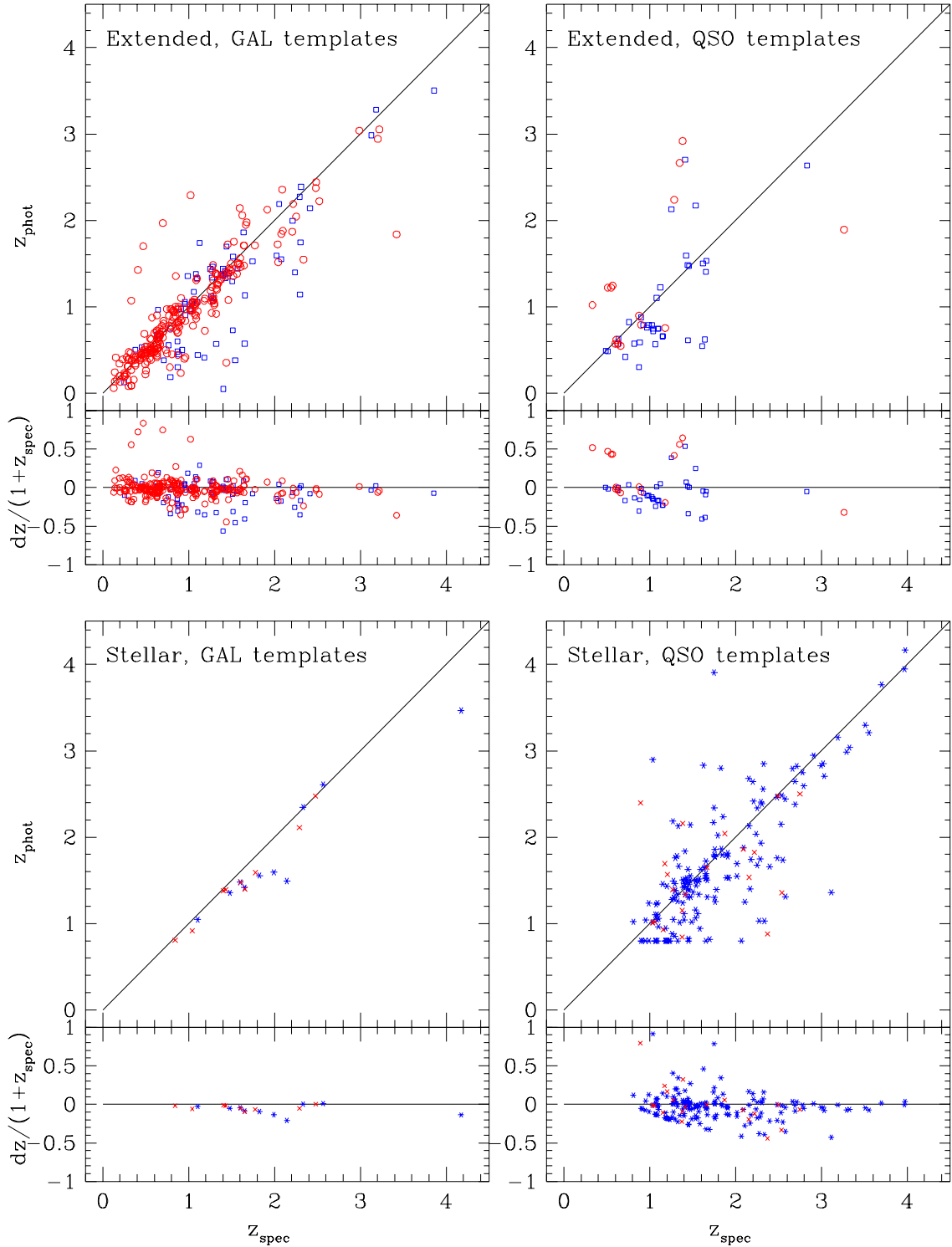} 
 \end{center}
\caption{
Spectroscopic redshift vs. photometric redshift
for objects with spectroscopic identification.
Blue squares and crosses and red circles and asterisks
represent broad-line and narrow-line AGNs.
Top panels show the results for extended objects
(squares and circles)
and left (right) panel is for objects fitted better with
GAL (QSO) template. Bottom panels are same as the upper panels for
stellar objects (crosses and asterisks).
The upper part of each panel shows spectroscopic
redshift vs. photometric redshift and bottom part is
${\rm d}z/(1+z_{\rm spec})$ as a function of $z_{\rm spec}$.}\label{fig:Hyperz_zsp_zph}
\end{figure*}

\begin{figure}
 \begin{center}
  \includegraphics[width=75mm]{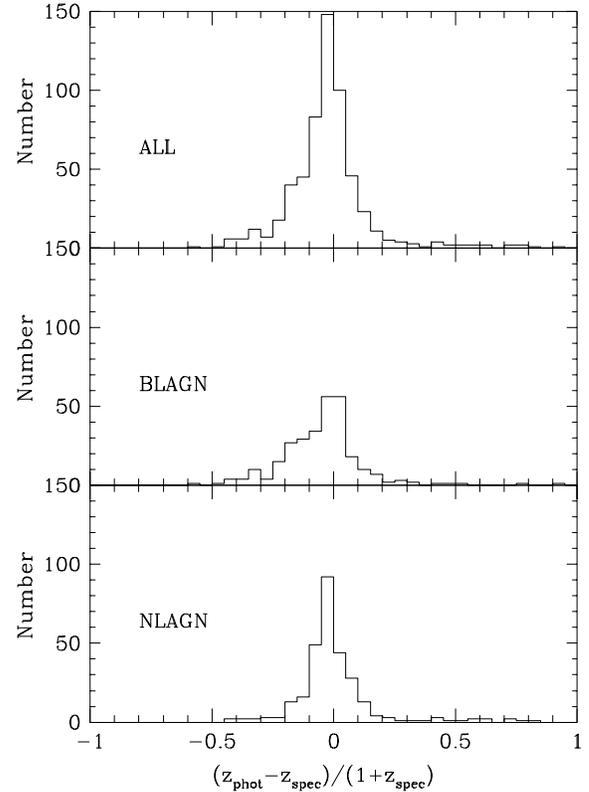} 
 \end{center}
\caption{
Distribution of ${\rm d}z/(1+z_{\rm spec})$ for all AGNs (top panel), 
broad-line AGNs (middle), and narrow-line AGNs (bottom).}\label{fig:Specz_Photz_hist}
\end{figure}

The reliability of the photometric redshifts is examined
by comparing them with the spectroscopic values.
The comparison of the photometric and spectroscopic redshifts
of spectroscopically-identified AGNs is shown in
Figure~\ref{fig:Hyperz_zsp_zph}. We separate the sample into
4 categories using the stellarity and the best-fit template. 
The difference between the spectroscopic and
photometric redshifts is defined as ${\rm d}z\equiv(z_{\rm phot}-z_{\rm spec})$,
and the bottom part of the panels show ${\rm d}z/(1+z_{\rm spec})$. Photometric
redshifts for objects that are fitted better with galaxy 
templates have a small scatter (shown in left panels), but the scatter is larger
for objects fitted better with QSO templates (shown in right panels).
The distribution of ${\rm d}z/(1+z_{\rm spec})$
is shown in Figure~\ref{fig:Specz_Photz_hist}. There is
no systematic difference between the spectroscopic
and photometric redshifts; the median of 
of ${\rm d}z / (1+z_{\rm spec})$ is $-0.026$ for all the AGNs in 
the total sample, and $-0.038$ and $-0.022$ for broad-line and
narrow-line AGNs, respectively. The scatter of the difference
is examined with the normalized median absolute deviation
(NMAD; $\sigma_z$) following \citet{brammer08},
\[
  \sigma_{\rm NMAD} = 1.48 \times {\rm median}\left( \left| \frac{ {\rm d}z - {\rm median}({\rm d}z) }{1+z_{\rm spec}} \right| \right).
\]
$\sigma_{\rm NMAD}$ corresponds to the standard deviation 
of a Gaussian distribution, and is less sensitive to 
the outliers than the usual definition of the standard 
deviation \citep{brammer08}.
For the total sample, $\sigma_{\rm NMAD}$ is 0.098,
which is larger than that of the photometric redshift
estimations for X-ray-selected AGNs with medium band filters
(\cite{salvato09}; \cite{cardamone10}; \cite{luo10}).
For example, \citet{luo10} reports $\sigma_{\rm NMAD}$ of 0.059
for a blind test sample. Because we do not train the
photometric redshift code with SEDs obtained in the
spectroscopic redshift sample, our $\sigma_{\rm NMAD}$
should be compared with their blind test results.
The larger $\sigma_{\rm NMAD}$ can be explained by the
lack of the photometric data with medium band filters.
The $\sigma_{\rm NMAD}$ 
for broad-line AGNs (0.109) is larger than that for narrow-line AGNs
(0.072), as expected from Figure~\ref{fig:Specz_Photz_hist}. 
This is because there is no strong feature in the
SEDs of the broad-line AGNs except for the break below Ly$\alpha$.
Since the break can only be covered by deep optical
observations for AGNs at $z>3$, the discrepancy between $z_{\rm spec}$
and $z_{\rm phot}$ is smaller for AGNs at $z>3$. 

\begin{figure*}
 \begin{center}
  \includegraphics[width=170mm]{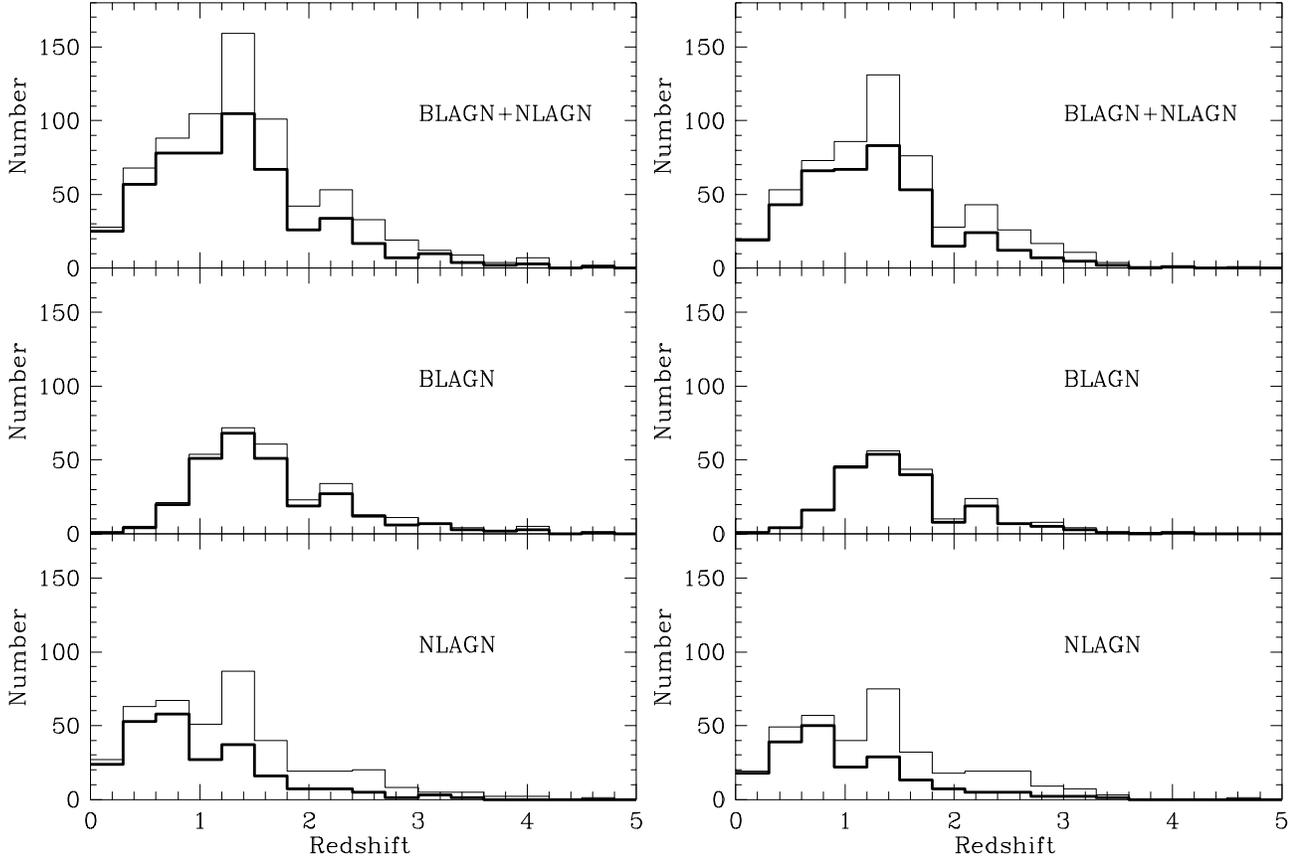} 
 \end{center}
\caption{
Redshift distributions of the AGNs in the soft-band (hard-band)
sample in the left (right) row. The thin solid
histogram indicates the redshift distribution of
spectroscopic plus photometric redshift samples, the
thick solid histogram shows spectroscopic samples only.
Shown from top to bottom are the redshift distributions of
all X-ray sources, broad-line AGNs, and narrow-line AGNs.}\label{fig:Redshift_hist}
\end{figure*}

Photometric redshift determination is crucial to understand
the nature of the X-ray-selected AGNs. 
The redshift distribution 
of the AGN sample is summarised in Figure~\ref{fig:Redshift_hist}.
For narrow-line AGNs, the redshift distribution of 
spectroscopically-identified objects is 
different from that of objects without spectroscopic identification:
they are located at higher redshifts on average.
As shown in the right panel of Figure~\ref{fig:SXDS_Cstar_i},
among the 310 primary candidates without spectroscopic
identification, 258 objects have extended morphology. 
Photometric redshift estimation suggests that they are
narrow-line AGNs at $z\ge1$. The green lines in the
figure indicate the redshift evolution of the $i'$-band
magnitude of 10Gyr (solid) and 2Gyr (dashed) single burst stellar
population with stellar mass of $\log M_{\rm *}=10.5$ (upper)
and $\log M_{\rm *}=12.0$ (lower).
The $i'$-band magnitudes of most of the spectroscopically-identified
narrow-line AGNs are enclosed within the green lines, and 
the $i'$-band magnitudes of narrow-line AGNs with photometric redshifts
are also enclosed within the lines.

The photometric redshift results suggest that
there are 45 broad-line AGN candidates that have not
been spectroscopically identified, and 86\% of the broad-line AGNs
are thought to be spectroscopically identified. 
Among broad-line AGN candidates without spectroscopic identification, 7 
objects have a photometric redshift above 3, where 
photometric redshift estimation is reliable even for
broad-line AGNs.

\section{Discussion}\label{sec:discussion}

\subsection{X-ray Luminosity and Absorption}\label{sec:Xrayprop}

\begin{figure}
 \begin{center}
  \includegraphics[width=75mm]{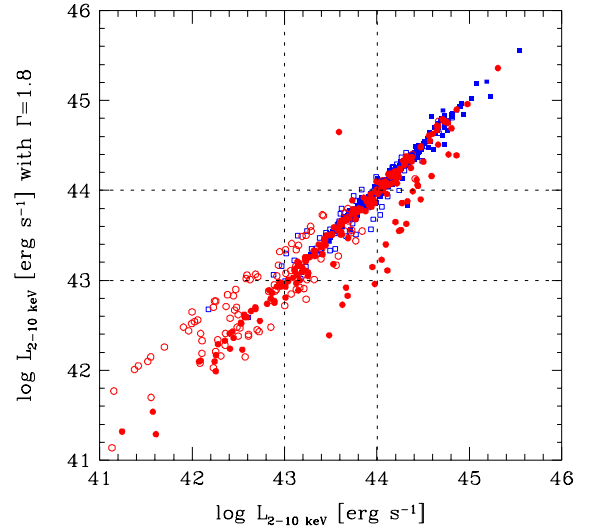} 
 \end{center}
\caption{
Best-estimated $L_{\rm 2-10keV}$ vs. $L_{\rm 2-10keV}$
estimated simply assuming a power-law with $\Gamma=1.8$.
Blue squares and red circles represent broad-line and
narrow-line AGNs, respectively.
Open symbols represent objects that are not detected
in the 2--10~keV band and $L_{\rm 2-10keV}$ is estimated
from the 0.5--2~keV count rate.}\label{fig:SXDS_lum_comp}
\end{figure}

\begin{figure*}
 \begin{center}
  \includegraphics[width=170mm]{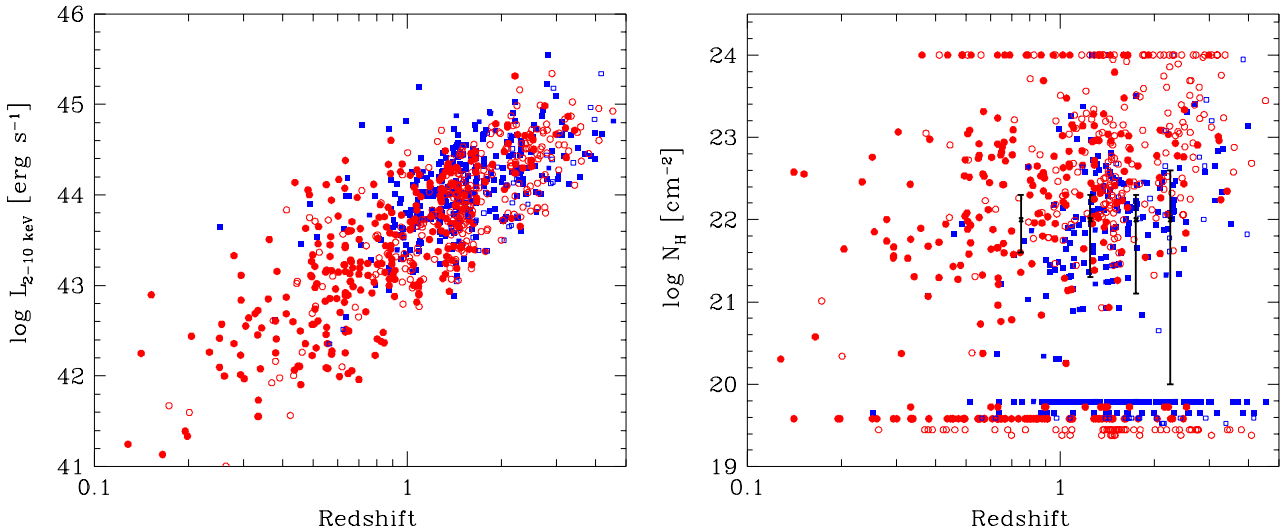} 
 \end{center}
\caption{
Left) Redshift vs. luminosity of
the broad-line (blue squares) and narrow-line AGNs (red circles).
Open symbols represent objects with
photometric redshift.
Right) Redshift vs. $\log N_{\rm H}$ (cm$^{-2}$).
Same symbols as in the left panel. $\log N_{\rm H}$
is derived from the HR2 value. AGNs only detected in 
the soft-(hard-)band are plotted below $\log N_{\rm H}=20.0$
(above $\log N_{\rm H}=24.0$) with arbitrary shift for clarity.
Error bars at $\log N_{\rm H}=22.0$ represent medians of uncertainties 
for objects with
$\log N_{\rm H}=21.5-22.5$ at $z=0.5-1.0$, $1.0-1.5$, $1.5-2.0$,
and $2.0-2.5$.}\label{fig:SXDS_red_lum}
\end{figure*}

In order to examine the physical properties of the X-ray AGNs,
at first, we evaluate the intrinsic 
2--10~keV luminosity, $L_{\rm 2-10keV}$ (erg s$^{-1}$), 
and the column density of photo-electric absorption to 
the nucleus, $N_{\rm H}$ (cm$^{-2}$), of each AGN.
In order to evaluate $L_{\rm 2-10keV}$, it is necessary to
correct for the effect of photo-electric absorption using the $N_{\rm H}$ value.
The number of X-ray photons 
of each AGN in the current sample is not sufficient to evaluate
$N_{\rm H}$ for individual AGN accurately with spectral fitting.
Thus we estimate $N_{\rm H}$ of each AGN from its HR2 value.
It needs to be noted
that the relation between the HR2 and $N_{\rm H}$ is non-linear
\citep{ueda08}, and the uncertainty associated with $N_{\rm H}$
can be large, especially around HR2 of $\pm1$. Furthermore, the
relation varies with redshift: at higher redshifts $N_{\rm H}$ 
is steeper function of HR2, and thus the uncertainty is larger
than the AGNs at low redshifts with same uncertainty of HR2.
Due to these limitations,
a statistical analysis, 
like that applied in \citet{hiroi12} and \citet{ueda08}, is
necessary to derive the distribution function of $N_{\rm H}$
and the hard X-ray luminosity function.
Therefore, in this paper, we use $L_{\rm 2-10keV}$ and
$N_{\rm H}$ to evaluate average properties of a population of
AGNs (Section 7.2) and to construct statistical sample of AGNs sub-divided
by their X-ray properties (Section 7.3).

In order to estimate $N_{\rm H}$ from HR2, we assume an
intrinsic shape of the X-ray spectrum of AGNs  
following \citet{ueda03}.
The intrinsic X-ray spectrum of AGNs is modeled by a combination of
a power-law component with
high-energy cut off  $E^{-\Gamma} \times \exp{(-E/E_{c})}$ and
a reflection component. A photon index
$\Gamma$ of 1.9 and cutoff energy $E_{c}$ of 300~keV are assumed.
We calculate the reflection component with the  ``pexrav'' \citep{magdziarz95}
model in the XSPEC package applying a solid angle of 2$\pi$, an inclination angle
of $\cos i = 0.5$, and solar abundance of all elements.
The strength of the reflection component is about 10\% of the direct
component just below 7.1~keV.
For the assumed intrinsic X-ray spectrum, we ignore any scattered light component and the luminosity dependence of the
strength of the reflection component for simplicity.
We only consider $\log N_{\rm H}$ up to 24.0. For Compton-thick
AGNs whose X-ray spectrum is affected by a scattered component,
$N_{\rm H}$ can be underestimated due to the soft X-ray spectrum of
that component. However, the number of 
such AGNs should be small in the X-ray flux range covered by 
the current sample \citep{ueda14}.
If HR2 is smaller than that expected for $\Gamma$=1.9, 
we determine the corresponding photon index by assuming no absorption.
If an object is only detected in the soft-band, we assign
$\Gamma$=1.9 and we neglect their $\log N_{\rm H}$.
For objects only detected in the hard-band, 
we assign $\log N_{\rm H}=24.0$ and $\Gamma$=1.9. 

$L_{\rm 2-10keV}$ is derived
from the observed 2--10~keV count-rate assuming the intrinsic
spectrum discussed above and correcting for the
effect of the photo-electric absorption with the best estimated $N_{\rm H}$.
For objects only detected in the 0.5--2~keV band, the count-rate
in that band is used instead of the 2--10~keV count-rate to derive 
$L_{\rm 2-10keV}$. 
The best-estimated $L_{\rm 2-10keV}$ are compared with 
$L_{\rm 2-10keV}$ estimated simply assuming typical apparent power-law
spectra of non-absorbed AGNs with $\Gamma=1.8$ in Figure~\ref{fig:SXDS_lum_comp}.
We use the count-rate in the 2--10~keV band in the estimation, but 
for objects only detected in the 0.5--2~keV band, we use the count-rate
in the 0.5--2~keV band and show them with open symbols. 
Because the observed 2--10~keV energy range 
covers higher energy range in the object rest-frame, for most of the
X-ray sources, the amounts of absorption correction are not large. 
Therefore, the statistical conclusions
derived using the best-estimated $L_{\rm 2-10keV}$ in later sections do not
significantly change even if we use the $L_{\rm 2-10keV}$ derived simply
with $\Gamma=1.8$.

The best-estimated $L_{\rm 2-10keV}$ and $N_{\rm H}$ are shown as a function
of redshift in the left and right panels of Figure~\ref{fig:SXDS_red_lum}.
In the right panel, we show the uncertainties of $\log N_{\rm H}$ 
for objects with $\log N_{\rm H}=22.0$. The uncertainties are determined
by medians of uncertainties for objects with
$\log N_{\rm H}=21.5-22.5$ at $z=0.5-1.0$, $1.0-1.5$, $1.5-2.0$,
and $2.0-2.5$. Uncertainty of each object is determined with the 
uncertainty of HR2.

The SXDS AGN sample covers wide range of the hard X-ray
luminosity, from $10^{42}$ to $10^{46}$ erg s$^{-1}$, and 
encloses the knee of the hard X-ray luminosity function at 
$z=0.4-4.0$ ($\log L_{\rm 2-10keV} \sim 44$ at $z\le1.0$ and 
$\log L_{\rm 2-10keV} \sim 44.6$ at $z\ge2.0$; \cite{ueda14}).
There is an apparent relation between the redshift and luminosity
due to the flux limit of the sample. 

The distribution of $N_{\rm H}$ is broadly consistent
with the optical classification on the basis of the existence of broad-emission lines
or by SED modeling: broad-line AGNs have smaller $N_{\rm H}$ on
average. In the redshift range below 0.6, almost all of the 
AGNs with $\log N_{\rm H} < 22$ lack
broad-emission lines. For these AGNs, it is likely that 
the broad-emission line component may be diluted by the host galaxy
component. In Table~\ref{tab:XrayProp}, columns 2 and 3 show the
 redshift and HR2 values assumed to derive the best-estimated $\log N_{\rm H}$,
$\Gamma$, and $\log L_{\rm 2-10 keV}$. Derived values are shown in columns 4-6.
$-$ in the column 4 means $\log N_{\rm H}$ is not given because
HR2 is smaller than expected for a power-law of $\Gamma$=1.9 or the object is
only detected in the soft-band.

\subsection{Stellar Mass of Narrow-line AGNs and M$_{\rm *}$-$L_{\rm X}$ Relation}

\begin{figure*}
 \begin{center}
  \includegraphics[width=170mm]{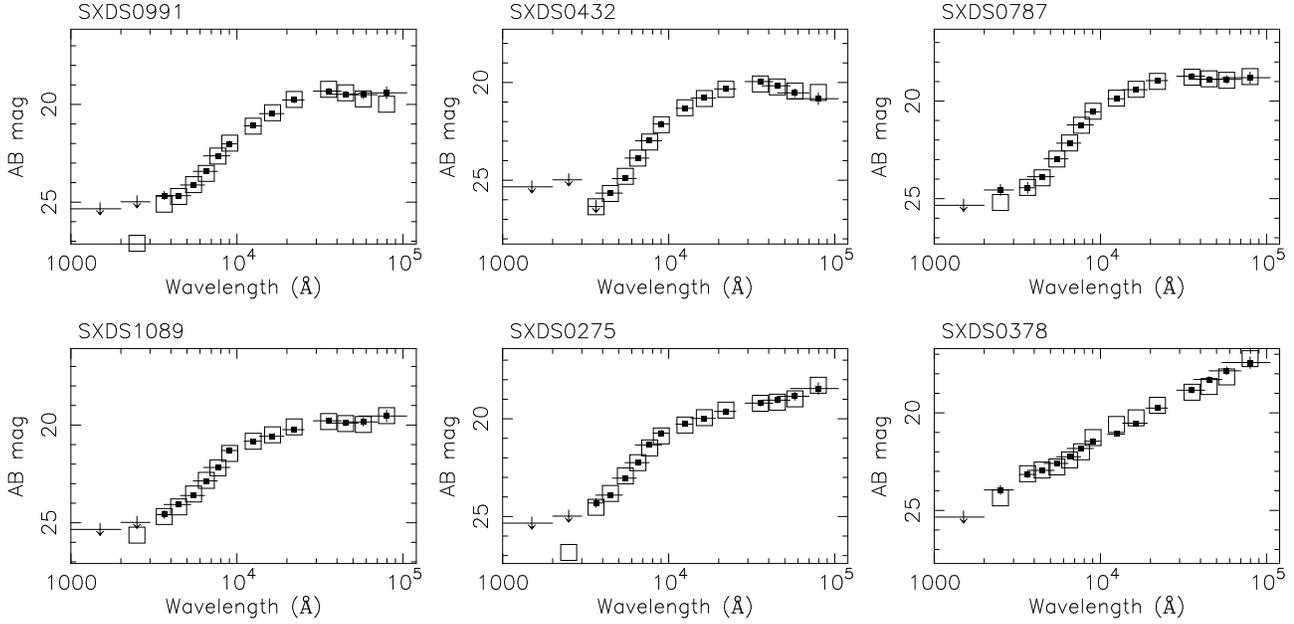} 
 \end{center}
\caption{
Sample SED fitting results for narrow-line AGNs at z$\sim$1
in order of the strength of excess component above 1{\micron}
from top-left to bottm-right panels.
Filled squares
indicate observed magnitudes at each observed wavelength
and arrows show upper limits. 
Open squares represent magnitudes of the best-fit
SED model.}\label{fig:SXDS_SEDfit_fix_sel}
\end{figure*}

We derive the stellar mass of host galaxies of 
narrow-line AGNs, $M_{*}$, by assuming that their UV to NIR SEDs are
dominated by the host galaxy component. 
The similarity of the SEDs of narrow-line AGNs to those of
non X-ray galaxies (e.g. \cite{kiuchi09}) supports this assumption.
The similarity will be discussed further in Section 7.3.
We apply SED template fitting
with a fixed redshift which is determined either spectroscopically 
or photometrically, and we use larger number of 
SED templates than those used in the photometric redshift
estimation in order to cover the variety of the SEDs of
galaxies accurately. Furthermore, the effect of hot dust
component above 1{\micron} wavelength range is considered
in the templates as described below.

We construct the SED templates using PEGASE.2 stellar
population synthesis code \citep{fioc97}. 
A Salpeter IMF with mass range of 0.1-120
$M_{\odot}$ is assumed. SED templates with burst
and continuous star formation histories are 
constructed with a simple exponentially declining star formation
rate proportional to $\exp (-t/\tau)$. $t$ represents the
age of the stellar system and $\tau$ represents the decay
time scale of the star formation. We use $\tau$ of
0.5Gyr and 2.0Gyr for burst and continuous models.
We use ages of 1.0, 2.0, 2.5, 3.0, 3.5, 4.0, 4.5, 5.0, 6.0, 8.0, and 10.0 Gyrs
for the burst model and 0.5, 1.0, 2.0, 3.0, 4.0, 5.0, 6.0 , 8.0, and 10.0 Gyrs
for the continuous model. We considered three metallicities, 
0.5, 1.0, and 2.0 times the solar metallicity.
We apply reddening with the Calzetti extinction
curve \citep{calzetti00} and consider color excess $E(B-V)$ with 0.1 mag steps
up to 0.9 mag. In addition to the simple exponentially
declining and continuous models, we construct templates by
combining the two models with a stellar mass ratio of
0.0, 1.0, and 50.0, representing the mixture of the bulge
and disk systems of galaxies.
In the wavelength range above 1{\micron}, a significant
fraction of narrow-line AGNs show an excess compared to the
galaxy templates due to hot dust component. 
In order to reproduce the excess, we construct galaxy
templates with an additional power-law component representing
the excess above 1{\micron}. Figure~\ref{fig:SXDS_SEDfit_fix_sel} shows
samples of the SED fitting results for narrow-line AGNs at $z\sim1$
in order of the strength of the excess component from the
top-left panel with no excess to the bottom-right panel with the
strongest excess. The excess components with
various strength
can be seen above observed wavelength of 2{\micron} are well
reproduced with the best-fit templates.

Furthermore, for galaxies at $z<0.6$, 
the 8.0{\micron} photometry is affected by the 8{\micron}
PAH feature, which is not included in the galaxy templates.
We do not use 8.0{\micron} photometry for galaxies at $z<0.6$.
We do not consider the constraints on the absolute magnitude
in the fitting.

\begin{figure}
 \begin{center}
  \includegraphics[width=75mm]{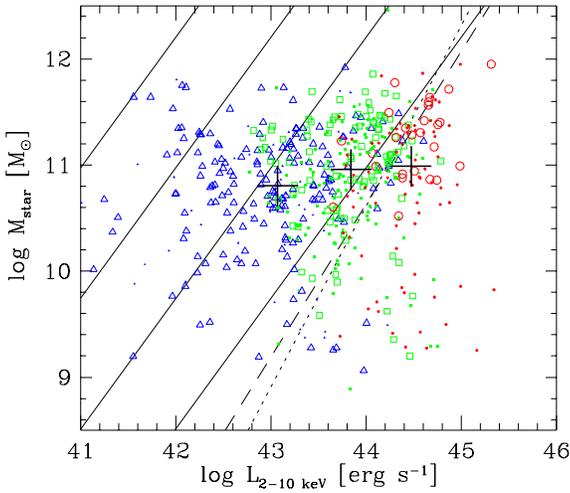} 
 \end{center}
\caption{
Estimated stellar mass 
vs 2--10~keV luminosity for narrow-line AGNs. Symbols are
coded with redshift; blue triangles are for
$z=0.1-1.0$, green squares are for $z=1.0-2.0$, 
and red circles are for $z>2.0$. Spectroscopically
identified narrow-line AGNs are plotted with
large open symbols, and photometric narrow-line
AGN candidates are plotted with small filled symbols.
Photometric candidates with phot-z flags of 0 or 1 are 
plotted. Big crosses indicate the median values in 
the three redshift ranges. Solid lines
indicate the stellar mass
and 2--10~keV luminosity relation assuming the 
$M_{\rm BH}$-$M_{\rm *}$ relation from \citet{bennert11}
and bolometric correction of \citet{marconi04}.
From bottom to top, an Eddington ratio of 1, 0.1, 0.01, and 0.001 are
assumed. Dotted and dashed lines indicate the relations
derived with the
$M_{\rm BH}$-$M_{\rm *,bulge}$ relations of \citet{sani11}
and \citet{marconi03}, respectively,
for an Eddigton ratio of 1.}\label{fig:SXDS_lum_mass}
\end{figure}

\begin{figure}
 \begin{center}
  \includegraphics[width=75mm]{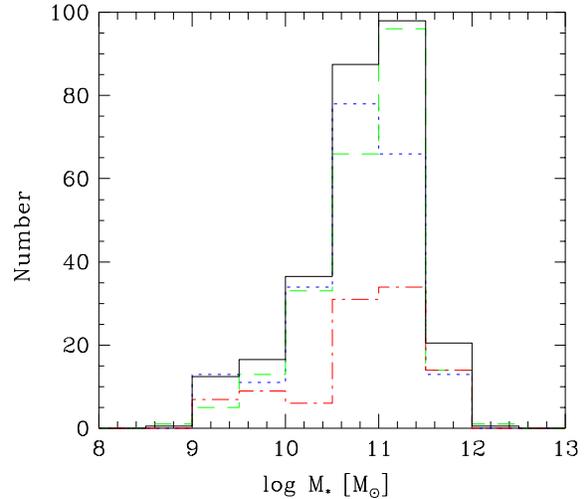} 
 \end{center}
\caption{
Distribution of $M_{*}$ for narrow-line AGNs.
The solid histogram is for the entire sample, and 
blue dotted, green dashed, and red dot-dashed histograms
represent the distribution of $0.1<z<1.0$, 
$1.0<z<2.0$, and $2.0<z$ narrow-line AGNs, respectively.
The numbers for the entire sample are divided by 2 for clarity.}\label{fig:SXDS_mass_hist}
\end{figure}

\begin{figure}
 \begin{center}
  \includegraphics[width=75mm]{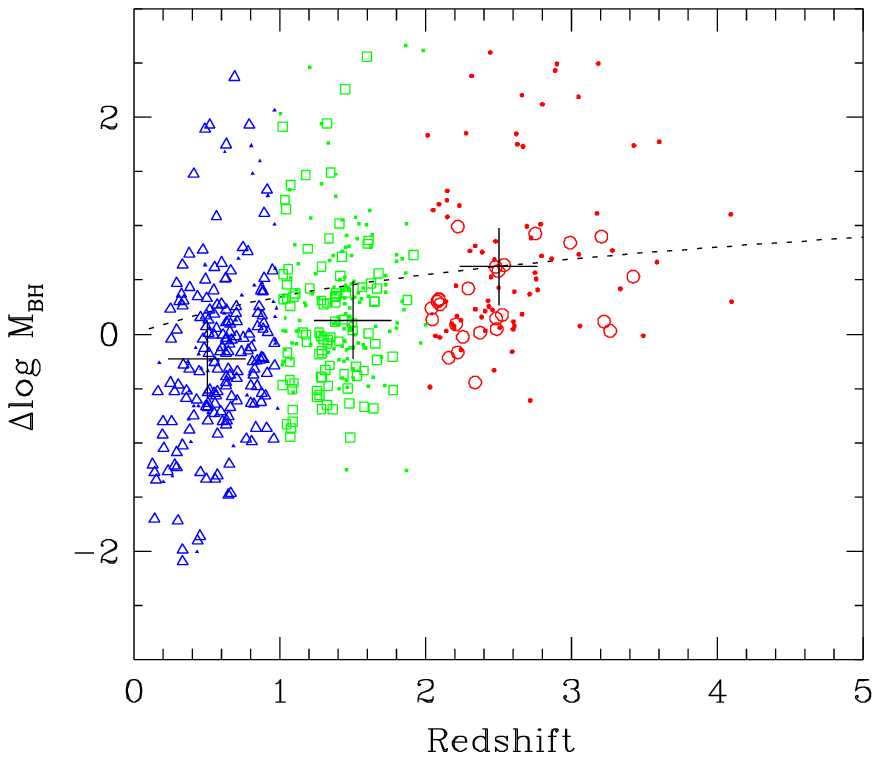} 
 \end{center}
\caption{
Difference between $M_{\rm BH}$ derived from 
$L_{\rm 2-10keV}$ and $\lambda_{\rm Edd}$-$L_{\rm bol}$
relation \citep{nobuta12} and $M_{\rm BH}$ derived from 
$M_{\rm *}$ and local $M_{\rm BH}$-$M_{\rm *}$
relation \citep{bennert11}. Symbols are
coded with redshift; blue triangles are for
$z=0.1-1.0$, green squares are for $z=1.0-2.0$,
and red circles are for $z>2.0$. Spectroscopically
identified narrow-line AGNs are plotted with
large open symbols, and photometric narrow-line
AGN candidates are plotted with small filled symbols.
Photometric candidates with phot-z flags of 0 or 1 are
plotted. Big crosses indicate the median values in 
the three redshift ranges. Dotted line indicate
the redshift dependence of the difference with 
cosmological evolution of the $M_{\rm BH}$-$M_{\rm *}$
relation with $(1+z)^{1.15}$ \citep{bennert11}.}\label{fig:SXDS_red_massratio}
\end{figure}

The resulting stellar masses of host galaxies
of narrow-line AGNs are shown in Figure~\ref{fig:SXDS_lum_mass}.
Although the SXDS sample covers a wide range in 
luminosity and redshift, the estimated $M_{\rm *}$ values are
remarkably constant. If we divide the sample into 3 redshift
bins with $0.1<z<1.0$ (low), $1.0<z<2.0$ (mid), and $2.0<z$ (high), 
the median $M_{\rm *}$ increases
by 0.18 dex from the low to the high redshift bins, 
although the median $L_{\rm 2-10keV}$ increases by 1.4 dex.
The distributions of the $M_{\rm *}$ in the 3 redshift
bins are compared with each other in Figure~\ref{fig:SXDS_mass_hist}.
The probabilities that a pair of distributions is drawn from 
the same sample are examined with the Kolmogorov-Smirnov test, and
they are 1.0\%, 4.7\%, and 28\% with the low-mid, low-high, and
mid-high redshift samples.
The mass range of the host galaxies is broadly consistent with
those for various X-ray selected samples
of narrow-line (e.g., \cite{akiyama05}; \cite{alonso08}; \cite{kiuchi09};
 \cite{brusa09}; \cite{yamada09}; \cite{xue10}; \cite{mainieri11}; \cite{rosario13})
and broad-line AGNs (e.g., \cite{jahnke09}; \cite{merloni10}; \cite{bongiorno14})
at redshifts up to 4. It needs to be noted that we consider
the absolute magnitude range of the galaxy template in the photometric
redshift determination. The absolute magnitude range can affect
the distribution of $M_{\rm *}$ of narrow-line AGNs without spectroscopic
redshifts, however, only 15 out of the 259 narrow-line AGNs without
spectroscopic redshifts are significantly affected by the faint
limit of the range and have absolute magnitude close to the 
limit.
Some narrow-line AGNs have $M_{\rm *}$ less 
than $10^{10}$ M$_{\odot}$. Their stellar mass 
estimation would be affected by scattered nuclear light
in the UV. Such contamination with a blue continuum 
possibly makes the SED appear similar to a younger stellar population with a 
smaller mass-to-light ratio, and the estimated stellar mass
can thus appear smaller even with the rest-frame NIR photometry constraints. 

The solid lines in Figure~\ref{fig:SXDS_lum_mass} indicate the
relationship between the stellar mass and X-ray luminosity
if we assume the local
black hole and host-galaxy stellar mass, $M_{\rm BH}$-$M_{\rm *}$,
relation from \citet{bennert11} and the AGN bolometric correction for 2--10~keV
X-ray luminosity from \citet{marconi04}. The lines from 
bottom to top correspond to an Eddington ratio of 
1, 0.1, 0.01, and 0.001.
\citet{bennert11} derive the relationship for broad-line AGNs
following \citet{haring04} with a Chabrier IMF \citep{chabrier02}.
We convert their relation to that with the Salpeter IMF,
considering that
$M_{\rm *}$ is 0.255 dex heavier with the Salpeter IMF.
We also plot the $M_{*}$ and $L_{\rm 2-10keV}$ relation
based on the black hole and bulge stellar mass, 
$M_{\rm BH}$-$M_{\rm *, bulge}$, relations
from \citet{sani11} and \citet{marconi03}
with dotted and dashed lines, respectively. 
They are consistent with the relation derived
with $M_{\rm BH}$-$M_{\rm *}$ relation of
\citet{bennert11} around $M_{\rm *}=10^{12}$ M$_{\odot}$,
and they show steeper dependence on $L_{\rm 2-10keV}$
than the relation. The difference below $M_{\rm *}=10^{12}$
M$_{\odot}$ can be explained with the difference
between $M_{\rm *}$ and $M_{\rm *, bulge}$. 
The constant $M_{\rm *}$ of narrow-line AGNs with wide
luminosity and redshift ranges implies
higher $\lambda_{\rm Edd}$ in the higher redshift bins;
significant fraction of narrow-line AGNs at $z>2$ possibly
have $L_{\rm 2-10keV}$ close to or above the Eddington limit.

On the other hand, the distribution of narrow-line AGNs
on the $M_{\rm *}$-$L_{\rm 2-10keV}$ plane
can be explained with the cosmological evolution of
the $M_{\rm BH}$-$M_{\rm *}$ relationship 
with higher $M_{\rm BH} / M_{\rm *}$ at higher redshifts
(e.g., \cite{mclure06}; \cite{peng06}; \cite{decarli10}; 
\cite{bennert11}; \cite{schramm13}), and
$\lambda_{\rm Edd}$ for the narrow-line AGNs at high-redshifts 
may be overestimated. Assuming that the $\lambda_{\rm Edd}$ 
and $L_{\rm bol}$ relation that is derived with broad-line
AGNs at $0.5<z<2.3$ in the same SXDS X-ray sample (Eq.(16) of \citet{nobuta12})
is applicable to the narrow-line AGNs, we estimate
$M_{{\rm BH},L}$ of the narrow-line AGNs from $L_{\rm 2-10keV}$ and
the bolometric correction of \citet{marconi04}. Figure~\ref{fig:SXDS_red_massratio}
shows the difference between the estimated
$M_{{\rm BH},L}$ and the $M_{{\rm BH},M_{*}}$, which is calculated from
$M_{\rm *}$ and the local $M_{\rm BH}$-$M_{\rm *}$ 
relationship from \citet{bennert11}. The difference implies
the cosmological evolution of the $M_{\rm BH}$-$M_{\rm *}$ relationship
with larger $M_{\rm BH}$ compared to $M_{\rm *}$ at higher
redshifts can explain the observed distribution on 
the $M_{\rm *}$-$L_{\rm 2-10keV}$ plane: 
the $0.1<z<1.0$ narrow-line AGNs have $-0.23$ dex smaller
$M_{\rm BH}$ than that expected from $M_{\rm *}$, and
the $z>2.0$ narrow-line AGNs have $0.63$ dex larger
$M_{\rm BH}$. If we simply fit the distribution, 
cosmological evolution of $M_{\rm BH}$-$M_{\rm *}$ relationship
with $(1+z)^{2.6}$ is required. If we take at face value, the
derived cosmological evolution of $M_{\rm BH}$-$M_{\rm *}$
relation is stronger than that in \citet{bennert11} shown with
dotted line in the figure. In order to derive the
cosmological evolution of $M_{\rm BH}$-$M_{\rm *}$ relationship
reliably, we need to consider the bias associated with the
luminosity-limited sample \citep{schulze14}. Such detailed
analysis is beyond the scope of this paper, and will be
discussed somewhere else.

\subsection{SEDs of X-ray Selected AGNs}\label{sec:SEDs}

\begin{figure*}
 \begin{center}
  \includegraphics[width=120mm]{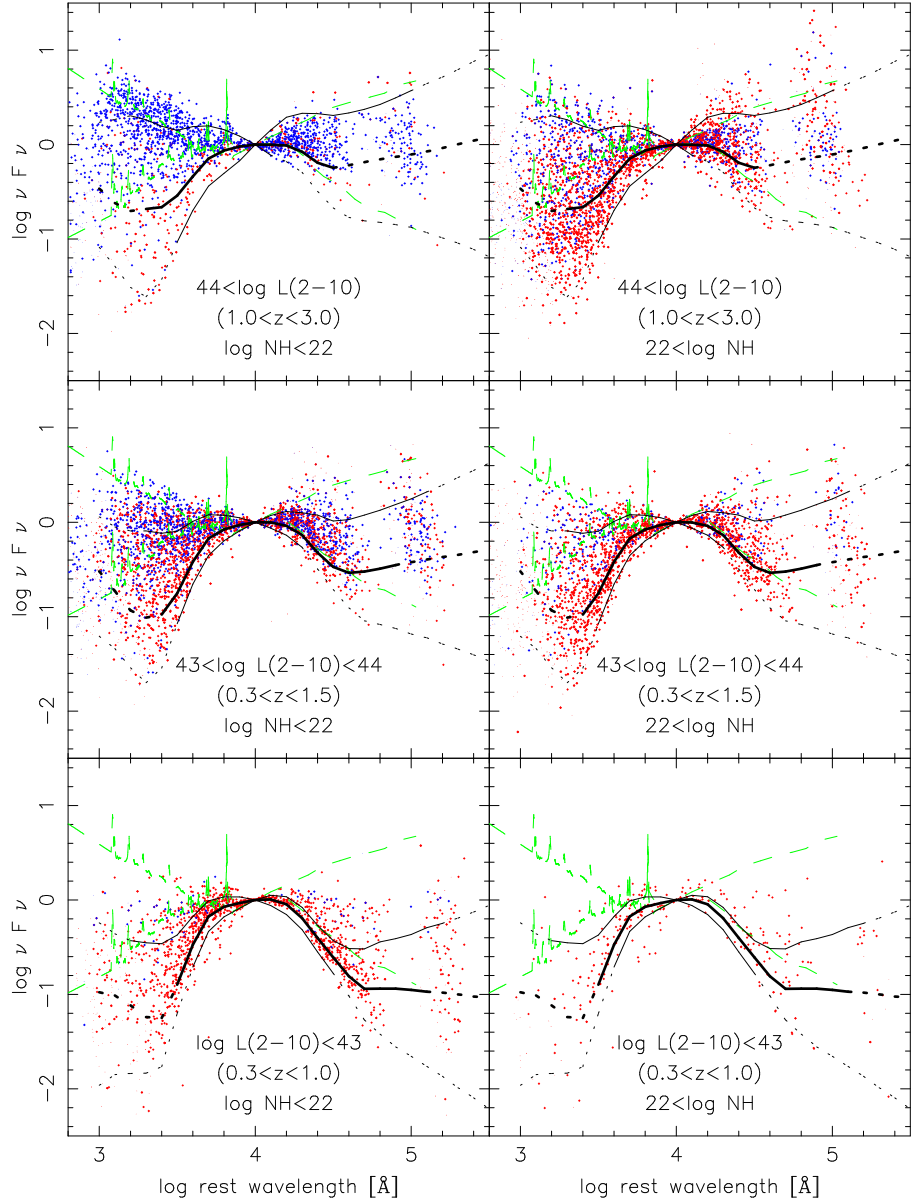} 
 \end{center}
\caption{
Rest-frame UV to MIR SEDs of X-ray AGNs. 
All SEDs are normalized at 1{\micron} in the rest-frame.
Broad-line and narrow-line AGNs are plotted with
blue and red symbols, respectively.
Small dots indicate upper limits.
The thick black solid and dotted lines represent
the median (50\%-ile) SED of galaxies in the matched
sample for each luminosity bin.
Dotted parts represent upper limits, i.e. less than 50\%
of the galaxies are detected in the wavelength range.
Thin black solid and dotted lines show the 10\%-ile (upper) and
90\%-ile (lower) SED of the galaxies, and dotted lines
represent upper limits. 
Green dashed lines indicate the range of the
template QSO SEDs
used in the photometric redshift determination.
The X-ray AGN sample is subdivided by X-ray luminosity
and $\log N_{\rm H}$;
from top to bottom, panels for AGNs with $44.0<\log L_{\rm 2-10keV}$,
$43.0<\log L_{\rm 2-10keV}<44.0$, and $\log L_{\rm 2-10keV}<43.0$,
and from left to right, $\log N_{\rm H}<22.0$ and $22.0< \log N_{\rm H}$.
The typical redshift range of each AGN sample is indicated in 
parentheses.}\label{fig:SED_plot_NH1}
\end{figure*}

\begin{figure*}
 \begin{center}
  \includegraphics[width=120mm]{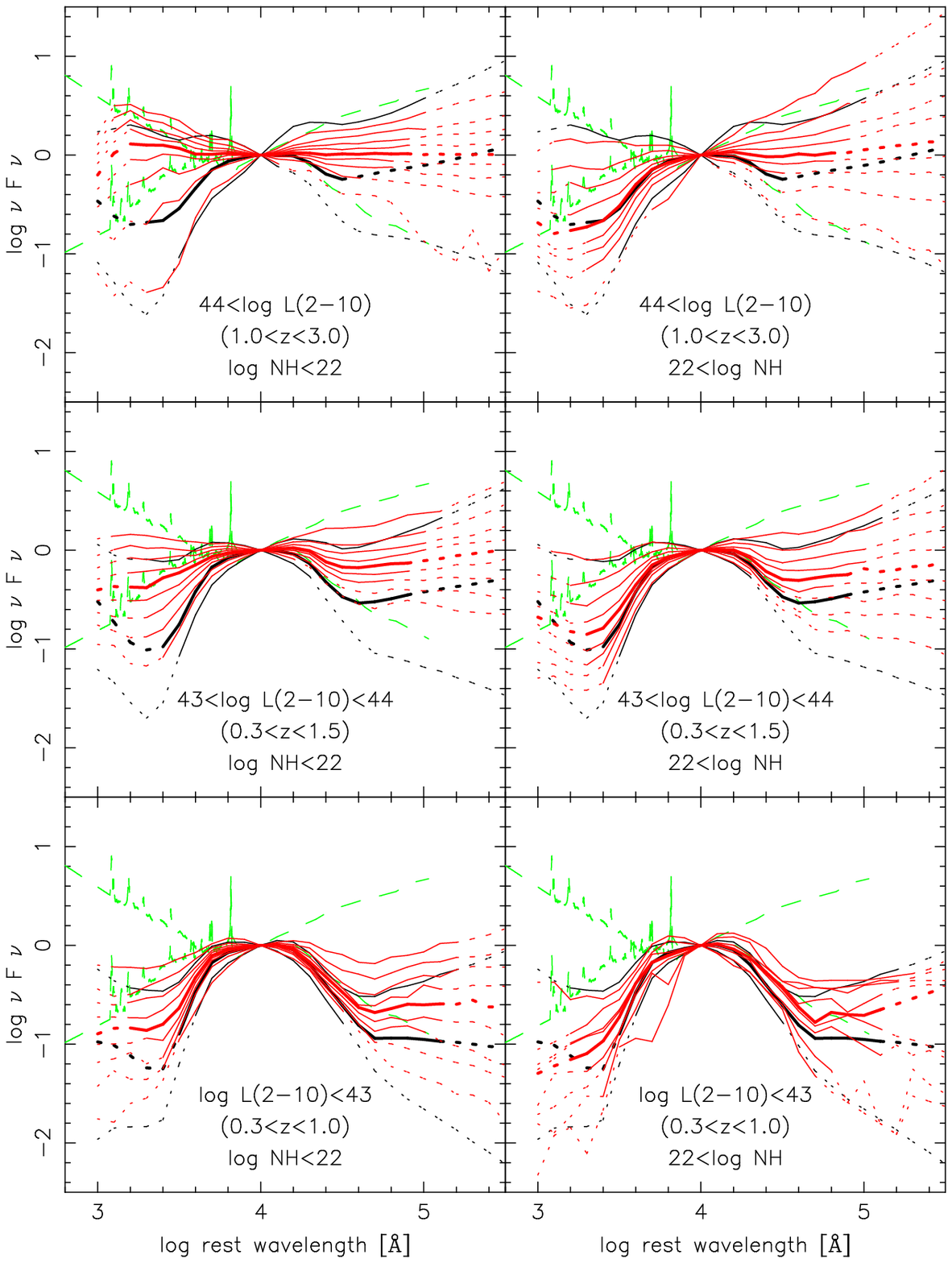} 
 \end{center}
\caption{
Rest-frame UV to MIR percentile SEDs of X-ray AGNs (red lines). 
All SEDs are normalized at 1{\micron} in the rest-frame. 
From top to bottom, 10\%-ile to 90\%-ile SEDs are shown with
10\% interval. Thick solid and dotted lines indicate
the median (50\%-ile) SED. Dotted lines represent
upper limits. 
The thick black solid and dotted lines represent
median SEDs of the galaxies in the matched sample
for each luminosity bin. 
The thin black solid and dotted lines show 10\%-ile and
90\%-ile SEDs. Dotted parts indicate upper limits.
Green dashed lines indicate the range of the
template QSO SEDs used in the photometric redshift determination.
The X-ray AGN sample is subdivided in the same way as
in Figure~\ref{fig:SED_plot_NH1}.}\label{fig:SED_plot_NH2}
\end{figure*}

Utilising the multi-wavelength photometric
data of the large sample of X-ray selected AGNs,
we examine the scatter of their UV-MIR
SEDs as a function of $L_{\rm 2-10keV}$ and $N_{\rm H}$, 
and compare the range of the scatter with 
that of galaxies without X-ray AGN activity. 
We choose galaxies which have the same mass and 
redshift distributions as the X-ray AGNs. Because X-ray AGNs
only appear among massive galaxies, it is important
to compare the properties of AGNs with those of
galaxies without X-ray AGN activity in the same mass range
(\cite{silverman09}; \cite{xue10}).

We construct a sample of galaxies without
X-ray AGN activity in the overlapping region of
the deep Suprime-cam optical and UDS NIR imaging data.
As a first step the photometric redshift is determined
with the FUV to MIR photometric data for each object
in the same way as for the X-ray AGNs as described
in Section 6. We do not apply a QSO template to the
galaxies and do not consider the constraints on the
absolute magnitude. Once photometric redshifts are determined,
we evaluate the stellar masses of the galaxies
applying the model described in Section 7.2.
In order to construct the sample of galaxies without X-ray
activity, we select galaxies that have similar 
stellar mass and redshift distribution of the X-ray
AGNs. For each X-ray AGN, we
select galaxies without X-ray detection within 
${\rm d}z=\pm0.2$ of the redshift of the AGN. 
We assume broad-line AGNs follow the same host galaxy
stellar mass distribution 
shown in Figure~\ref{fig:SXDS_mass_hist}.
Within the redshift range, we randomly 
select 14 galaxies for each AGN following the
mass distribution of 1 in the range $9.5<\log M_{*}<10.0$,
2 in the range $10.0<\log M_{*}<10.5$, 5 in the range $10.5<\log M_{*}<11.0$, 
5 in the range $11.0<\log M_{*}<11.5$, and 1 in the range $11.5<\log M_{*}<12.0$.
The mass distribution roughly reproduces that
of the total sample of the narrow-line AGNs.
In the redshift range between 1.0 and 2.0, 
there are 
12248 ($9.5<\log M_{*}<10.0$),
9140 ($10.0<\log M_{*}<10.5$),
5899 ($10.5<\log M_{*}<11.0$),
2557 ($11.0<\log M_{*}<11.5$),
and 310 ($11.5<\log M_{*}<12.0$)
galaxies that are
not detected in X-ray in the mass range shown in the
parenthesis.
We allow overlap in the selection, and a galaxy 
can be selected multiple times.
We refer to this sample of galaxies as the matched sample, hereafter.

The SEDs of X-ray AGNs are shown with 
dots in Figure~\ref{fig:SED_plot_NH1}. 
From top to bottom, the sample is subdivided
by X-ray luminosity: $44.0<\log L_{\rm 2-10keV}$, 
$43.0<\log L_{\rm 2-10keV}<44.0$, and $\log L_{\rm 2-10keV}<43.0$,
and from left to right, the sample is subdivided by
X-ray absorption to the nucleus:
$\log N_{\rm H}<22.0$ and $22.0< \log N_{\rm H}$. 
Blue and red symbols
represent broad- and narrow-line AGNs, respectively.
All of the SEDs are normalized at 1{\micron}, 
and each photometric point is located at
its rest-frame effective wavelength. 
It is clear that the luminous and less-absorbed
AGNs show SEDs within the range of the 
SEDs of the optically-selected QSOs (green dashed lines).

Less-luminous or absorbed AGNs show redder UV-optical SEDs, 
which are similar to those of galaxies in the matched sample.
The thick black solid and dotted lines in the figure
show the median (50\%-ile) value of the 1{\micron}-normalized
SEDs of the galaxies in the matched sample. We construct
the matched sample for each luminosity bin. In order to determine
the median value at each wavelength, we linearly 
interpolate the photometric data points of each galaxy
in the $\log \nu F \nu$ - $\log \lambda$ plane. In the percentile
calculation, we include photometric data points which are upper limits. 
In the wavelength
range where less than 50\% of galaxies in the sample 
are detected, the derived median value is an upper limit, and
we plot this with the dotted line. 
The thin black solid and dotted
lines represent 10\%-ile and 90\%-ile from top to bottom
at each wavelength range.

In order to quantitatively compare the 
range of the SEDs, we also construct the 
percentile SEDs of X-ray AGNs and show them in 
Figure~\ref{fig:SED_plot_NH2}. The SED of
each AGN is interpolated in the same way as for
the galaxies, and is normalized at 1.0{\micron}. 
At each wavelength we derive from brightest 10\%-ile to 90\%-ile
SEDs with 10\% interval. If the fraction of
detected objects at a certain wavelength 
is less than the corresponding percentile,
we regard the percentile SED as an upper limit
and plot the SED with dotted lines.

At wavelengths below 1{\micron} where
the SEDs of broad-line AGNs are dominated
by a blue power-law component, 
most of the broad-line AGNs have bluer
SEDs than the galaxies in the matched sample, 
especially in the upper-left highest
luminosity and lowest absorption bin:
80\% of the X-ray AGNs have bluer SEDs than 
the median SED of the galaxies in the matched sample. 
The almost constant median SED of the most luminous 
less-absorbed AGNs is consistent with the median
pre-host subtraction SEDs of broad-line AGNs derived 
with the COSMOS dataset (\cite{elvis12, hao14}). 
If we compare 90\%-ile range of \citet{elvis12}, 
the upper envelope is consistent, but the current sample
has larger scatter in the lower envelope 
due to the inclusion of narrow-line AGNs.
The SEDs of X-ray AGNs become redder
with lower luminosity or higher $N_{\rm H}$.
In the lowest luminosity bin, even with
the low $N_{\rm H}$, the median SED of 
the AGNs is similar to that of galaxies in 
the matched sample. As can be seen in Figure~\ref{fig:SED_plot_NH1},
almost all of them are narrow-line AGNs,
and they do not show strong broad-emission 
lines in their observed spectra.

The median
SEDs of the absorbed ($\log N_{\rm H}>22.0$) 
X-ray AGNs have similar SEDs to those
of the galaxies in the matched sample.
For example, the rest-frame
$U-V$ color of the median SED of
AGNs with $\log N_{\rm H}>22.0$ and $43.0<\log L_{\rm 2-10keV}<44.0$
is only 0.05 mag bluer than the galaxies in the matched sample.
The similarity of the $U-V$ colors of the AGN host galaxies
and the galaxies in the matched sample is consistent
with the results obtained in the {\it Chandra} Deep Fields 
\citep{xue10}, and does not support the 
distribution of AGN host galaxies in a distinct
region of color-magnitude space
(e.g., \cite{nandra07}).

In the wavelength range above 1{\micron}, 
the SEDs of the AGNs are thought to be dominated by
$\le$ 1,000K hot thermal emission from a dusty torus, 
and are roughly constant on the $\nu L_{\nu}$ plane.
On the other hand, the SEDs of galaxies without X-ray AGN activity
are dominated by the stellar continuum, which shows a
1.6{\micron} bump and continuous decrease at longer wavelengths. PAH 8{\micron} and cold-dust
emissions from star-forming regions contribute in the
wavelength range longward of 5{\micron}.
As can be seen in Figure~\ref{fig:SED_plot_NH1},
AGNs with $44<\log L_{\rm 2-10keV}$ are distributed
around a constant SED above 1{\micron} for both the
$\log N_{\rm H}<22.0$ and $>22.0$ samples,
and the SEDs are consistent with the 
observed range of broad-line AGNs shown with green dashed lines.
For AGNs with lower luminosity, the contribution from
the hot dust component decreases and the SEDs
are bluer and closer to those of galaxies peaked
at around 1.6{\micron}.

The galaxies in the matched sample do not show an
excess component above 1{\micron} in the
lowest luminosity bin, which samples galaxies in the
low redshift range,
shown in the bottom panels. On the contrary
in the highest luminosity bin, which samples galaxies
in the higher redshift range, shown in the top panels, 
the median SED of the galaxies in the matched sample 
show a constant SED in this wavelength range. 
As a result, even though in the middle and high luminosity 
ranges more than 70-80\% of X-ray AGNs show excess
component compared to the median SED of the galaxies
in the matched sample, the 10\%-ile SEDs of X-ray AGNs
are similar to those of the galaxies in the matched sample.

Several AGN selection methods based on the detection of 
hot thermal emission by multi-band photometric data 
are proposed with the
{\it Spitzer}
(\cite{lacy04}; \cite{ivison04}; \cite{stern05};
\cite{donley07}; \cite{daddi07}; \cite{fiore08};
\cite{donley12}),
{\it AKARI}
(\cite{takagi10}; \cite{hanami12}), 
and {\it Wide-field Infrared Survey Explorer (WISE)}
(\cite{mateos12}) dataset.
The similarity of the SEDs of AGNs 
with $\log L_{\rm 2-10keV}>43$ to those of
the galaxies in the matched sample suggests that
AGN selection based on detecting $>1.6${\micron} 
excess emission above the stellar continuum
can be contaminated by galaxies without X-ray AGN activity at $z>1$
(e.g. \cite{georgakakis10}; \cite{alexander11}; \cite{donley12}).
It needs to be noted that the SEDs above rest-frame
3{\micron} in the current study are based on broad-band 
photometry only at 8 and 24 {\micron}, and detailed
continuum structure can be missed.

The constant SEDs of galaxies without X-ray AGN activity at $z>1$ can
be caused by high star-formation activity
associated with massive galaxies whose star-formation
activity is strongly affected by dust.
The cosmological evolution 
of the $M_{\rm *}$ -
star-formation rate relation suggests
galaxies with $\log M_{\rm *}=10.5$ at $z=1.0$
and $z=2.0$ have 7 and 24 times higher star-formation 
rate than those at $z=0.0$ on average \citep{elbaz11}.
Furthermore, star-formation activity in massive
galaxies is strongly affected by dust \citep{pannella14}.
\citet{donley07} show SEDs of IR-excess galaxies can be
reproduced with a SED of local star-forming galaxy with 
additional reddening with $A_{V}=0.8$ mag.  
A larger contribution from thermal dust emission 
than for local massive galaxies can explain 
the constant SEDs of the non-X-ray galaxies at $z>1$ 
\citep{georgakakis10}.
Such evolution of SEDs of galaxies may affect
the discussion of MIR SEDs of torus components of 
the broad-line AGNs \citep{lusso13}.

\section{Summary}

We report the multi-wavelength identification of the 
X-ray sources found in the SXDS field using deep imaging data
covering the wavelength range between Far-UV to Mid-IR. 
Among the 1045 X-ray sources detected in the 7 FoVs of
{\it XMM-Newton} \citep{ueda08}, we consider 945 
X-ray sources that are detected above likelihood 7
in either of the soft- or hard-band and covered by 
the deep Suprime-cam imaging observations.
We select a primary counterpart for each X-ray source
applying the likelihood ratio method to $R$-band, 
3.6{\micron}, Near-UV, and 24{\micron} source catalogs.
We also consider the matching catalogs with
AGN candidates selected
in 1.4GHz radio and $i'$-band variability surveys. 

Once candidates of Galactic stars, ULXs in a nearby galaxy, 
and clusters of galaxies are removed there are 896 AGN
candidates in the sample.
We conduct spectroscopic observations of 596 and 851 
primary counterparts in the optical and NIR wavelengths, respectively, and
spectroscopically identify 65\%, 597,  of the total sample.
For the remaining X-ray AGN candidates, we evaluate their 
photometric redshift with photometric data in 15 bands. 
In the photometric redshift determination, we apply
SED templates of QSOs and galaxies separately to each 
primary counterpart, and
select the best-fit photometric redshift considering
the $i'$-band stellarity. 

Utilising the multi-wavelength photometric
data of the large sample of X-ray selected AGNs, 
we examine $M_{*}$ of the host galaxies of
the narrow-line AGNs. The median $M_{*}$ in the
three redshift bins, $0.1<z<1.0$, $1.0<z<2.0$, and
$2.0<z$, only increases 0.18 dex from the low to the
high redshift bins, although the median $L_{\rm 2-10keV}$
increases 1.4 dex in the redshift range. If we consider
the $M_{\rm BH}$-$M_{*}$ relation in the local Universe,
the constant $M_{*}$ implies higher $\lambda_{\rm Edd}$
for luminous AGNs at high redshifts. On the other hand, if we assume
the $\lambda_{\rm Edd}$ and $L_{\rm bol}$ relation 
derived for broad-line AGNs, we can estimate
$M_{\rm BH}$ of the narrow-line AGNs. The constant $M_{*}$
can also be explained with strong redshift 
evolution of $M_{\rm BH}$-$M_{*}$
relation from that determined in the local Universe.

Finally, the scatter of the rest-frame UV-MIR SEDs
of X-ray AGNs is examined as a function of $L_{\rm 2-10keV}$
and $N_{\rm H}$. The scatter is compared with the
scatter of the SEDs of the galaxies without X-ray detection.
The comparison sample of galaxies are selected to have
redshift and $M_{*}$ distributions matched to the X-ray AGNs.
The UV-NIR SEDs of X-ray AGNs with $\log N_{\rm H}>22.0$ 
are remarkably similar to those of the galaxies in the matched 
sample. In the NIR-MIR wavelength range, the median
SEDs of X-ray AGNs are redder than those of the galaxies
in the matched sample, and luminous AGNs show constant median SED
in the wavelength range, suggesting the existence of the hot
dust component. However, the scatter of the SEDs of
the luminous X-ray AGNs broadly overlap with that of the 
galaxies in the matched sample.
The galaxies in the matched sample of the luminous AGNs
also show a more constant SED than the galaxies in the matched 
sample of the less-luminous AGNs, which are at lower redshifts
than the luminous AGNs on average. A higher star-formation 
rate in massive galaxies at $z=1-3$ can explain
the constant SED of the galaxies in the matched sample.
Due to the similarity of the SEDs of X-ray AGNs to those
of the galaxies in the matched sample, it is expected that
AGN selection criteria looking for excess in the NIR-MIR
wavelength range can be significantly contaminated by 
non-AGN galaxies. 


\begin{table*}
  \caption{X-ray source content of the sample}\label{tab:number_summary}
  \begin{center}
    \begin{tabular}{lrrrl}
\hline
                               & N$_{\rm total}$ & N$_{\rm soft}$ & N$_{\rm hard}$ & COMMENT \\
\hline
X-ray sources                  & 1045            & 866            & 645            & Listed in \citet{ueda08} \\
X-ray sample                   &  945            & 781            & 584            & within deep Suprime-cam image coverage \\
Covered by 1.4GHz radio survey &  789            & 655            & 497            & Covered in \citet{simpson06} \\
Clusters of galaxies candidates &   14            &  14            &   1            & See Table~\ref{tab:CLUSTER} \\
Galactic star candidates       &   31            &  30            &   4            & See Table~\ref{tab:STAR} \\ 
ULX candidates                 &    4            &   4            &   3            & See Section~\ref{sec:ULX} \\
AGN candidates                 &  896            & 733            & 576            & \\ \hline
                               &                 &                &                & \\ \hline
\multicolumn{5}{l}{Observation summary for the 896 AGN candidates} \\ \hline
Optical spec. observed         &  597            & 523            & 399            & \\
NIR spec. observed             &  851            & 704            & 548            & \\
Spec. identified               &  586            & 514            & 397            & \\
Photz flag=1                   &    1            &   1            &   0            & \\
Photz flag=2                   &   10            &  10            &   7            & \\
Photz flag=3                   &   12            &   6            &   7            & \\ 
\hline
    \end{tabular}
  \end{center}
\end{table*}


\begin{longtable}{lccccl}
\caption{Multiwavelength dataset}\label{tab:imagelist}
\hline
Obs./Inst. &
Band &
\multicolumn{2}{c}{Detection Limits} &
Correction &
NOTE \\
   &
   &
Aperture &
Total &
   &
    \\
      &
      &
(mag) &
(mag) &
(mag) &
      \\
\multicolumn{1}{c}{(1)} &
\multicolumn{1}{c}{(2)} &
\multicolumn{1}{c}{(3)} &
\multicolumn{1}{c}{(4)} &
\multicolumn{1}{c}{(5)} &
\multicolumn{1}{l}{(6)} \\
\endfirsthead
\hline
\endhead
\hline
\endfoot
\hline
\multicolumn{6}{l}{\hbox to 0pt{\parbox{180mm}{\footnotesize
\par\noindent
Notes. Column (3) is 2 $\sigma$ aperture magnitude
limit measured in a fixed aperture.
The diameters of the apertures are
$7.\!^{\prime\prime}6$ for {\it GALEX} FUV and NUV images,
$2.\!^{\prime\prime}0$ for U-band, optical, and NIR images,
$3.\!^{\prime\prime}8$ for IRAC images, and
$6.\!^{\prime\prime}0$ for MIPS images. 
Column (4) gives the 3$\sigma$  limits for the
total magnitudes,
for {\it GALEX} FUV and for {\it Sptitzer} MIPS data the 4$\sigma$ magnitude limits.
They are determined from the aperture magnitude limit and
the aperture correction for stellar objects. These upper limits
are used in the photometric redshift determination and
in Table~\ref{tab:Multi1}. Applied aperture corrections
are shown in the right side in the parenthesis.
Column (5) is the correction for Galactic 
extinction.
}}}
\endlastfoot
\hline

{\it GALEX}            & FUV  & 26.64 & 25.52(4$\sigma$,0.36) & 0.18 & Deep Imaging Survey \\
                       &      &       &                       &      & \\
{\it GALEX}            & NUV  & 26.18 & 25.15(3$\sigma$,0.59) & 0.17 & Deep Imaging Survey \\
                       &      &       &                       &      & \\
Subaru/Suprime         & $U$  & 27.39 & 26.46(3$\sigma$,0.49) & 0.11 & South FoV only \\
CTIO/Mosaic-II         & $U$  & 26.86 & 26.29(3$\sigma$,0.13) & 0.11 & \\
CFHT/Mega-cam          & $U$  & 26.65 & 26.11(3$\sigma$,0.10) & 0.11 & CFHTLS wide \\
INT/WFCam              & $U$  &       &                       & 0.11 & Only used for bright objects \\
                       &      &       &                       &      & \\
Subaru/Suprime         & $B$  & 28.50-28.83 & 27.91-28.24(3$\sigma$,0.15) & 0.09 & \\
INT/WFCam              & $G$  &       &                       &      & Only used for bright objects \\
                       &      &       &                       &      & \\
Subaru/Suprime         & $V$  & 28.09-28.22 & 27.55-27.68(3$\sigma$,0.10) & 0.07 & \\
                       &      &       &                       &      & \\
Subaru/Suprime         & $R$  & 27.95-28.18 & 27.38-27.61(3$\sigma$,0.13) & 0.07 & \\
INT/WFCam              & $R$  &       &                       &      & Only used for bright objects. \\
                       &      &       &                       &      & \\
Subaru/Suprime         & $i'$ & 27.91-28.10 & 27.30-27.49(3$\sigma$,0.17) & 0.06 & \\
INT/WFCam              & $i'$ &       &                       &      & Only used for bright objects. \\
                       &      &       &                       &      & \\
Subaru/Suprime         & $z'$ & 26.83-27.08 & 26.25-26.50(3$\sigma$,0.14) & 0.04 & \\
INT/WFCam              & $z'$ &       &                       &      & Only used for bright objects. \\
                       &      &       &                       &      & \\
UKIRT/WFCAM            & $J$  & 26.06 & 25.40(3$\sigma$,0.22) & 0.02 & UKIDSS/UDS-DR8 \\
                       &      &       &                       &      & \\
UKIRT/WFCAM            & $H$  & 25.48 & 24.82(3$\sigma$,0.22) & 0.02 & UKIDSS/UDS-DR8 \\
                       &      &       &                       &      & \\
UKIRT/WFCAM            & $K$  & 25.96 & 25.30(3$\sigma$,0.22) & 0.01 & UKIDSS/UDS-DR8 \\
                       &      &       &                       &      & \\
{\it Spitzer}/IRAC     & 3.6\micron & 23.54 & 22.77(3$\sigma$,0.33) & 0.01 & SWIRE \\
{\it Spitzer}/IRAC     & 3.6\micron & 25.26 & 24.49(3$\sigma$,0.33) & 0.01 & SpUDS \\
                       &            &       &                       &      & \\
{\it Spitzer}/IRAC     & 4.5\micron & 22.87 & 22.07(3$\sigma$,0.36) & 0.00 & SWIRE \\
{\it Spitzer}/IRAC     & 4.5\micron & 24.83 & 24.03(3$\sigma$,0.36) & 0.00 & SpUDS \\
                       &            &       &                       &      & \\
{\it Spitzer}/IRAC     & 5.8\micron & 21.27 & 19.97(4$\sigma$,0.54) & 0.00 & SWIRE \\
{\it Spitzer}/IRAC     & 5.8\micron & 22.99 & 21.69(4$\sigma$,0.54) & 0.00 & SpUDS \\
                       &            &       &                       &      & \\
{\it Spitzer}/IRAC     & 8.0\micron & 21.17 & 19.91(4$\sigma$,0.66) & 0.00 & SWIRE \\
{\it Spitzer}/IRAC     & 8.0\micron & 23.19 & 21.77(4$\sigma$,0.66) & 0.00 & SpUDS \\
                       &            &       &                       &      &  \\
{\it Spitzer}/MIPS     & 24\micron  & 20.21 & 18.22(4$\sigma$,1.23) & 0.00 & SWIRE \\
{\it Spitzer}/MIPS     & 24\micron  & 22.06 & 20.07(4$\sigma$,1.23) & 0.00 & SpUDS \\
\end{longtable}


\begin{landscape}
\setlength{\topmargin}{30mm}
\setlength{\tabcolsep}{6pt}
\setlength{\textheight}{160mm}
\begin{longtable}{lllrrrccccccll}
\caption{Identification of the X$-$ray sources}\label{tab:XrayID}
\hline
ID &
\multicolumn{2}{c}{X$-$ray} &
ERR &
ML2 &
ML34 &
\multicolumn{4}{c}{Primary Flag} &
\multicolumn{2}{c}{Matching} &
\multicolumn{2}{c}{Counterpart} \\
 &
\multicolumn{1}{c}{RA} &
\multicolumn{1}{c}{Dec.} &
 ($^{\prime\prime}$) &
 &
 &
\multicolumn{1}{c}{O} &
\multicolumn{1}{c}{U} &
\multicolumn{1}{c}{I} &
\multicolumn{1}{c}{M} &
\multicolumn{1}{c}{R} &
\multicolumn{1}{c}{V} &
\multicolumn{1}{c}{RA} &
\multicolumn{1}{c}{Dec.} \\
\multicolumn{1}{c}{(1)} &
\multicolumn{1}{c}{(2)} &
\multicolumn{1}{c}{(3)} &
\multicolumn{1}{c}{(4)} &
\multicolumn{1}{c}{(5)} &
\multicolumn{1}{c}{(6)} &
\multicolumn{1}{c}{(7)} &
\multicolumn{1}{c}{(8)} &
\multicolumn{1}{c}{(9)} &
\multicolumn{1}{c}{(10)} &
\multicolumn{1}{c}{(11)} &
\multicolumn{1}{c}{(12)} &
\multicolumn{1}{c}{(13)} &
\multicolumn{1}{c}{(14)} \\
\endfirsthead
\hline
ID &
\multicolumn{1}{c}{RA} &
\multicolumn{1}{c}{Dec.} &
ERR &
ML2 &
ML34 &
O &
U &
I &
M &
R &
V &
\multicolumn{1}{c}{RA} &
\multicolumn{1}{c}{Dec.} \\
\endhead
\hline
\endfoot
\hline
\multicolumn{14}{l}{\hbox to 0pt{\parbox{180mm}{\footnotesize
\par\noindent
Notes. Column (1) is the X-ray name. Columns (2) and (3) are the
X-ray coordinates. Column (4) is $\sigma_{{\rm X}i}$ from 
Ueda et al. (2008). Column
(5) and (6) are the maximum likelihoods in the 0.5--2~keV and 2--10~keV
bands. Column (7)--(10) show whether the primary counterpart is
selected as the primary candidate in $R$-band (col.(7)), {\it GALEX}
NUV band (col.(8)), {\it Spitzer} 3.6{\micron} (col.(9)), and {\it Spitzer}
24{\micron} (col.(10)) identifications. 1 means the object is selected 
as the primary candidate of the identification in the band. 
$-1$ means the object is selected as the primary candidate, but
the likelihood is less than 1.0. Columns (11) and (12) indicate
the primary counterpart is selected as counterpart of 1.4GHz radio source
and variability AGN candidate, respectively.
Columns (13) and (14) are coordinates of the primary counterpart.
(This table is available in its entirety in a machine-readable form
in the online journal. A portion is shown here for guidance regarding
its form and content.)
}}}
\endlastfoot
\hline
SXDS0001 &  02:15:24.9 & $-$04:55:43.3 & 2.40 & 19.9 & 5.1 & 1 & 1 & 1 & 1 & 0 & 0 &  02:15:24.9 & $-$04:55:41.2 \\ 
SXDS0002 &  02:15:24.9 & $-$04:54:07.9 & 2.20 & 33.7 & 6.2 & 1 & 1 & 1 & 0 & 0 & 0 &  02:15:24.9 & $-$04:54:07.8 \\ 
SXDS0003 &  02:15:27.2 & $-$05:02:00.7 & 2.90 & 0.0 & 7.5 & 0 & 1 & 0 & 0 & 0 & 0 &  02:15:26.9 & $-$05:02:04.3 \\ 
SXDS0004 &  02:15:29.0 & $-$04:57:46.5 & 3.00 & 14.1 & 0.0 & 1 & 1 & 1 & 0 & 0 & 0 &  02:15:28.9 & $-$04:57:47.1 \\ 
SXDS0005 &  02:15:29.4 & $-$04:53:27.5 & 1.90 & 11.8 & 8.6 & 1 & 0 & 1 & 1 & 0 & 1 &  02:15:29.3 & $-$04:53:25.2 \\ 
SXDS0008 &  02:15:37.1 & $-$04:56:56.5 & 1.20 & 41.6 & 5.9 & 1 & 1 & 1 & 0 & 0 & 1 &  02:15:37.1 & $-$04:56:57.2 \\ 
SXDS0010 &  02:15:38.8 & $-$05:00:52.2 & 1.30 & 72.1 & 11.0 & 1 & 1 & 1 & 1 & 0 & 0 &  02:15:38.7 & $-$05:00:50.3 \\ 
SXDS0012 &  02:15:41.6 & $-$05:04:13.5 & 2.40 & 7.4 & 8.3 & $-$1 & 0 & $-$1 & 0 & 0 & 0 &  02:15:41.6 & $-$05:04:11.0 \\ 
SXDS0015 &  02:15:43.9 & $-$05:07:15.8 & 0.30 & 830.0 & 150.0 & 1 & 1 & 0 & 0 & 0 & 0 &  02:15:43.9 & $-$05:07:16.3 \\ 
SXDS0016 &  02:15:44.0 & $-$04:55:25.8 & 2.30 & 11.7 & 4.4 & $-$1 & 0 & $-$1 & 0 & 0 & 0 &  02:15:44.0 & $-$04:55:22.2 \\ 
\end{longtable}
\end{landscape}


\begin{longtable}{rcrrl}
\caption{Candidate clusters of galaxies}\label{tab:CLUSTER}
\hline
ID &
HR2 &
z$_{\rm spec}$ &
z$_{\rm phot}$ &
NOTE \\
(1) &
(2) &
(3) &
(4) &
(5) \\
\endfirsthead
\hline
\endhead
\hline
\endfoot
\hline
\multicolumn{5}{l}{\hbox to 0pt{\parbox{140mm}{\footnotesize
\par\noindent
Notes. Column (1) is the X-ray source name. 
Column (2) is HR2. Sources only detected
in the soft-band are indicated with HR2 of $-1.00$.
Column (3) is spectroscopic redshift: 9.99 
means no spectroscopic redshift is available.
Column (4) is photometric redshift. 
Column (5) describes the corresponding name 
in Finoguenov et al. (2010). 
}}}
\endlastfoot
\hline

SXDS0047 & $-$0.79 &  9.99 &  1.00 &  \\ 
SXDS0280 & $-$0.95 &  9.99 &  0.45 & SXDF79XGG \\ 
SXDS0295 & $-$0.53 &  9.99 &  0.60 &  \\ 
SXDS0434 & $-$0.44 &  9.99 &  0.62 &  \\ 
SXDS0453 & $-$1.00 &  0.20 &  0.20 & SXDF16XGG \\ 
SXDS0514 & $-$0.54 &  0.65 &  0.65 & SXDF69XGG \\ 
SXDS0552 & $-$0.82 &  0.52 &  0.52 & SXDF34XGG,Radio \\ 
SXDS0621 & $-$0.98 &  0.69 &  0.69 & SXDF04XGG,SXDF85XGG,Radio \\ 
SXDS0647 & $-$0.64 &  9.99 &  0.46 & SXDF35XGG \\ 
SXDS0829 & $-$0.82 &  0.87 &  0.88 & SXDF46XGG \\ 
SXDS0876 & $-$0.53 &  0.38 &  0.38 & SXDF01XCG \\ 
SXDS0889 & $-$0.70 &  0.20 &  0.38 & SXDF01XCG \\ 
SXDS0913 & $-$0.88 &  9.99 &  0.47 &  \\ 
SXDS1176 & $-$0.74 &  0.33 &  0.33 & SXDF36XGG \\ 

\end{longtable}


\begin{longtable}{ccccccccc}
\caption{Journal of FOCAS/Subaru Observations}\label{tab:FOCAS_FOVs}
\hline
Mask Name &
\multicolumn{2}{c}{FoV Center} &
Period &
Exptime &
Grism &
Filter &
$\lambda$ Coverage &
Note \\
 &
RA &
Dec. &
 &
(s) &
 &
 &
({\AA}) &
 \\
\multicolumn{1}{c}{(1)} &
\multicolumn{1}{c}{(2)} &
\multicolumn{1}{c}{(3)} &
\multicolumn{1}{c}{(4)} &
\multicolumn{1}{c}{(5)} &
\multicolumn{1}{c}{(6)} &
\multicolumn{1}{c}{(7)} &
\multicolumn{1}{c}{(8)} &
\multicolumn{1}{c}{(9)} \\
\endfirsthead
\hline
\endhead
\hline
\endfoot
\hline
\multicolumn{9}{l}{\hbox to 0pt{\parbox{140mm}{\footnotesize
\par\noindent
Notes. Column (1) is the name of the FoV, and 
columns (2) and (3) are coordinates of the FoV
center. Column (4) is the observation period.
Column (5) is the integration time. Column (6)
and (7) indicate the name of grism and filter
used in the observation. Column (8) shows the
typical wavelength coverage.
}}}
\endlastfoot
\hline

150A\_20 & 02:17:57.60 & $-$05:02:25.1 & 2001/01 & 5400 & 150  & SY47 & 4700 9400 & \\
150A\_21 & 02:17:39.00 & $-$05:01:35.4 & 2001/11 & 7200 & 150  & SY47 & 4700 9400 & \\
150A\_22 & 02:18:28.80 & $-$05:00:32.4 & 2001/11 & 5400 & 150  & SY47 & 4700 9400 & \\
150A\_23 & 02:18:21.10 & $-$05:05:27.0 & 2001/11 & 1800 & 150  & SY47 & 4700 9400 & \\
150A\_01 & 02:17:09.67 & $-$05:13:17.9 & 2003/10 & 7200 & 150  & SY47 & 4700 9400 & \\
150A\_02 & 02:19:40.20 & $-$04:50:54.5 & 2003/12 & 3600 & 150  & SY47 & 4700 9400 & \\
150A\_03 & 02:17:16.41 & $-$04:58:49.6 & 2003/10 & 7200 & 150  & SY47 & 4700 9400 & \\
150A\_04 & 02:19:23.44 & $-$05:05:26.8 & 2003/12 & 3600 & 150  & SY47 & 4700 9400 & \\
150A\_05 & 02:16:54.54 & $-$04:48:08.3 & 2003/12 & 3600 & 150  & SY47 & 4700 9400 & \\
150A\_06 & 02:17:23.92 & $-$05:28:55.2 & 2003/12 & 3600 & 150  & SY47 & 4700 9400 & \\
150A\_07 & 02:19:31.70 & $-$04:58:42.6 & 2003/12 & 3600 & 150  & SY47 & 4700 9400 & \\
150A\_08 & 02:20:06.97 & $-$04:56:13.6 & 2003/12 & 4800 & 150  & SY47 & 4700 9400 & \\
150A\_09 & 02:19:23.09 & $-$04:50:37.3 & 2003/12 & 3600 & 150  & SY47 & 4700 9400 & \\
150A\_10 & 02:18:11.78 & $-$05:02:23.2 & 2003/12 & 3600 & 150  & SY47 & 4700 9400 & \\
150A\_11 & 02:17:34.52 & $-$05:17:15.6 & 2003/12 & 3600 & 150  & SY47 & 4700 9400 & \\
150A\_12 & 02:17:59.68 & $-$04:54:43.2 & 2003/12 & 4800 & 150  & SY47 & 4700 9400 & \\
150A\_13 & 02:20:22.53 & $-$05:02:12.1 & 2003/12 & 3600 & 150  & SY47 & 4700 9400 & \\
150A\_14 & 02:19:33.89 & $-$05:14:18.7 & 2003/12 & 3600 & 150  & SY47 & 4700 9400 & \\
150A\_15 & 02:16:18.00 & $-$05:10:32.7 & 2003/10 & 5400 & 150  & SY47 & 4700 9400 & \\
150A\_16 & 02:18:26.74 & $-$04:46:35.3 & 2003/12 & 3600 & 150  & SY47 & 4700 9400 & \\
150A\_17 & 02:17:07.19 & $-$04:44:21.3 & 2003/12 & 4800 & 150  & SY47 & 4700 9400 & \\
150A\_18 & 02:18:38.77 & $-$05:05:14.4 & 2003/12 & 5400 & 150  & SY47 & 4700 9400 & \\
150A\_19 & 02:17:35.15 & $-$05:03:11.8 & 2003/12 & 3600 & 150  & SY47 & 4700 9400 & \\
150B\_01 & 02:18:08.28 & $-$05:01:32.3 & 2003/12 & 3600 & 150  & SY47 & 4700 9400 & \\
150C\_01 & 02:19:40.70 & $-$05:06:41.3 & 2003/10 & 3600 & 150  & SY47 & 4700 9400 & \\
150C\_05 & 02:17:44.60 & $-$04:51:05.9 & 2003/10 & 4800 & 150  & SY47 & 4700 9400 & \\
150C\_07 & 02:17:20.50 & $-$05:06:19.8 & 2003/10 & 6000 & 150  & SY47 & 4700 9400 & \\
150C\_08 & 02:17:20.80 & $-$05:12:07.8 & 2003/10 & 4800 & 150  & SY47 & 4700 9400 & \\
150C\_10 & 02:16:53.50 & $-$04:57:08.5 & 2003/10 & 3600 & 150  & SY47 & 4700 9400 & \\
300A\_10 & 02:19:01.33 & $-$04:31:07.2 & 2003/10 & 7200 & 300B & SY47 & 4700 9200 & \\
300C\_05 & 02:17:51.50 & $-$05:29:30.0 & 2003/12 & 7200 & 300B & SY47 & 4700 9200 & \\
300C\_06 & 02:18:25.92 & $-$05:32:47.3 & 2003/12 & 7200 & 300B & SY47 & 4700 9200 & \\
300C\_07 & 02:20:22.32 & $-$04:50:23.9 & 2003/12 & 7200 & 300B & SY47 & 4700 9200 & \\
300E\_01 & 02:18:09.35 & $-$05:03:03.2 & 2004/10 &14400 & 300B & NONE & 3800 9200 & \\
         &             & $ $           & 2004/12 & 7200 & 150  & SY47 & 4700 9400 & \\
300E\_02 & 02:17:24.23 & $-$04:53:28.5 & 2004/10 & 9000 & 300B & NONE & 3800 9200 & \\
300E\_03 & 02:17:34.08 & $-$05:11:11.7 & 2004/11 & 5400 & 150  & SY47 & 4700 9400 & \\
300E\_04 & 02:17:43.84 & $-$05:08:31.6 & 2004/11 &10800 & 300B & NONE & 3800 9200 & \\
300E\_05 & 02:17:17.19 & $-$05:00:15.3 & 2004/11 &10800 & 300B & NONE & 3800 9200 & \\
300E\_06 & 02:18:12.63 & $-$04:53:22.8 & 2004/11 &10800 & 300B & NONE & 3800 9200 & \\
300E\_07 & 02:18:44.59 & $-$05:00:22.7 & 2004/11 &10800 & 300B & NONE & 3800 9200 & \\
300E\_08 & 02:17:42.10 & $-$04:49:54.6 & 2004/11 &18000 & 300B & NONE & 3800 9200 & \\
300E\_09 & 02:17:57.53 & $-$04:56:47.0 & 2004/11 & 9000 & 300B & NONE & 3800 9200 & \\
300G\_01 & 02:17:18.07 & $-$05:08:26.3 & 2005/11 & 9000 & 300B & NONE & 3800 9200 & \\
300G\_02 & 02:18:33.27 & $-$05:18:31.1 & 2005/11 & 7200 & 300B & NONE & 3800 9200 & \\
300G\_04 & 02:16:28.60 & $-$05:09:35.9 & 2005/11 &10800 & 300B & NONE & 3800 9200 & \\
300G\_05 & 02:16:11.08 & $-$05:10:14.0 & 2005/11 & 7200 & 300B & NONE & 3800 9200 & \\
300G\_06 & 02:16:51.87 & $-$04:45:07.1 & 2005/11 &12600 & 300B & NONE & 3800 9200 & \\
300G\_08 & 02:16:51.21 & $-$04:50:36.5 & 2005/11 &16200 & 300B & NONE & 3800 9200 & \\
300G\_10 & 02:19:10.45 & $-$05:12:59.1 & 2005/11 & 7200 & 300B & NONE & 3800 9200 & \\
300G\_12 & 02:17:47.25 & $-$04:33:39.3 & 2005/11 & 7200 & 300B & NONE & 3800 9200 & \\

\end{longtable}


\begin{longtable}{ccccccc}
\caption{Journal of the FMOS/Subaru observations}\label{tab:FMOS_FOVs}
\hline
FoV Name &
\multicolumn{2}{c}{FoV Center} &
Period &
Exptime &
MODE &
Note \\
 &
RA &
Dec. &
 &
(s) &
 &
 \\
\multicolumn{1}{c}{(1)} &
\multicolumn{1}{c}{(2)} &
\multicolumn{1}{c}{(3)} &
\multicolumn{1}{c}{(4)} &
\multicolumn{1}{c}{(5)} &
\multicolumn{1}{c}{(6)} &
\multicolumn{1}{c}{(7)} \\
\endfirsthead
\hline
\endhead
\hline
\endfoot
\hline
\multicolumn{7}{l}{\hbox to 0pt{\parbox{140mm}{\footnotesize
\par\noindent
Notes. Column (1) Name of the FoV, columns (2) and (3) central coordinates of the FoV, 
column (4) observation period, column (5) total integration time, column (6)
observation mode, and column (7) notes. 
}}}
\endlastfoot
\hline

SXDS00 & 02:17:32.6 & $-$05:00:34 & 2009/10 &  9000 & LR & IRS1 only\\
SXDS01 & 02:17:18.0 & $-$05:04:19 & 2009/12 &  5400 & LR & IRS1 only\\
SXDS02 & 02:17:32.5 & $-$05:10:50 & 2009/12 & 15300 & LR & \\
SXDS03 & 02:17:45.4 & $-$05:05:23 & 2010/09 & 11700 & LR & \\
SXDS05 & 02:17:32.9 & $-$04:39:00 & 2010/11 &  6300 & LR & \\
SXDS06 & 02:17:43.3 & $-$04:38:20 & 2010/11 &  3600 & LR & \\
SXDS07 & 02:17:40.4 & $-$05:18:17 & 2010/11 & 35100 & LR & \\
SXDS08 & 02:19:26.8 & $-$04:59:37 & 2010/11 & 16200 & LR & \\
SXDS09 & 02:16:35.6 & $-$05:01:44 & 2010/11 & 18900 & LR & \\
SXDS10 & 02:17:39.2 & $-$04:38:25 & 2010/11 & 11700 & LR & \\
SXDS11 & 02:17:41.1 & $-$05:02:24 & 2010/11 & 14400 & LR & \\
SXDS12 & 02:17:59.2 & $-$04:51:00 & 2010/12 & 13500 & LR & IRS1 only \\
SXDS13 & 02:18:26.9 & $-$05:18:03 & 2010/12 &  1800 & LR & \\
       &            &             & 2011/02 &  2700 & LR & \\
       &            &             & 2011/02 &  1800 & LR & \\
SXDS16 & 02:16:59.7 & $-$05:00:19 & 2011/10 & 10800 & LR & \\
SXDS17 & 02:17:48.7 & $-$05:15:38 & 2011/10 & 19800 & LR & \\
SXDS18 & 02:18:42.2 & $-$04:38:29 & 2011/10 & 18000 & LR & \\
SXDS19 & 02:17:19.4 & $-$05:21:00 & 2011/10 & 18000 & LR & \\
SXDS20 & 02:19:11.8 & $-$05:07:26 & 2011/10 & 18900 & LR & \\
SXDS21 & 02:17:40.7 & $-$05:01:30 & 2011/10 & 17100 & LR & \\
       &            &             & 2011/10 &  9000 & LR & \\
       &            &             & 2011/11 &  3600 & LR & IRS1 only \\
       &            &             & 2011/11 &  5400 & LR & \\
SXDS22 & 02:16:19.3 & $-$04:58:58 & 2011/12 &  9000 & LR & \\
       &            &             & 2011/12 & 14400 & LR & IRS1 only \\
SXDS23 & 02:18:52.5 & $-$05:02:11 & 2011/12 &  3600 & LR & \\
       &            &             & 2011/12 & 14400 & LR & IRS1 only \\
SXDS24 & 02:17:34.0 & $-$05:18:25 & 2011/12 & 11700 & LR & IRS1 only \\

\end{longtable}


\begin{longtable}{rc}
\caption{Candidate galactic stars}\label{tab:STAR}
\hline
ID &
HR2 \\
\multicolumn{1}{c}{(1)} &
\multicolumn{1}{c}{(2)} \\
\endfirsthead
\hline
\endhead
\hline
\endfoot
\hline
\multicolumn{2}{l}{\hbox to 0pt{\parbox{110mm}{\footnotesize
\par\noindent
Notes. Column (1) is the X-ray source name. Column (2) shows
HR2. Sources only detected in the soft-band are
indicated with HR2 of $-1.00$. 
}}}
\endlastfoot
\hline

SXDS0020 & $-1.00$ \\ 
SXDS0039 & $-1.00$ \\ 
SXDS0120 & $-0.98$ \\ 
SXDS0125 & $-0.93$ \\ 
SXDS0191 & $-1.00$ \\ 
SXDS0245 & $-0.90$ \\ 
SXDS0248 & $-0.82$ \\ 
SXDS0297 & $-0.99$ \\ 
SXDS0302 & $-1.00$ \\ 
SXDS0365 & $+0.32$ \\ 
SXDS0448 & $-0.67$ \\ 
SXDS0469 & $-0.74$ \\ 
SXDS0532 & $-0.94$ \\ 
SXDS0545 & $-0.90$ \\ 
SXDS0601 & $-0.93$ \\ 
SXDS0633 & $-1.00$ \\ 
SXDS0686 & $-0.63$ \\ 
SXDS0692 & $-1.00$ \\ 
SXDS0693 & $-0.96$ \\ 
SXDS0731 & $-1.00$ \\ 
SXDS0783 & $-1.00$ \\ 
SXDS0816 & $-0.92$ \\ 
SXDS0822 & $-0.98$ \\ 
SXDS0832 & $-0.80$ \\ 
SXDS0844 & $-0.97$ \\ 
SXDS0923 & $-0.92$ \\ 
SXDS0974 & $-0.91$ \\ 
SXDS0985 & $-1.00$ \\ 
SXDS0995 & $-0.99$ \\ 
SXDS1062 & $-1.00$ \\ 
SXDS1181 & $-0.98$ \\ 

\end{longtable}


\begin{landscape}
\setlength{\topmargin}{30mm}
\setlength{\tabcolsep}{6pt}
\setlength{\textheight}{160mm}
\setlength{\textwidth}{230mm}
\begin{longtable}{lrrrrrrrrrrrrrrrr}
\caption{Multiwavelength properties of the X$-$ray AGN candidates I}\label{tab:Multi1}
\hline
ID &
\multicolumn{2}{c}{FUV}   &
\multicolumn{2}{c}{NUV}   &
\multicolumn{2}{c}{$U$}   &
\multicolumn{2}{c}{$B$}   &
\multicolumn{2}{c}{$V$}   &
\multicolumn{2}{c}{$R$}   &
\multicolumn{2}{c}{$i'$}  &
\multicolumn{2}{c}{$z'$}  \\
\multicolumn{1}{c}{} &
\multicolumn{2}{c}{(mag)} &
\multicolumn{2}{c}{(mag)} &
\multicolumn{2}{c}{(mag)} &
\multicolumn{2}{c}{(mag)} &
\multicolumn{2}{c}{(mag)} &
\multicolumn{2}{c}{(mag)} &
\multicolumn{2}{c}{(mag)} &
\multicolumn{2}{c}{(mag)} \\
\multicolumn{1}{c}{(1)} &
\multicolumn{2}{c}{(2)} &
\multicolumn{2}{c}{(3)} &
\multicolumn{2}{c}{(4)} &
\multicolumn{2}{c}{(5)} &
\multicolumn{2}{c}{(6)} &
\multicolumn{2}{c}{(7)} &
\multicolumn{2}{c}{(8)} &
\multicolumn{2}{c}{(9)} \\
\endfirsthead
\endhead
\hline
\endfoot
\hline
\multicolumn{17}{l}{\hbox to 0pt{\parbox{180mm}{\footnotesize
\par\noindent
Notes. Column (1) is the X-ray source name.
Columns (2)--(9) show the measured total magnitude and
associated error in from FUV to $z'$-band. 
99.99 in the error column means the corresponding magnitude
is an upper limit. 99.99 in the total magnitude means
the object is not covered in that band.
(This table is available in its entirety in a machine-readable form
in the online journal. A portion is shown here for guidance regarding
its form and content.)
}}}
\endlastfoot
\hline
SXDS0001 & 23.40 &  0.07 & 21.66 &  0.05 & 21.13 &  0.05 & 20.62 &  0.01 & 19.99 &  0.01 & 19.30 &  0.01 & 19.00 &  0.01 & 18.72 &  0.01 \\ 
SXDS0002 & 25.34 & 99.99 & 23.36 &  0.09 & 22.99 &  0.06 & 22.61 &  0.01 & 22.17 &  0.01 & 21.55 &  0.01 & 21.07 &  0.01 & 20.83 &  0.01 \\ 
SXDS0003 & 24.39 &  0.10 & 23.73 &  0.10 & 23.84 &  0.10 & 23.39 &  0.01 & 22.82 &  0.01 & 22.60 &  0.02 & 22.66 &  0.01 & 22.51 &  0.02 \\ 
SXDS0004 & 23.04 &  0.05 & 22.18 &  0.05 & 18.98 &  0.05 & 20.93 &  0.01 & 99.99 & 99.99 & 19.21 &  0.00 & 17.94 &  0.00 & 17.55 &  0.00 \\ 
SXDS0005 & 25.34 & 99.99 & 24.98 & 99.99 & 25.36 &  0.26 & 24.76 &  0.02 & 23.91 &  0.02 & 22.86 &  0.02 & 22.00 &  0.01 & 21.20 &  0.01 \\ 
SXDS0008 & 23.09 &  0.06 & 22.13 &  0.06 & 22.28 &  0.05 & 22.43 &  0.01 & 21.96 &  0.01 & 22.02 &  0.01 & 22.04 &  0.01 & 21.97 &  0.01 \\ 
SXDS0010 & 22.13 &  0.05 & 21.16 &  0.05 & 20.94 &  0.05 & 20.85 &  0.01 & 20.64 &  0.01 & 20.56 &  0.01 & 20.60 &  0.01 & 20.61 &  0.01 \\ 
SXDS0012 & 25.34 & 99.99 & 24.98 & 99.99 & 24.42 &  0.12 & 24.12 &  0.01 & 24.01 &  0.02 & 23.70 &  0.02 & 23.54 &  0.01 & 23.11 &  0.02 \\ 
SXDS0015 & 21.10 &  0.05 & 20.67 &  0.05 & 18.33 &  0.05 & 20.95 &  0.01 & 99.99 & 99.99 & 19.48 &  0.00 & 18.31 &  0.00 & 17.86 &  0.00 \\ 
SXDS0016 & 25.34 & 99.99 & 24.98 & 99.99 & 23.52 &  0.07 & 22.01 &  0.01 & 21.01 &  0.01 & 20.87 &  0.01 & 20.87 &  0.01 & 20.82 &  0.01 \\ 
\end{longtable}
\end{landscape}


\begin{landscape}
\setlength{\topmargin}{30mm}
\setlength{\tabcolsep}{6pt}
\setlength{\textheight}{160mm}
\setlength{\textwidth}{230mm}
\begin{longtable}{lrrrrrrrrrrrrrrrr}
\caption{Multiwavelength properties of the X$-$ray AGN candidates II}\label{tab:Multi2}
\hline
ID &
\multicolumn{2}{c}{$J$}   &
\multicolumn{2}{c}{$H$}   &
\multicolumn{2}{c}{$K$}   &
\multicolumn{2}{c}{$3.6\mu$m}  &
\multicolumn{2}{c}{$4.5\mu$m}  &
\multicolumn{2}{c}{$5.8\mu$m}  &
\multicolumn{2}{c}{$8.0\mu$m}  &
\multicolumn{2}{c}{$24\mu$m} \\
\multicolumn{1}{c}{} &
\multicolumn{2}{c}{(mag)} &
\multicolumn{2}{c}{(mag)} &
\multicolumn{2}{c}{(mag)} &
\multicolumn{2}{c}{(mag)} &
\multicolumn{2}{c}{(mag)} &
\multicolumn{2}{c}{(mag)} &
\multicolumn{2}{c}{(mag)} &
\multicolumn{2}{c}{(mag)} \\
\multicolumn{1}{c}{(10)} &
\multicolumn{2}{c}{(11)} &
\multicolumn{2}{c}{(12)} &
\multicolumn{2}{c}{(13)} &
\multicolumn{2}{c}{(14)} &
\multicolumn{2}{c}{(15)} &
\multicolumn{2}{c}{(16)} &
\multicolumn{2}{c}{(17)} &
\multicolumn{2}{c}{(18)} \\
\endfirsthead
\endhead
\hline
\endfoot
\hline
\multicolumn{17}{l}{\hbox to 0pt{\parbox{180mm}{\footnotesize
\par\noindent
Notes. Column (10) is the X-ray source name.
Columns (11)--(18) show measured total magnitude and
associated error in from $J$- to 24{\micron}-band. 
99.99 in the error column means the corresponding magnitude
is upper limit. 99.99 in the total magnitude column means
the object is not covered in that band.
(This table is available in its entirety in a machine-readable form
in the online journal. A portion is shown here for guidance regarding
its form and content.)
}}}
\endlastfoot
\hline
SXDS0001 & 99.99 & 99.99 & 99.99 & 99.99 & 99.99 & 99.99 & 99.99 & 99.99 & 99.99 & 99.99 & 99.99 & 99.99 & 99.99 & 99.99 & 15.91 &  0.05 \\ 
SXDS0002 & 99.99 & 99.99 & 99.99 & 99.99 & 99.99 & 99.99 & 99.99 & 99.99 & 99.99 & 99.99 & 99.99 & 99.99 & 99.99 & 99.99 & 17.32 &  0.05 \\ 
SXDS0003 & 99.99 & 99.99 & 99.99 & 99.99 & 99.99 & 99.99 & 99.99 & 99.99 & 99.99 & 99.99 & 99.99 & 99.99 & 99.99 & 99.99 & 99.99 & 99.99 \\ 
SXDS0004 & 99.99 & 99.99 & 99.99 & 99.99 & 99.99 & 99.99 & 99.99 & 99.99 & 17.25 &  0.01 & 99.99 & 99.99 & 18.13 &  0.01 & 18.22 & 99.99 \\ 
SXDS0005 & 99.99 & 99.99 & 99.99 & 99.99 & 99.99 & 99.99 & 19.34 &  0.02 & 19.63 &  0.04 & 20.39 &  0.36 & 19.96 &  0.31 & 17.92 &  0.05 \\ 
SXDS0008 & 99.99 & 99.99 & 99.99 & 99.99 & 99.99 & 99.99 & 20.47 &  0.04 & 20.18 &  0.06 & 19.97 & 99.99 & 19.47 &  0.19 & 18.22 & 99.99 \\ 
SXDS0010 & 99.99 & 99.99 & 99.99 & 99.99 & 99.99 & 99.99 & 18.91 &  0.01 & 18.64 &  0.01 & 18.47 &  0.02 & 18.13 &  0.01 & 17.43 &  0.05 \\ 
SXDS0012 & 99.99 & 99.99 & 99.99 & 99.99 & 99.99 & 99.99 & 21.34 &  0.03 & 21.09 &  0.03 & 21.69 & 99.99 & 21.77 & 99.99 & 18.22 & 99.99 \\ 
SXDS0015 & 99.99 & 99.99 & 99.99 & 99.99 & 99.99 & 99.99 & 99.99 & 99.99 & 99.99 & 99.99 & 99.99 & 99.99 & 99.99 & 99.99 & 15.56 &  0.05 \\ 
SXDS0016 & 99.99 & 99.99 & 99.99 & 99.99 & 99.99 & 99.99 & 20.40 &  0.01 & 20.04 &  0.01 & 19.55 &  0.04 & 18.70 &  0.02 & 17.41 &  0.05 \\ 
\end{longtable}
\end{landscape}


\begin{landscape}
\setlength{\topmargin}{30mm}
\setlength{\tabcolsep}{6pt}
\setlength{\textheight}{160mm}
\begin{longtable}{lrccrcrlc}
\caption{Multiwavelength properties of the X$-$ray AGN candidates III}\label{tab:Multi3}
\hline
ID &
\multicolumn{2}{c}{$VLA$} & 
\multicolumn{1}{c}{$Si$} & 
\multicolumn{2}{c}{Spec.$z$}  & 
\multicolumn{3}{c}{Phot.$z$} \\
\multicolumn{1}{c}{} &
\multicolumn{2}{c}{($\mu$Jy)} &
\multicolumn{1}{c}{} &
\multicolumn{1}{c}{$z_{\rm spec}$}  & 
\multicolumn{1}{c}{flag}  & 
\multicolumn{1}{c}{$z_{\rm phot}$}  & 
\multicolumn{1}{c}{type} &
\multicolumn{1}{c}{flag} \\
\multicolumn{1}{c}{(19)} &
\multicolumn{2}{c}{(20)} &
\multicolumn{1}{c}{(21)} &
\multicolumn{1}{c}{(22)} &
\multicolumn{1}{c}{(23)} &
\multicolumn{1}{c}{(24)} &
\multicolumn{1}{c}{(25)} &
\multicolumn{1}{c}{(26)} \\
\endfirsthead
\endhead
\hline
\endfoot
\hline
\multicolumn{6}{l}{\hbox to 0pt{\parbox{120mm}{\footnotesize
\par\noindent
Notes. Column (19) is the X-ray name. Column (20) is 1.4 GHz flux measured 
with VLA observation from Simpson et al (2012); 
0 in the first column means the object is outside of the VLA coverage.
1(0) in the second column indicates the object 
is (not) detected. Column (21) is stellarity index determined
in the $i'$-band image; 1 means stellar
object, and 0 means extended object. Columns (22) and (23) are the spectroscopic
redshift and spectroscopic type; BLA and NLA represent broad-line
and narrow-line AGNs, respectively. NID indicates the object is observed
in the optical but can not be identified with the optical spectrum. 
NSP indicates the object is not observed in the optical.
Columns (24) -- (26) are
photometric redshift, photometric type, and flag for photometric
redshift. GAL (QSO) is the photometric type showing whether  the SED of the
object is fitted better with the galaxy (QSO) templates.
(This table is available in its entirety in a machine-readable form
in the online journal. A portion is shown here for guidance regarding
its form and content.)
}}}
\endlastfoot
\hline
SXDS0001 &     0.0 & 0 & 0 & 0.505 & NLA &  0.480 & GAL & 0 \\ 
SXDS0002 &     0.0 & 0 & 0 & 0.700 & NLA &  0.644 & GAL & 0 \\ 
SXDS0003 &     0.0 & 0 & 0 & 9.999 & NSP &  0.622 & GAL & 0 \\ 
SXDS0004 &     0.0 & 0 & 0 & 9.999 & NSP &  0.496 & GAL & 2 \\ 
SXDS0005 &     0.0 & 0 & 0 & 1.050 & NLA &  0.782 & GAL & 0 \\ 
SXDS0008 &     0.0 & 0 & 1 & 1.033 & NLA &  1.010 & QSO & 0 \\ 
SXDS0010 &     0.0 & 0 & 1 & 1.224 & BLA &  0.800 & QSO & 0 \\ 
SXDS0012 &     0.0 & 0 & 0 & 9.999 & NSP &  1.344 & GAL & 0 \\ 
SXDS0015 &     0.0 & 0 & 0 & 0.098 & NLA &  0.598 & GAL & 2 \\ 
SXDS0016 &     0.0 & 0 & 1 & 3.512 & BLA &  3.298 & QSO & 0 \\ 
\end{longtable}
\end{landscape}


\begin{longtable}{lcrccc}
\caption{X-ray properties of AGN candidates}\label{tab:XrayProp}
\hline
ID &
\multicolumn{1}{c}{$z$} &
\multicolumn{1}{c}{HR2} &
\multicolumn{1}{c}{$\log N_{\rm H}$} &
\multicolumn{1}{c}{Photon Index}  &
\multicolumn{1}{c}{$\log L_{\rm 2-10keV}$} \\
\multicolumn{1}{c}{} &
\multicolumn{1}{c}{} &
\multicolumn{1}{c}{} &
\multicolumn{1}{c}{(erg s$^{-1}$)} &
\multicolumn{1}{c}{} &
\multicolumn{1}{c}{(cm$^{-2}$)} \\
\multicolumn{1}{c}{(1)} &
\multicolumn{1}{c}{(2)} &
\multicolumn{1}{c}{(3)} &
\multicolumn{1}{c}{(4)} &
\multicolumn{1}{c}{(5)} &
\multicolumn{1}{c}{(6)} \\
\endfirsthead
\endhead
\hline
\endfoot
\hline
\multicolumn{6}{l}{\hbox to 0pt{\parbox{120mm}{\footnotesize
\par\noindent
Notes. Column (1) is the X-ray name. Columns (2) and (3) are 
redshift and HR2 used to calculate the best-estimated 
X-ray properties. HR2 of $-1.00$ ($1.00$) means the source
is only detected in the soft-(hard-)band. 
Column (4) is $\log N_{\rm H}$, 
column (5) is $\Gamma$, and 
column (6) is $\log L_{\rm 2-10keV}$.
$-$ in the column (4) means $\log N_{\rm H}$ is not
given because HR2 is smaller than expected for a power-law
of $\Gamma=1.9$. $\Gamma$ of 1.9 means the value is fixed.
Details of determination of the X-ray properties are described
in Section 7.1.
(This table is available in its entirety in a machine-readable form
in the online journal. A portion is shown here for guidance regarding
its form and content.)
}}}
\endlastfoot
\hline
SXDS0001 & 0.505 & $-0.42$ & 21.9 & 1.9 & 43.2  \\ 
SXDS0002 & 0.700 & $-0.64$ & 20.8 & 1.9 & 43.5  \\ 
SXDS0003 & 0.622 & $-0.34$ & 22.1 & 1.9 & 43.6  \\ 
SXDS0004 & 0.496 & $-1.00$ & $-$  & 1.9 & 42.7  \\ 
SXDS0005 & 1.050 & $-0.24$ & 22.5 & 1.9 & 44.1  \\ 
SXDS0008 & 1.033 & $-0.62$ & 21.3 & 1.9 & 43.7  \\ 
SXDS0010 & 1.221 & $-0.72$ & $-$  & 2.1 & 44.1  \\ 
SXDS0012 & 1.344 & $-0.36$ & 22.5 & 1.9 & 44.0  \\ 
SXDS0015 & 0.098 & $-0.62$ & 20.8 & 1.9 & 42.3  \\ 
SXDS0016 & 3.512 & $-0.60$ & 21.9 & 1.9 & 44.5  \\ 
\end{longtable}


We warmly thank the staff members of the Subaru telescope for
their support during the observations. We would like to thank the
anonymous referee for valuable comments that improved this paper.
MA is supported by JSPS Grant-in-Aid for Young Scientist (B) (18740118),
and MA and HH are supported by 
Grant-in-Aid for Scientific Research (B) (21340042).
HH thanks the Grant-in-Aids for Challenging Exploratory Research 
(24650145) from JSPS, which partially supported this research.

This publication makes use of data products from the Two 
Micron All Sky Survey, which is a joint project of the University of
Massachusetts and the Infrared Processing and Analysis Center/California
Institute of Technology, funded by the National Aeronautics and Space Administration
and the National Science Foundation.

Based on observations obtained with MegaPrime/MegaCam, 
a joint project of CFHT and CEA/IRFU, at the Canada-France-Hawaii 
Telescope (CFHT) which is operated by the National Research 
Council (NRC) of Canada, the Institut National des 
Science de l'Univers of the Centre National de la Recherche 
Scientifique (CNRS) of France, and the University of Hawaii. 
This work is based in part on data products produced at 
Terapix available at the Canadian Astronomy Data Centre 
as part of the Canada-France-Hawaii Telescope Legacy Survey, 
a collaborative project of NRC and CNRS.

\end{document}